\documentclass[11pt,a4paper]{article}
\pdfoutput=1
\usepackage{lscape}
\usepackage{caption}
\usepackage{jheppub}
\usepackage[utf8]{inputenc}
\usepackage{soul,xcolor}
\usepackage{cancel}
\usepackage{makecell}
\usepackage{diagbox}
\usepackage{makecell}

\usepackage{graphicx}
\usepackage{tikz}
\usetikzlibrary{snakes}
\usepackage{physics}

\usepackage[normalem]{ulem}
\newsavebox{\foobox}

\usepackage{multirow}
\usepackage{dcolumn}
\newcommand\Tstrut{\rule{0pt}{2.6ex}}         
   
\usepackage{ulem}
\usepackage{graphicx}
\usepackage{comment}
\usepackage{bm,array}
\usepackage{graphics}
\usepackage{epsf}
\usepackage{color}
\usepackage{amsmath}
\usepackage{amssymb}
\usepackage{latexsym}
\usepackage{slashed}
\usepackage{float}
\usepackage{cases}
\usepackage{multirow}
\usepackage{framed}
\def\be{\begin{equation}}
\def\ee{\end{equation}}
\def\ba{\begin{array}}
\def\ea{\end{array}}
\usepackage{amsmath}
\usepackage{amssymb}
\usepackage[utf8]{inputenc}
\def\sQ3{\widetilde{Q}_3}
\def\sU3{\widetilde{U}_3}
\def\sD3{\widetilde{D}_3}

\def\hsm{h^0}

\def\d{\partial}

\def\mhsm{m_{h^0}}

\newcommand{\Huzr}{h_u}

\def\vev{{\it vev}}
\def\vevs{{\it vevs}}

\def\beq{\begin{equation}}
\def\eeq{\end{equation}}
\def\beqa{\begin{eqnarray}}
\def\eeqa{\end{eqnarray}}

\allowdisplaybreaks

\def\micromegas{{\tt micrOMEGAs}}
\def\feynrules{{\tt FeynRules}}

\def\pythia8{{\tt PYTHIA8}}

\def\z3nmssm{$Z_3$-NMSSM}
\newcommand{\bea}{\begin{eqnarray}}
\newcommand{\eea}{\end{eqnarray}}

\title{{Interplay among gravitational waves, dark matter and collider signals in the singlet scalar extended type-II seesaw model}}
\author[a]{Purusottam Ghosh,}
\author[b,c]{Tathagata Ghosh}
\author[b,c]{and Subhojit Roy}
\affiliation[a]{School Of Physical Sciences, Indian Association for the Cultivation of Science, 2A and 2B, Raja
S.C. Mullick Road, Kolkata 700032, India}
\affiliation[b]{Harish-Chandra Research Institute, A CI of Homi Bhabha National
Institute, Chhatnag Road, Jhunsi, Prayagraj (Allahabad) 211019, India}
\affiliation[c]{Regional Centre for Accelerator-based Particle Physics, Harish-Chandra Research Institute, \\ Prayagraj (Allahabad) 211019, India}
\emailAdd{pghoshiitg@gmail.com, tathagataghosh@hri.res.in, subhojitroy@hri.res.in}
\preprint{HRI-RECAPP-2022--013}
\abstract{We study the prospect of simultaneous explanation of tiny neutrino masses, dark matter (DM), and the observed baryon asymmetry of the Universe in a $Z_3$-symmetric complex singlet scalar extended type-II seesaw model.
The complex singlet scalar plays the role of DM.
Analyzing the thermal history of the model, we identify the region of the parameter space that can generate a first-order electroweak phase transition (FOEWPT) in the early Universe, and the resulting stochastic gravitational waves (GW) can be detected at future space/ground-based GW experiments.
First, we find that light triplet scalars do favor an FOEWPT. In our study, we choose the type-II seesaw part of the parameter space in such a way that light triplet scalars, especially the doubly charged ones, evade the strong bounds from their canonical searches at the Large Hadron Collider (LHC). However, the relevant part of the parameter space, where FOEWPT can happen only due to strong SM doublet-triplet interactions, is in tension with the SM-like Higgs decay to a pair of photons, which has already excluded the bulk of this parameter space. 
On the other hand, the latest spin-independent DM direct detection constraints from XENON-1T and PANDA-4T eliminate a significant amount of parameter space relevant for the dark sector assisted FOEWPT scenarios,
and it is only possible when the complex scalar DM is significantly underabundant.
In short, we conclude from our analysis that the absence of new physics at the HL-LHC and/or various DM experiments in the near future will severely limit the prospects of detecting a stochastic GW at
future GW experiments and will exclude the possibility of electroweak baryogenesis within this model.}
\begin{document}
\maketitle
%
\section{Introduction}
\label{Introduction}
The discovery of the Higgs Boson at the Large Hadron Collider (LHC) around 125 GeV~\cite{ATLAS:2012yve, CMS:2012qbp} was a breakthrough moment in our understanding of the laws of nature, completing the particle content of the Standard Model (SM). Despite the unparalleled successes of the SM, it cannot be a complete theory of nature as it fails to explain several phenomena in nature. Among many other issues, the SM cannot explain either the observed baryon asymmetry of the Universe (BAU) or tiny neutrino masses, in addition to not providing a good dark matter (DM) candidate.

In order to generate baryon asymmetry in any model, one needs to satisfy the three well-known Sakharov conditions~\cite{Sakharov:1967dj}.
One of the conditions, i.e., the out-of-equilibrium condition can be achieved 
if the electroweak phase transition (EWPT) in the SM had been a strong first-order one.
However, after the discovery of the 125 GeV Higgs, one can conclude that it is only a smooth crossover. Therefore, the SM cannot satisfy all the Sakharov's conditions. Going beyond the SM (BSM), with the addition of new scalars, one can modify the scalar potential so that a first-order electroweak phase transition (FOEWPT) can occur in the early Universe. Then one can generate the observed BAU via the electroweak baryogenesis (EWBG) mechanism~\cite{Trodden:1998ym, Anderson:1991zb, Huet:1995sh, Morrissey:2012db}.

On the other hand, several astrophysical and cosmological data, including rotation curves
of galaxies, the bullet cluster, gravitational lensing, and the anisotropy of the cosmic microwave background~\cite{Hu:2001bc}, provide strong motivation for the presence of DM in the Universe.
By analyzing the anisotropies in the CMB data~\cite{Hu:2001bc}, WMAP~\cite{WMAP:2006bqn} and PLANCK~\cite{Planck:2018vyg} experiments one can find that around a quarter of the energy density of the Universe is composed of non-baryonic, non-luminous matter with only known interaction through gravity.
Apart from its abundance, which is precisely measured by the PLANCK experiment to be $\Omega_{\rm DM}h^2 = 0.120\pm 0.001$~\cite{Planck:2018vyg}, the nature of DM and its non-gravitational interactions are still unknown to us. 


At the same time, the evidence of neutrino oscillations~\cite{Kajita:2016cak, McDonald:2016ixn} has opened up a new frontier in particle physics, requiring small neutrino masses, and consequently, offering another evidence for phenomena beyond the SM. Various seesaw mechanisms, such as type-I~\cite{Minkowski:1977sc,Yanagida:1979as,Mohapatra:1979ia}, type-II ~\cite{Konetschny:1977bn,Magg:1980ut,Lazarides:1980nt,Schechter:1980gr,Cheng:1980qt,Bilenky:1980cx} etc., are the most popular way to generate tiny neutrino masses. In this article, we focus on type-II seesaw only, where the SM scalar sector is extended by a $SU(2)$ triplet ($\Delta$) with hypercharge $Y=2$. Neutrinos obtain masses after the electroweak symmetry breaking (EWSB) when the neutral component of the triplet acquires an induced vacuum expectation value ($\vev$).

We extend the type-II seesaw model by an extra complex singlet ($S$) and further impose a discrete $Z_3$ symmetry on the model, due to which $S$ becomes  a stable DM candidate. The presence of the triplet and the complex singlet in the model and their interaction with the SM Higgs doublet can modify the Higgs potential in such a way that a strong first-order electroweak phase transition (SFOEWPT) is feasible in the early Universe. 
Hence, by proposing this economical extension of the SM, we attempt to alleviate three major shortcomings of it.


The non-observation of DM in its direct search experiments has severely constrained the simplest weakly interacting massive particle (WIMP)-like DM scenarios.
For instance, DM direct searches impose stringent limits on the parameter space of the simplest SM Higgs portal scalar DM~\cite{Casas:2017jjg,Bhattacharya:2017fid}.
However, the situation can drastically change in the presence of new DM annihilation channels, and one can evade those constraints. In this model, DM annihilation due to the presence of portal interaction of DM with the triplet sector and the $Z_3-$symmetry induced DM semi-annihilation channels allow a significant amount of permissible parameter space for DM that satisfies the observed DM relic density constraint from the PLANCK experiment~\cite{Planck:2018vyg}, the DM direct detection (DMDD) limits from XENON-1T~\cite{Aprile:2018dbl} and PANDAX-4T~\cite{PandaX-4T:2021bab}, and the DM indirect search bounds from FERMI-LAT~\cite{Fermi-LAT:2017bpc} and MAGIC~\cite{MAGIC:2016xys}.

A natural consequence of a first-order phase transition (FOPT) in the early Universe is the generation of relic gravitational waves (GW) via the nucleation of bubbles of the broken phase~\cite{Apreda:2001us, Grojean:2004xa, Weir:2017wfa, Alves:2018oct, Alves:2018jsw, Alves:2019igs, Alves:2020bpi, Chatterjee:2022pxf, Caprini:2019egz, Witten:1984rs,Hogan:1986qda, Ellis:2018mja, Alanne:2019bsm}. Such a GW signal can be detected in LISA~\cite{LISA:2017pwj}, or other proposed ground-based and space-borne experiments, viz.,
ALIA~\cite{Gong:2014mca}, Taiji~\cite{Hu:2017mde}, TianQin~\cite{TianQin:2015yph}, aLigo+~\cite{Harry:2010zz}, Big Bang Observer (BBO)~\cite{Corbin:2005ny} and  Ultimate(U)-DECIGO~\cite{Kudoh:2005as}, if the amplitude is high enough. These studies receive more attention after the  observation of GW from LIGO and VIRGO collaborations~\cite{LIGOScientific:2016aoc, LIGOScientific:2017vwq, LIGOScientific:2018mvr, LIGOScientific:2020ibl}. 
Note that, the first evidence of stochastic GWs has been announced very recently by NanoGrav~\cite{NANOGrav:2023gor} and EPTA~\cite{EPTA:2023fyk} collaborations.
This discovery provides a new opportunity to explore physics scenarios beyond the SM.
Certain regions of the parameter space of the present model can be probed via various future GW experiments due to the production of the GW as a result of an FOEWPT in the scalar sector.


In this paper, we study the interplay among the three sectors mentioned above, i.e. the production of GW resulting from FOEWPT, the DM prospect, and the LHC probes of this model. We perform a dedicated scan of the motivated region of the parameter space. Thereafter, we impose various theoretical constraints, such as boundedness of the scalar potential from below, electroweak vacuum stability at zero temperature, perturbativity, unitarity, as well as the relevant experimental constraints coming from the neutrino sector, flavor sector, electroweak precision study, dark sector and the heavy Higgs searches at the LHC. We allow minimal mixing between SM-like Higgs and the triplet-like neutral scalar to comply with the Higgs signal strength bounds. We further study the dependencies of the decay width of the SM Higgs into a photon pair on the model parameters.
We then discuss the overall phenomenology of a complex scalar DM in the presence of the scalar triplet and the SM doublet and the novel effect of $Z_3-$symmetry induced semi-annihilation processes. We split the study of the FOPT of this model into two parts.
First, we examine the parameter space, where an FOEWPT is facilitated by the triplet in conjunction with the SM doublet. Next, we study the feasibility of generating an FOEWPT with the help of the dark sector. Thereafter, we establish a correlation between the precise measurement of the observed Higgs boson properties with the detection prospect of the produced GW when the doublet-triplet interactions are strong. Then we analyze the connection between a large DM direct detection spin-independent (DMDDSI) cross-section and an FOEWPT due to significant interaction between DM and the SM Higgs doublet. To illustrate such correlations, we present a few benchmark points in this work and discuss the patterns of phase transitions in each case followed by thorough discussions on the detection prospect at various GW detectors, the signal-to-noise ratio (SNR) for LISA and possible complementing searches at the HL-LHC and future DM direct-detection experiments.


The paper is organized as follows: In section~\ref{sec:TFW}, we briefly discuss the theoretical framework of the model, including the interaction pieces of the Lagrangian, the WIMP DM production mechanism and its possible detection prospects, the study of EWPT and the production of GW from it. The relevant theoretical and experimental constraints on the model parameter space are outlined in section~\ref{constraints}. 
The motivated region of parameter space for this study has been discussed in section~\ref{sec:motivatedregion}. In section~\ref{results}, we present the results of this work. 
Finally, we summarize the outcome of our discussion in section~\ref{subsec:summary}. Various tadpole relations, field-dependent masses of scalar and fermionic degrees of freedom with thermal correlations (daisy coefficients),  various Feynman diagrams for DM annihilation and various constraints are presented in the appendix.
\section{The theoretical framework}
\label{sec:TFW}

\subsection{The Model}
\label{sec:model}
As discussed above, despite many successes of the SM, it cannot explain the smallness of neutrino masses, the observed matter-antimatter asymmetry and accommodate a particle DM candidate among other things. To address these three issues, we consider a model by introducing an additional discrete $Z_3$ symmetry and enlarging the particle content of the model by adding a $SU(2)_L$ scalar triplet, $\Delta$, with hypercharge, $Y=2$, and a complex scalar singlet, $S$, which transforms non-trivially under $Z_3$~\cite{Yang:2021buu}. The quantum numbers of the BSM scalars in our model along with the SM Higgs doublet under the extended gauge group $SU(3)_C\times SU(2)_L \times U(1)_Y \times Z_3$ are tabulated in Table \ref{tab:tab1}.

\begin{table}[h]
\resizebox{\linewidth}{!}{
 \begin{tabular}{|c|c|c|c|}
\hline \multicolumn{2}{|c}{Fields}&  \multicolumn{1}{|c|}{ $\underbrace{ SU(3)_C \otimes SU(2)_L \otimes U(1)_Y}$ $\otimes  Z_3  $} \\ \hline
Complex Scalar DM & $S=\frac{1}{\sqrt{2}}(h_s + i a_s)$ & ~~1 ~~~~~~~~~~~1~~~~~~~~~~~~0~~~~~~~~~$e^{i\frac{2\pi}{3}}$ \\
\hline
\hline
Scalar Triplet & $\Delta=\left(\begin{matrix} \frac{\Delta^+}{\sqrt{2}} & \Delta^{++} \\ \frac{1}{\sqrt{2}}\big(h_t + i a_t \big) & -\frac{\Delta^+}{\sqrt{2}} \end{matrix}\right)$ & ~~1 ~~~~~~~~~~~3~~~~~~~~~~2~~~~~~~~~~1 \\
\hline
Higgs doublet & $H=\left(\begin{matrix} G^+ \\ \frac{1}{\sqrt{2}}\big(h_d + i a_d \big) \end{matrix}\right)$ & ~~1 ~~~~~~~~~~~2~~~~~~~~~~1~~~~~~~~~~1 \\
\hline
\end{tabular}
}
\caption{ 
Charge assignments of the scalar fields in the model under the gauge group $\mathcal{G} \equiv \mathcal{G}_{\rm SM} \otimes Z_3$  where $\mathcal{G}_{\rm SM}\equiv SU(3)_C \otimes SU(2)_L \otimes U(1)_Y$. Hypercharge ($Y$) of the field is obtained by using the relation: $Q= I_3 +\frac{Y}{2}$, where 
  $I_3$ is the third component of isospin and $Q$ is the electromagnetic charge. }
    \label{tab:tab1}
\end{table}

The part of the Lagrangian relevant for our study in this paper is given below:
\bea
\label{totallag}
\mathcal{L} \supset \left(D^{\mu}H\right)^{\dagger}\left(D_{\mu}H\right) +  {\rm Tr} \left[\left(D^{\mu}\Delta\right)^{\dagger}\left(D_{\mu}\Delta\right)\right]-V(\Delta,H)+\mathcal{L}^{\rm Yukawa}+\mathcal{L}^{\rm DM} ,
\eea

where 
\bea
D_{\mu}H &=& \Big(\partial_\mu - i g_2 \frac{\sigma^a}{2} W_{\mu}^a-i g_1 \frac{Y_{H}}{2} B_{\mu}\Big)H , \nonumber \\
D_{\mu}\Delta &=& \partial_{\mu}\Delta - i g_2 \left[\frac{\sigma^{a}}{2} W_{\mu}^a ,\Delta\right] - {i g_1} \frac{Y_\Delta}{2} B_{\mu}\Delta ~.
\eea
The most general scalar potential involving the SM Higgs doublet and the scalar triplet can be written as~\cite{Arhrib:2011uy,Yang:2021buu}:
\bea
V(H,\Delta)&=& -\mu_{H}^2 (H^\dagger H)+ \lambda_{H} (H^\dagger H)^2 ~ \nonumber \\
&& + \mu_{\Delta}^2 {\rm Tr}\left[\Delta^{\dagger}\Delta\right]  + \lambda_{1} \left(H^{\dagger}H\right){\rm Tr}\left[\Delta^{\dagger}\Delta\right] + \lambda_2 \left({\rm Tr}[\Delta^{\dagger}\Delta]\right)^2 \nonumber \\
& & +\lambda_3 ~{\rm Tr} [\left(\Delta^{\dagger}\Delta\right)^2] + \lambda_4 ~\left(H^{\dagger}\Delta\Delta^{\dagger}H \right)+ \left[\mu\left(H^T i \sigma^2 \Delta^{\dagger}H\right)+h.c.\right]~~.
\label{scalpot}
\eea
Please note that in the above potential $\mu_{\Delta}^2 >0$ and the Higgs triplet cannot trigger any spontaneous symmetry breaking (SSB). So, similar to the SM, the EWSB happens when the doublet, $H$, develops a vacuum expectation value (VEV), 
$\langle {H} \rangle = {v_d}/{\sqrt{2}}$. However, the presence of the cubic term $H^T i \sigma^2 \Delta^{\dagger}H$ induces a non-vanishing VEV for the triplet, $\langle{\Delta}\rangle=v_t/\sqrt{2}$, after the EWSB.
 Then the scalar fields, $H$ and $\Delta$ can be represented as:
\bea
H=\left(\begin{matrix} G^+ \\ \frac{h_d+v_d+i a_d}{\sqrt{2}} \end{matrix}\right)~, ~~\Delta=\left(\begin{matrix} \frac{\Delta^+}{\sqrt{2}} & \Delta^{++} \\ \frac{h_t+v_t+i a_t}{\sqrt{2}} & -\frac{\Delta^+}{\sqrt{2}} \end{matrix}\right).
\eea
Note that $\sqrt{v_d^2+ 2 v_t^2} = v =246$ GeV and in the alignment limit $v_t << v_d$. Minimizing the scalar potential at the vacuums ($v_d$ and $v_t$), the bare masses of the scalar potential can be expressed in terms of other free parameters and can obtain the following conditions:
\bea
\mu_H^2 &=& -\frac{2 M_\Delta^2 v_t^2}{v_d^2}+\frac{1}{2} v_t^2 (\lambda _1+\lambda _4)+\lambda_H v_d^2  \nonumber \\
\mu_\Delta^2 &=& M_\Delta^2-v_t^2 (\lambda_2+\lambda_3)-\frac{1}{2} v_d^2 (\lambda_1+\lambda_4) ~~~~{\rm with}~~~M_{\Delta}^2=\frac{\mu  v_d^2}{\sqrt{2} v_t}.
\eea
\noindent {\bf{The field composition of $V(H,\Delta):$}}
The two CP even states ($h_d$ and $h_t$) are mixed up after the EWSB and give rise to two physical eigenstates ($h^0$, $H^0$) under the orthogonal transformation. The physical and interaction states are related as,
\bea
h^0 = h_d ~\cos\theta_t + h_t ~\sin\theta_t  ~, ~~~~
H^0 = -h_d~ \sin\theta_t + h_t ~\cos\theta_t
\eea
where $\theta_t$ is the mixing angle defined by
\bea
\label{thetat}
\tan2\theta_t &=& \frac{\sqrt{2}\mu v_d -(\lambda_1+\lambda_4)v_d ~v_t}{M_\Delta^2-\frac{1}{4}\lambda_H v_d^2 + (\lambda_2+\lambda_3) v_t^2 }~.
\eea
The corresponding mass eigenvalues of these physical eigenstates are given by
\bea
m_{h^0}^2&=& \left(M_\Delta^2+2 v_t^2 (\lambda_2+\lambda_3)\right) \sin ^2\theta_t + 2 \lambda_H v_d^2 \cos ^2\theta_t-\frac{v_t \sin2\theta_t \left(2 M_\Delta^2 - v_d^2 (\lambda_1+\lambda_4)\right)}{v_d}   \nonumber \\
m_{H^0}^2&=& \left(M_\Delta^2+2 v_t^2 (\lambda_2+\lambda_3)\right) \cos ^2\theta_t + 2 \lambda_H v_d^2 \sin^2\theta_t+\frac{v_t \sin2\theta_t \left(2 M_\Delta^2 - v_d^2 (\lambda_1+\lambda_4)\right)}{v_d} .\nonumber \\
\label{MCPE}
\eea

\noindent With the mass hierarchy $m_{h^0} < m_{H^0}$, the lighter state, $h^0$ acts like the SM higgs with mass, $m_{h^0}=125$ GeV. 

Similarly, the mixing between two CP odd states ($a_d$, $a_t$) leads to one massless Goldstone mode which is associated with the gauge boson $Z$ in the SM  and one massive pseudo scalar, $A^0$ and its mass ($m_{A^0}$) is given by,
\bea
m_{A^0}^2&=& \frac{M_\Delta^2 \left(4 v_t^2+v_d^2\right)}{v_d^2}~~ .
\label{MCPO}
\eea
In the singly charged scalar sector, the orthogonal rotation of $G^\pm$ and $\Delta^\pm$ fields lead to one massless Goldstone state absorbed by the longitudinal components of the SM gauge boson $W^\pm$ and one massive charged scalar eigenstate, $H^\pm$. The mass of $H^\pm$ is given by,
\bea
m_{H^\pm}^2 &=& \frac{\left(2 v_t^2+v_d^2\right) \left(4 M_\Delta^2 - \lambda_4 v_d^2\right)}{4 v_d^2}  .
\label{MSC}
\eea
The doubly charged scalar ($H^{\pm\pm} \equiv \Delta^{\pm\pm}$) mass is given by,
\bea
m_{H^{\pm\pm}}^2 &=& M_\Delta^2-\lambda_3 v_t^2-\frac{\lambda_4 v_d^2}{2}   ~~.
\label{MDC}
\eea
Therefore the scalar potential, $V(H,\Delta)$ contains seven massive physical states: two CP even states ($h^0,~H^0$), one CP odd state $(A^0)$, two singly charged scalar ($H^\pm$) and two doubly charged scalar ($H^{\pm\pm}$). The scalar potential, $V(H,\Delta)$ has seven independent free parameters as 
\bea
\{M_\Delta,~v_t,~\lambda_H (\simeq 0.129),~\lambda_{1},~\lambda_2,~\lambda_3,~\lambda_4 \}
\label{tripar}
\eea
whereas the physical masses and mixing angle can be represented in terms of free parameters using above equations. 

In the limiting case, $v_t^2/v_d^2 <<1$, some interesting mass relations are followed by:
\bea
\label{trplmassdiff}
\big(m_{H^{\pm\pm}}^2-m_{H^\pm}^2 \big) \approx -\frac{\lambda_4 v_d^2}{4}~; ~~ \big( m_{H^\pm}^2-m_{A^0}^2 \big) \approx -\frac{\lambda_4 v_d^2}{4} ~;~~
m_{H^0}  \approx m_{A^0} \approx M_\Delta .
\eea
Depending on the sign of the quartic coupling, $\lambda_4$, there are two types of mass hierarchy among different components of the triplet scalar:
\begin{itemize}
    \item when $\lambda_4$ is negative:  $m_{H^{\pm\pm}} > m_{H^\pm} > m_{H^0,A^0}$,
    \item when $\lambda_4$ is positive:  $m_{H^{\pm\pm}} < m_{H^\pm} < m_{H^0,A^0}$.
\end{itemize}
In this work, we particularly focus on the first scenario where $\lambda_4$ is negative. The mass difference between $H^{\pm\pm}$ and $H^\pm$ defined here by $\Delta m$ as: $\Delta m=m_{H^{\pm\pm}}-m_{H^\pm}$. \\~\\
\noindent{\bf Neutrino masses:}
The Yukawa Lagrangian involving the SM lepton doublets ($L$) and scalar triplet ($\Delta$) to generate neutrino masses can be written as~\cite{FileviezPerez:2008jbu},
\begin{equation}
    -\mathcal{L}^{\rm Yukawa} = (y_L)_{\alpha\beta} ~\overline{L_\alpha^c}~i \sigma^2 \Delta ~L_\beta + h.c.
\end{equation}
where $y_L$ is a $3 \times 3$ symmetric complex matrix. The heavy triplet scalar helps to
generate neutrino masses via the familiar type-II seesaw mechanism. Neutrino masses are generated once the neutral triplet scalar gets induced  non-zero \textit{vev} $v_t$. The light neutrino mass in the flavor basis is given by~\cite{FileviezPerez:2008jbu}  
\begin{equation}
    (m_{\nu})_{\alpha \beta} =  \sqrt{2}(y_L)_{\alpha \beta}{v_t},
\end{equation}
where $\{\alpha,\beta\}$ be the flavour indices and $m_{\nu}$ be the $3 \times 3$ light neutrino mass matrix. In order to obtain the physical neutrino masses, the mass matrix, $ m_{\nu}$ can be diagonalised using the Pontecorvo-Maki-Nakagawa-Sakata (PMNS) matrix, $U_{\text{PMNS}}$, as
\begin{equation}
    U_{\text{PMNS}}^T~ m_\nu~ U_{\text{PMNS}} = \text{diag}(m_1, m_2, m_3),
\end{equation}
where \begin{equation}
    U_{\text{PMNS}} = \begin{pmatrix}
    c_{12}c_{13} & s_{12}c_{13} & s_{13}e^{-i\delta} \\
    -c_{23}s_{12}-s_{23}s_{13}c_{12}e^{i\delta} & 
c_{23}c_{12}-s_{23}s_{13}s_{12}e^{i\delta} & s_{23}c_{13} \\
    s_{23}s_{12}-c_{23}s_{13}c_{12}e^{i\delta} & 
-s_{23}c_{12}-c_{23}s_{13}s_{12}e^{i\delta} & c_{23}c_{13}
    \end{pmatrix}
    \begin{pmatrix}
    1 & 0 & 0 \\
    0 & e^{i\alpha_2} & 0 \\
    0 & 0 & e^{i \alpha_3}
    \end{pmatrix}.
    \label{eq:PMNS}
\end{equation}
$m_i~(i=1,2,3)$ is the eigenvalue of the light neutrino mass eigenstate $\nu_i$.
Here $c_{ij}=\cos\theta_{ij}$ and $s_{ij}=\sin\theta_{ij}$ (with $i,j = 1, 2, 3$ and $i \neq j$) are three mixing angles (with $0 \leq \theta_{ij} \leq \pi/2$) in neutrino sector and $\delta$ denotes the Dirac CP-phase (with $0 \leq \delta \leq 2\pi$) responsible for CP violation in neutrino oscillations. Here $\alpha_{2,3}$ be the Majorana phases confined to the range $[0,\pi]$. The parameters of neutrino oscillations are related to the mass-squared differences defined as $\Delta m_{sol}^2 \equiv \Delta m_{21}^2 = m_2^2 - m_1^2$,  $\Delta m_{atm}^2 \equiv \Delta m_{31}^2 = m_3^2 - m_1^2$ for $m_3> m_2 >m_1$ (Normal Hierarchy (NH)) and $\Delta m_{atm}^2 \equiv \Delta m_{32}^2 = m_3^2 - m_2^2$ for $m_2> m_1 >m_3$ (Inverted Hierarchy (IH)) as well as with the three mixing angles ($\theta_{ij}$).~\\

\noindent {\bf{The DM Lagrangian:}}
The $Z_3$ symmetry invariant Lagrangian for the scalar singlet, $S$ which act as a stable DM candidate, can be written as~\cite{Yang:2021buu}
\bea
\mathcal{L}^{\rm DM}=\frac{1}{2}|\partial_{\mu}S|^2-V^{\rm DM} , \nonumber
\eea
where $V^{\rm DM}$ is the scalar potential involving the DM, $S$. The scalar potential reads
\bea
V^{\rm DM}&=&  \mu_{S}^2 ~ |S|^2 + \lambda_{S} |S|^4 + \frac{\mu_3}{3!} \Big(S^3 + {S^*}^3 \Big)\nonumber \\
&&+ \lambda_{S H} H^\dagger H  (S^* S) 
+ \lambda_{S \Delta}~ {\rm Tr}\left[\Delta^{\dagger}\Delta\right]\big(S^* S \big) ~~~~.
\eea
\noindent 
The complex scalar field, $S$  can be a stable DM particle under the so-called freeze-out mechanism. After EWSB the mass of the DM turns out to be:
\bea
\label{DMmassT0}
m_S^2= \mu_S^2 + \frac{1}{2} \lambda_{S H} v_d^2 + \frac{1}{2} \lambda_{S \Delta} v_t^2 \equiv m_{\rm DM}^2  ~.
\eea
The free parameters involved in the dark sector are following 
\bea
\{ ~ m_{S} (\equiv m_{\rm DM}),~ \mu_3,~  \lambda_{S H},~\lambda_{S \Delta},~\lambda_{S}\} .
\eea
In the following subsection, we examine in detail how the aforementioned free parameters play a role in the phenomenology of DM.
\subsection{The Dark Matter}
\label{darkmatter}
In this subsection, we discuss the phenomenology of the scalar DM of this model in the presence of the scalar triplet and the SM scalar doublet.
The singlet-like DM interacts with the doublet ($H$) and the triplet ($\Delta$) sectors of the model through the scalar portal interactions $\lambda_{S H}(S^* S)(H^\dagger H)$  and $\lambda_{S \Delta}(S^* S){\rm Tr}[\Delta^\dagger \Delta]$, respectively.
These interactions are essential for the DM annihilation in the early Universe.
At the same time, the cubic term ( $\mu_3(S^3 + {S^*}^3)$) in the dark scalar potential has a novel feature of DM semi-annihilation.
At the early time of the Universe, DM is in thermal equilibrium with the bath particles
via the portal interactions as well as the cubic interactions in the dark sector. 
During that period of time, the interaction rate between DM and thermal bath particles (TBPs) ($\Gamma_{\rm DM - TBP}=n_S \langle {\sigma v}_{S S \to {\rm TBPs}} \rangle$) dominates over the Hubble expansion rate ($\mathcal{H}$). 
With the expansion of the Universe, DM freezes out from the thermal bath when  $\Gamma_{\rm DM -TBP} < \mathcal{H}$ yielding the DM density of the current Universe. 
This phenomenon is often referred to as the WIMP miracle~\cite{Kolb:1990vq}.
In this setup, the number density of DM is mainly governed by several number changing processes which are classified  as: 
\begin{align*}
 {\rm DM ~annihilation}: \hspace{0.3cm} & (i)~ S~ S^* ~\to {\rm SM~~SM} ~~ \\
  & (ii)~ S~ S^* ~\to {\rm X~~Y}; ~~~\{\rm X,Y\}=\{H^0,A^0,H^\pm, H^{\pm\pm}\} \\
 {\rm DM ~semi-annihilation}:  \hspace{0.3cm}&~~~~~S~ S ~\to  S~\xi; ~~~~~~~\{\xi\}=\{h^0,H^0\}  ~,
\end{align*}
where the Feynman diagrams of these  number changing processes of DM are shown in figures ~\ref{Feyn_diag1}, \ref{Feyn_diag2} and \ref{Feyn_diag3}, respectively, in appendix~\ref{feynmandiag}. The evolution of DM number density ($n_S$) in terms of more convenient variables, such as the co-moving number density ($Y_S=\frac{{n_S}}{s}$) and dimensionless parameter $x=\frac{m_S}{T}$, is described by the Boltzmann equation(BEQ) in the following form~\cite{Hektor:2019ote,Yang:2021buu}:
\begin{eqnarray}
\label{comving}
\frac{dY_{S}}{dx}&=&- \frac{ s ~\langle \sigma v \rangle_{S~S\to SM~SM}}{ \mathcal{H} ~x}(Y_{S}^2-{Y_{S}^{eq}}^2) ~\Theta\Big( m_S - m_{SM} \Big) \nonumber \\
&& -\frac{s}{\mathcal{H}~x}\sum_{X,Y}\langle \sigma v \rangle_{S~S\to X~ Y}(Y_{S}^2-{Y_{X}^{eq}} {Y_{Y}^{eq}}) ~\Theta\Big(2 m_S -m_X -m _Y \Big)\nonumber \\
&& -  \frac{1}{2}  \frac{ s ~\langle \sigma v \rangle_{S~S\to S SM}}{ \mathcal{H} ~x}(Y_{S}^2-Y_{S}Y_{S}^{eq})~\Theta\Big( m_S -m_{SM}\Big)~.
\end{eqnarray}
Depending on the mass hierarchies among the DM and various scalars of this model, different number changing processes start to contribute to DM number density. 
This fact has been illustrated in the BEQ using the theta function ($\Theta$).  The co-moving equilibrium  density of $i$th particle is defined as: $Y_i^{eq}=\frac{45}{4 \pi^4}\frac{g_i}{g_{*s}}\Big( \frac{m_i}{T}\Big)^2  K_2[\frac{m_i}{T}]$
where $g_i$ is the degree of freedom of the $i-$th species and
$g_{*s}$ is the internal relativistic degree of freedom associated with entropy. $K_2$ is the modified Bessel function of the second kind.
The entropy density and Hubble expansion rates are defined as: $s=g_{*s} \frac{2 \pi^2}{45}T^3$ and $\mathcal{H}=\sqrt{\frac{\pi^2 g_*}{90}}\frac{1}{M_{\rm Pl}} T^2$, respectively with $g_*=106.7$ and $M_{\rm Pl}=2.4\times 10^{18}$ GeV. $\langle \sigma v \rangle_{S S \to i j}$ is the thermal average cross-section of the corresponding number changing process: $S S \to i j$~\cite{Kolb:1990vq}.
Solving the above BEQ, today's DM relic density is given by~\cite{Kolb:1990vq},
\bea
\label{relic}
\Omega_{S}h^2&=& 2.752 \times 10^{8} ~\Big( \frac{m_S}{\rm GeV}\Big) ~ Y_S(x\to \infty) .
\eea

The  DM abundance ($\Omega_S h^2$) followed from equations~\ref{comving} and \ref{relic} depends on the following independent free parameters which are entered in thermal average cross sections:
\begin{align}
 &\{m_{S},~\mu_3,~\lambda_{S H},~\lambda_{S \Delta}\}~~~~~~~~~~{\rm from ~dark~ sector}, \nonumber \\
 &\{~\lambda_1,~\lambda_2,~\lambda_3,~\lambda_4~\}~~~~~~~~~~~~~~~{\rm from ~scalar ~triplet~ sector.} \label{eq:DMpar}
\end{align}
 The dark sector self coupling, $\lambda_S$ does not affect the computation of the relic density.
We will discuss the role of the above parameters on DM density in section~\ref{DMpar}. Note that, the present day DM relic density is constrained from the Planck observation~\cite{Planck:2018vyg}  which will be discussed in the subsection~\ref{subsec:DMconstraints}.

 \begin{figure}[htb!]
   \begin{center}
    \begin{tikzpicture}[line width=0.6 pt, scale=2.1]
\draw[dashed] (-1.8,1.0)--(-0.8,0.5);
\draw[solid] (-1.8,-1.0)--(-0.8,-0.5);
\draw[dashed] (-0.8,0.5)--(-0.8,-0.5);
\draw[dashed] (-0.8,0.5)--(0.2,1.0);
\draw[solid] (-0.8,-0.5)--(0.2,-1.0);
\node at (-2.1,1.1) {$S$};
\node at (-2.1,-1.1) {$n$};
\node at (-0.5,0.07) {$h^0,H^0$};
\node at (0.5,1.2) {$S$};
\node at (0.5,-1.2) {$n$};
     \end{tikzpicture}
 \end{center}
\caption{Feynman diagram for spin-independent DM-nucleon scattering process for the scalar DM. }
\label{DD}
 \end{figure}
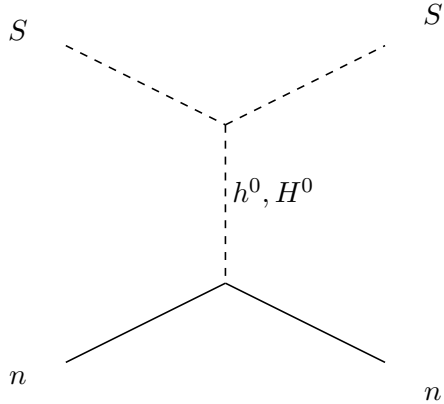

There are several attempts have been made to detect DM in laboratory experiments. DMDD is one of them where incoming  DM flux scatters with the nuclei in the target crystals and the recoil of the target nuclei can be searched for a DM signal.
The effective spin-independent DM-Nucleon scattering cross-section ($\sigma_{\rm DD}^{\rm SI}$) mediated via $t-$channel $CP-$even scalars ($h^0$ and $H^0$) exchange diagrams, as shown in figure~\ref{DD}, read as ($q \to 0$):
\begin{equation}
\sigma_{\rm DD}^{\rm SI} =\Big( \frac{\Omega_S h^2}{\Omega_{\rm DM}h^2}\Big)~\Big(\frac{1}{4\pi} \bigg(\frac{f_{n} \mu_{n}}{m_{S}}\bigg)^{2}\bigg(\frac{m_{n}}{v_{d}}\bigg)^{2}
~\bigg[\frac{\lambda_{D1}}{m_{h^0}^{2}}+\frac{\lambda_{D2}}{m_{H^0}^{2}}\bigg]^{2} \Big),
\label{eq:sigmadd}
\end{equation}
where $\Big( {\Omega_S h^2}/{\Omega_{\rm DM}h^2}\Big) \equiv \xi_{\rm DM}$ is the fractional DM density. The coupling $\lambda_{D1}$  and $\lambda_{D2}$ are defined as
\bea
\lambda_{D1}&=&  \Big(~~\lambda_{S H} ~v_d \cos\theta_t + \lambda_{S \Delta} v_t \sin\theta_t \Big)\cos\theta_t ~  \xrightarrow{\theta_t \to 0}~ \simeq \lambda_{S H} ~v_d \nonumber \\
\lambda_{D2}&=& \Big(-\lambda_{S H} ~v_d \sin\theta_t + \lambda_{S \Delta} v_t \cos\theta_t \Big)\sin\theta_t  \xrightarrow{\theta_t \to 0} ~\simeq 0
\eea
The reduced mass of DM-nucleon system is defined by $\mu_n$  $\big(=\frac{m_S m_n}{m_S+m_n}\big)$ with $m_n=0.946$ GeV and the nucleon form factor $f_n=0.28$~\cite{Alarcon:2012nr}. For the limiting case: $\theta_t \to 0$, $\sigma_{\rm DD}^{\rm SI}$ solely depends on the the Higgs portal DM coupling, $\lambda_{SH}$ and the DM mass, $m_S$. 
The null results from the various DM direct search experiments such as XENON-1T~\cite{Aprile:2018dbl} and PANDA-4T~\cite{PandaX-4T:2021bab} put a strong constraint on the DMDD cross-sections which can be further expressed in terms of the model parameters.

At the same time, DM can also be detected at various indirect search experiments such as the space-based telescopes, Fermi Large Area Telescope (Fermi-LAT)~\cite{Fermi-LAT:2017bpc} and the ground-based
telescopes, Major
Atmospheric Gamma-ray Imaging Cherenkov (MAGIC)~\cite{MAGIC:2016xys}. These experiments are looking for the gamma-ray flux which is produced via the production of the SM particles
either through DM annihilation or via decay in the local Universe. The photons which are emitted from WIMP-like DM lie in the gamma-ray regime and behave as ideal messengers of DM indirect detection (DMID). The total gamma-ray flux due to DM annihilation into the SM charge pairs($\psi \overline{\psi}$) in a specific energy range is given by~\cite{MAGIC:2016xys}
\begin{eqnarray}
\label{eq:id}
\Phi_{\psi \overline{\psi}}&=&\frac{1}{4\pi}  \frac{\langle \sigma v \rangle _{SS \to \psi \overline{\psi}}}{2 m_{S}^2} \int_{E_{\rm min}}^{E_{\rm max}} \frac{dN_\gamma}{dE_\gamma} dE_{\gamma} \underbrace{\int dx ~\rho_S^2\big(r(b,l,x)\big)}_{J} ~,
\end{eqnarray}
where $\psi:=\{\mu^-,\tau^-,b,W^-\}$. The $J$ factor contains the astrophysical information about the DM distribution in the
galaxy and $\frac{dN_\gamma}{dE_\gamma}$ is the energy spectrum of incoming photons from DM annihilation. Measuring the gamma-ray flux and using the standard astrophysical inputs, the indirect search experiments put limits on the thermal average cross-section of  DM annihilation into different SM charge pairs like $\mu^-\mu^+,~\tau^-\tau^+,~b\overline{b},~W^-W^+$.
In order to compare the experimental bound with the theoretically estimated thermal average cross-section of the corresponding DM annihilation channel, one can express the above equation~\ref{eq:id}~as:
\begin{eqnarray}
\langle \sigma v \rangle_{SS \to \psi\overline{\psi}}^{\rm eff}&=& \big(\Omega_S/\Omega_{\rm DM} \big)^2\langle \sigma v \rangle_{SS \to \psi\overline{\psi}}= \xi_{\rm DM}^2  \langle \sigma v \rangle_{SS \to \psi\overline{\psi}} ~~.
\end{eqnarray}
Here the thermal average cross-sections for DMID are scaled with the fractional DM density as $\xi_{\rm DM}^2$. In this setup the thermal average annihilation cross-sections,$\langle \sigma v \rangle_{SS \to \psi\overline{\psi}}$  are mostly dominated by $s$-channel SM Higgs ($h^0$) exchange diagram and varies as $\propto \lambda_{SH}^2/m_S^2$ (for  $\sin\theta_t \to 0$). Non-observation of any DM signal in such indirect detection experiments puts a constraint on the thermal average cross-section of DM annihilation into different SM charge pairs. Those constraints are further expressed in terms of the model parameters.

So far, we have talked about the analytical form of DM abundance, DMDDSI cross-section and DMID cross-section which will help to understand the numerical results. For the numerical analysis, we have used publicly available numerical packages. We first implemented the model in $\feynrules$~\cite{Alloul:2013bka} and then
the outputs are fed into $\micromegas$~\cite{Belanger:2006is}  to obtain DM abundance, $\Omega_{S} h^2$ and DMDDSI and DMID cross-sections.

\subsection{Study of EWPT and the production of GW}
\label{subsec:ewpt-nmssm}

In this section, we discuss the theoretical formulation of EWPT in our model and examine the possibility of the EWPT being a strong first-order one, which in turn may facilitate EWBG. We also outline the production of stochastic GW from FOPT.

Note that in the present study of EWPT, we assume that no spontaneous or explicit \cancel{$CP$} occurs in the Higgs sector. However, one can incorporate a $Z_3$ symmetric $CP$-violating dimension-7 operator, $y_t \, \overline{Q} \Tilde{H} \, \big( 1 + i c \frac{S^3}{\Lambda^3}\big) \, t_R \, + \, h.c.$, in this model to generate $CP$- violation which is essential for EWBG to take place. 
We do not perform a detailed  calculation of EWBG in this paper since  the focus of the work is to study the correlation/interplay
of the prospects of detecting stochastic GW from an SFOEWPT in an almost unconstrained parameter
space at various future proposed GW experiments, the collider signals of this parameter space, and the prospects of detecting DM at various DM direct detection (DMDD) experiments. {However, we should mention that the detailed baryogenesis calculation for an equivalent $Z_2$-symmetric $CP$-violating dimension-6 operator has been performed in the literature~\cite{Vaskonen:2016yiu, Ellis:2022lft}. One can follow the same prescription for our $Z_3$ symmetric $CP$-violating dimension-7 operator given above.} We leave such an EWBG calculation for a future study.
%
\subsubsection{Effective Higgs potential at finite-temperature}
\label{subsec:effpot}
The tree-level scalar potential of the model (see equation~\ref{scalpot}) relevant for the study of EWPT, in terms of the $CP$-even Higgs fields \{$h_d$, $h_t$ and $h_s$\}, can be written as 
\begin{align}
\label{Vtree}
V_0(h_d, h_t, h_s) = &\, \frac{1}{4}( \lambda_H h_d^4 + (\lambda_2 + \lambda_3) h_t^4  + \lambda_S h_s^4 + (\lambda_1 + \lambda_4)h_d^2 h_t^2 +\lambda_{SH} h_d^2 h_s^2 + \lambda_{S \Delta} h_t^2 h_s^2) \nonumber\\
& + \frac{\mu_3}{6\sqrt{2}} h_s^3 - \frac{\mu}{\sqrt{2}} h_d^2 h_t  - \frac{\mu_H^2}{2}h_d^2 + \frac{\mu_{\Delta}^2}{2}h_t^2 +\frac{\mu_S^2}{2}h_s^2 .
\end{align}
The vacuum expectation values ($\vevs$) of the scalar fields are assumed to be real at all temperatures. As we have discussed earlier that at $T=0$, the minimum of the scalar potential is $\langle h_d \rangle \equiv v_d$, $\langle h_t \rangle \equiv v_{t}$ and  $\langle h_s \rangle \equiv 0$ where $v = \sqrt{v_d^2 + v_{t}^2} = 246$~GeV.
The zero temperature tree-level potential (as described in equation~\ref{Vtree}) gets quantum corrections from all fields that couple to $h_d$, $h_t$ and $h_s$. These corrections can be written in terms of the well-known Coleman-Weinberg (CW)~\cite{Coleman:1973jx} one-loop potential in the following form:
\begin{align}
\label{VCW}
V_{\rm CW}(h_d, h_t, h_s, T=0) = \sum_{i} (-1)^{2s_i} n_i\frac{m_i^4(h_d, h_t, h_s)}{64\pi^2}\left[\ln \frac{m_i^2 (h_d, h_t, h_s,)}{Q^2}-C_i\right] \; \ ,
\end{align}
where `$Q$' is the renormalization scale which we set at $Q = 246$ GeV . The sum `$i$' runs over all particles in the model. The constants $C_i$ depend on the choice of the renormalization scheme. In this work, we consider $\overline{\text{MS}}$ on-shell scheme, whereby for the transverse polarizations of the gauge bosons $C_i = \frac{1}{2}$ and for their longitudinal polarizations and for all other particles (scalars and fermions) $C_i = \frac{3}{2}$, i.e., $C_{W^\pm,Z}=5/6$ and for other species $C_{i}=3/2$. Here, $m_i$ is the field-dependent mass of the $i$-th species (see Appendix \ref{field-dependent-masses} for the details), and $n_i$ and $s_i$ are their associated number of degrees of freedom and spin, respectively. The number of degrees of freedom of all the fields in the model relevant for equation~\ref{VCW} is given below:
\begin{equation}
    n_{h_d,h_t,h_s,a_d,a_t,a_s,G^0}=1, \, \, n_{H_{\Delta}^\pm,H_{\Delta}^{\pm\pm},G^\pm} = 2, \, \, n_{Z} = 3, \, \, n_{W^\pm} = 6, \, \, n_{t}=12 \, \,.
\end{equation} 
In this work, we work in the Landau gauge as the ghost bosons decouple in this gauge and we do not need
 to consider them in our calculation.

Note that the location of the electroweak minimum in the field space for the tree-level potential is shifted by the one-loop CW potential.
Thus, suitable counter-terms need to add to the effective potential to ensure that the minimum of the effective potential coincides
with that of the tree-level potential in the $\overline{\text{MS}}$ renormalization scheme. The added counter-terms to the potential are parameterized as,
\begin{align}
\label{VCT}
V_{\rm CT}= \delta_{\mu_H^2} h_d^2+\delta_{\mu_{\Delta}^2} h_t^2 + \delta_{\lambda_H} h_d^4 +\delta_{\lambda_2} h_t^4 + \delta_{\lambda_{S\Delta}} h_d^2 h_t^2 + \delta_{\lambda_{SH}} h_d^2 h_s^2\; .
\end{align}
Various coefficients in $V_{\rm CT}$ (of equation~\ref{VCT}) can be found from the on-shell renormalization conditions at zero temperature:
\begin{equation}
\frac{\partial (V_\text{CW} + V_\text{CT})}{\partial \phi_{_i}} =0 \, \, \, , \, \, \,\\
\frac{\partial^2 (V_\text{CW} + V_\text{CT})}{\partial \phi_{_i} \partial \phi_{_i}} =0 \, \, \, ,
\end{equation}
where $\phi_{_i}, \phi_{_j} = \{h_d, h_t, h_s\}$. All the derivatives are taken at the true EW minima, i.e., $h_d = v_d$, $h_t = v_t$ and $h_s =0. $
The various coefficients of the counter-term potentials are presented in the appendix~\ref{CTcoeff}.
Note that, in the Landau gauge that we opt for this work, the masses of the Goldstone modes vanish at the physical minimum. As a result, when physical masses and coupling coefficients are estimated from derivatives of the loop-corrected effective potential, it leads to divergences~\cite{Martin:2014bca,Elias-Miro:2014pca} and it can be handled by using an infrared regulator by changing the masses of the Goldstone modes, $m_G^2 \to m_G^2 + \mu_{\rm IR}^2$. In this work, we adopt the approach which is considered in reference~\cite{Baum:2020vfl, Chatterjee:2022pxf} i.e., for numerical calculation it is sufficient to set $\mu_{\rm IR}^2 = 1\,{\rm GeV}^2$.

At finite-temperatures, 
the effective potential receives additional corrections and that at the one-loop level 
are given by~\cite{Dolan:1973qd, Weinberg:1974hy, Kirzhnits:1976ts} 
\begin{align}
\label{Vthermal}
 V_{\rm th}(h_d, h_t, h_s, T) = \frac{T^4}{2\pi^2}\, \sum_i n_i J_{B,F}\left( \frac{ m_i^2(h_d, h_t, h_s)}{T^2}\right)\;,
\end{align}
where $n_i$ are the numbers of degrees of freedom for particles as discussed earlier. The thermal functions $J_{B(F)}$ are for bosonic (fermionic) particles defined as
\beq
J_{B/F}(y^2)=
\pm {\rm Re}
\int_0^{\infty}
		x^2 \ln
		\left(
			1 \mp \exp^{-\sqrt{x^2 + y^2}}
		\right)
{\rm d}{x} \, ,
\label{eq:jbjf}
\eeq
where the lower (upper) sign corresponds to fermions (bosons), $\beta \equiv 1/\text{T}$ and the sum includes all the particles as described in equation~\ref{VCW}.

Note that, at high temperature the treatment of the perturbative expansion of the effective potential no longer remain valid. The quadratically divergent contributions from the so-called non-zero Matsubara modes need to be re-summed through the 
insertion of thermal masses in the one-loop propagator. The corrections are only to the Matsubara zero-modes, i.e., to the longitudinal polarization states of the vector bosons and to all the scalars. Since the thermal contributions 
to the transverse modes are suppressed due to gauge
symmetry~\cite{Espinosa:1992kf}. To make the expansion reliable, we adopt the well-known Parwani method~\cite{Parwani:1991gq} and we denote the thermally improved field-dependent masses as $M_i^2$, where $M_i^2 = \text{eigenvalues}[m_i^2 + \Pi(T^2)]$ where $\Pi(T^2) = c_{ij} T^2$ and $c_{ij}$'s are the so-called daisy coefficients.  Note that the gauge symmetries
plus the discrete $Z_3$-symmetry of the present model set the off-diagonal terms of the $\Pi(T^2)$ matrix
to zero. Thus, the Daisy coefficients are listed in equations~\ref{eq:daisy-coeff}.\\
Including the Coleman-Weinberg and the thermal corrections, the temperature-dependent
effective potential at one-loop order is given by
\beq
\label{Vtot}
V_\text{eff}(T)= V_0 + V_{\rm CW}(M_i^2) + V_{\rm CT} + V_{\rm th}(M_i^2, T) \,.
\eeq
The EWPT may now be investigated using this potential, with the minimum of the potential being tracked as a function of temperature. Note that the total one-loop effective potential (equation~\ref{Vtot}) carries explicit gauge dependence. Since, the important quantities of the study of phase transitions such as the locations of the extrema of $V_\text{eff}(T)$, as well as the ratio
$\phi_c(T_c)/T_c$, both are gauge-dependent~\cite{Dolan:1973qd,Nielsen:1975fs,Fukuda:1975di,Laine:1994zq,Baacke:1993aj,Baacke:1994ix, Garny:2012cg, Espinosa:2016nld, Patel:2011th, Arunasalam:2021zrs, Lofgren:2021ogg}.~\footnote{
Using the Nielsen identities~\cite{Nielsen:1975fs,Fukuda:1975di}, we can get gauge-independent variables of the effective potential.} 
In addition, the one-loop effective potential of equation~\ref{Vtot} explicitly depends on the choice of the renormalization scale ($Q$), which might have a more significant impact than the gauge uncertainty~\cite{Laine:2017hdk}.

The production of the stochastic GW from FOPTs will be discussed in the next subsection.
Later, we shall use this to identify a suitable region of parameter space of the model that can be investigated using various future GW experiments. 

\subsubsection{Production of stochastic  GW from FOPT}\label{GW_section}

In this $Z_3$-symmetric singlet scalar extended type-II seesaw model, multi-step FOPTs may occur in the early Universe, potentially generating a stochastic background of GW. The amplitude of this stochastic background, in contrast to GW from a binary system, is a random quantity that is unpolarized, isotropic and has a Gaussian distribution~\cite{Allen:1997ad}. As a result, the two-point correlation function can characterize this, and it is proportional to the power spectral density $\Omega_{\text{GW}}{\rm h}^2$. The ``cross-correlation'' method between two or more detectors can be used to detect this kind of stochastic GW~\cite{Caprini:2015zlo, Cai:2017cbj,Caprini:2018mtu, Romano:2016dpx, Christensen:2018iqi}.

To determine the GW signals arising from FOPTs and evaluate their evolution with temperature we require the knowledge of the following set of portal
parameters: $T_n, \alpha, \beta/H_n, v_w$. Here, the bubble nucleation temperature, $T_n$, is the approximate time at which one bubble per Hubble volume forms due to a phase transition. A dimensionless quantity, `$\alpha$', is defined as the ratio of the energy released from the phase transition to the total radiation energy density at the time when the phase transition is complete. Thus, it relates to the energy budget of the FOPT and is given by~\cite{Espinosa:2010hh}
\beq
\alpha = \frac{\rho_{\text{vac}}}{\rho^*_{\text{rad}}} = \frac{1}{\rho^*_{\text{rad}}}\left[T\frac{\d \Delta V(T)}{\d T} - \Delta V(T)\right]\Bigg|_{T_*},
\eeq
where $T_*=T|_{t_*}$. $t_*$ is the time when the FOPT completes. In the absence of substantial reheating effects, $T_* \simeq T_n$. The total radiation energy density of the plasma background is $\rho^*_{\text{rad}} = g_* \pi^2 T_*^4/30$, where $g_*$ is the number of relativistic degrees of freedom at $T=T_*$. In this work, we consider $g_* \sim 100$. The difference between the potential energies at the false minimum and the true minimum is denoted by $\Delta V(T) = V_{\text{false}}(T)-V_{\text{true}}(T)$.
The other two parameters are the parameter $\beta$, which roughly indicates the inverse time duration of the phase transition, and the parameter $v_w$, which is the bubble wall velocity.

The tunnelling probability from the false vacuum to the true vacuum per unit time per unit volume is given by~\cite{Turner:1992tz}
\beq
\Gamma(T)\simeq T^4 \left( \frac{S_3\left(T \right)}{2\pi T} \right)^{3/2}e^{-S_3\left(T \right)/T},
\eeq
where the Euclidean action $S_3\left(T \right)$ of the background field ($\phi$), in the spherical polar coordinate, is given by
\begin{equation}\label{eq:64}
S_3\left(T \right)=4\pi \int dr \hspace{1mm}r^{2} \left[ \dfrac{1}{2} \left(\partial_{r} \vec{\phi} \right)^2 +V_{eff}\right]\,\,,
\end{equation}
with $V_{eff}$ being the finite temperature total effective potential defined in equation~\ref{Vtot}, and $\vec{\phi}$ denoting the three components $\vev$ of the scalar fields.
The condition for the formation of a bubble of critical size can be found by extremizing this Euclidean action. 
For this study,  we use the publicly available toolbox $\mathbf{CosmoTransitions}$~\cite{Wainwright:2011kj}
to find bounce solutions of the above Euclidean action. It uses a path deformation method and this computation is the most technically challenging part. The bubble nucleation rate ($\Gamma$) pet unit volume at temperature T is given by $\frac{\Gamma(T)}{V} \propto T^4 e^{-S_3/T}$. The nucleation temperature is usually determined by solving
the following equation of the nucleation rate,
\beq
\int_{T_n}^{\infty}\frac{dT}{T}\frac{\Gamma(T)}{H(T)^4} \simeq 1. 
\eeq
This equation states that the nucleation probability of a single bubble within one horizon volume is approximately 1,
and it reduces to the condition $S_3(T)/T \approx 140$, solving which one can obtain the nucleation temperature $T_n$~\cite{Apreda:2001us}. This is the highest temperature for which $S_3/ T \lesssim 140$. For all $T >0$, if $S_3/T > 140$, the corresponding transition to the true vacuum or nucleation of bubbles of critical size of the broken phase does not happen. This occurs because either the barrier between the two local minima is too high or the distance between them is too large in the field space, resulting in a low tunnelling probability. For $T < T_c$, if this condition is not fulfilled, the system would remain trapped at the metastable or false minimum. Thus, in spite of the presence of a deeper minimum of the potential at the zero temperature the Universe remains at the false minimum. This phenomenon suggests that, rather than just establishing the existence of a critical temperature for an FOEWPT, it is more important to investigate the successful nucleation of a bubble of the broken electroweak phase. 

The inverse time duration of the phase transition, $\beta$, is obtained from the relation
\beq
\beta = -\frac{d S_3}{dt}\Bigr|_{t_*} \simeq \dfrac{\dot{\Gamma}}{\Gamma} = H_*T_* \frac{d(S_3/T)}{dT}\Bigr|_{T_*},
\eeq
where $H_*$ is the Hubble rate at $T_*$. For a strong GW signal, the ratio needs to be low which implies a relatively slow phase transition.
 
Having defined the portal we need to know two more parameters to calculate the GW power spectrum coming from an FOPT in the early Universe. When a phase transition occurs in a thermal plasma, the released energy is shared between the plasma's kinetic energy, which causes a bulk motion of the fluid in the plasma resulting in GW, and heating the plasma. The quantity $\kappa_v$ is the fraction of latent heat energy converted into the bulk motion of the fluid, which takes the form~\cite{Espinosa:2010hh, Chiang:2019oms} 
\begin{equation}\label{eq:76}
\kappa_v \simeq \left[ \dfrac{\alpha}{0.73+0.083\sqrt{\alpha}+\alpha}\right]\,\,.
\end{equation}
Finally, we need to know $\kappa_{\text{turb}}$, which is a fraction of $\kappa_v$ that is used to generate Magneto-Hydrodynamic-Turbulence (MHD) in the plasma.
The value of $\kappa_\text{turb}$ is unknown but it is expected that $\kappa_\text{turb} \approx (5\sim 10) \, \kappa_v$~\cite{Hindmarsh:2015qta}, and we consider this fractional value to be 0.1 in all our GW calculations.
Now we have all the relevant parameters at our disposal, using which we can calculate the GW energy density spectrum.

It is generally accepted that during a cosmological FOPT, the GW can be produced by three distinct processes: bubble wall collisions, long-standing sound waves in the plasma, and 
MHD turbulence. For bubble collisions, the GW is generated by the stress-energy tensor located at the wall, and the mechanism is referred to as 
the ``envelope approximation"\footnote{In fact, the ``Envelope approximation" is actually two approximations. In the first approximation, the expanding bubble's stress-energy tensor is believed to be non-zero only in an infinitesimally thin shell on the bubble surfaces. In the second approximation, when two bubbles interact, it is assumed that this stress-energy tensor vanishes instantly, leaving just the `envelopes' of the bubbles to interact~\cite{Kosowsky:1992rz, Kosowsky:1991ua, Kosowsky:1992vn, Caprini:2019egz}.}.
However, for a phase transition proceeding in a thermal plasma, bubble collisions' contribution to the total GW energy density is believed to be  negligible~\cite{Bodeker:2017cim}. On the other hand, the bulk motion of the plasma gives rise to velocity perturbations in it, resulting in the generation of sound waves in a medium made up of relativistic particles. The sound waves receive the majority of the energy released during the phase transition~\cite{Hindmarsh:2013xza, Giblin:2013kea, Giblin:2014qia, Hindmarsh:2015qta}. This relatively long-living acoustic production of GW is often regarded as the most dominant one~\footnote{Here, we are presuming that a bubble expanding in plasma can achieve a relativistic terminal velocity. The energy in the scalar field is negligible in this situation and the most significant contributions to the signal are predicted to come from the fluid's bulk motion in the form of sound waves and/or MHD. This is the non-runaway bubble in a plasma (NP) scenario~\cite{Caprini:2015zlo, Schmitz:2020syl}.}. 
Percolation may also cause turbulence in the plasma, particularly MHD turbulence since the plasma is completely ionized. This corresponds to the third 
source~\cite{Caprini:2006jb, Kahniashvili:2008pf, Kahniashvili:2008pe, Kahniashvili:2009mf, Caprini:2009yp, Kisslinger:2015hua} of GW production. Thus, summing the contributions of the above two processes (i.e. only the contributions from the sound wave and the MHD turbulence) one obtains the overall GW intensity $\Omega_{\text{GW}}{\rm h}^2$ from the SFOPT for a given frequency $f$,
\beq
\label{TotalGW}
\Omega_{\text{GW}}h^2 \simeq \Omega_{\text{sw}}h^2+\Omega_{\text{turb}}h^2 ,
\eeq
where $h \approx 0.673$~\cite{DES:2017txv}.

The contribution to the total GW power spectrum from the sound waves, $\Omega_{\text{sw}}h^2$, in equation~\ref{TotalGW} can be modelled by the following fit formula~\cite{Hindmarsh:2019phv}
\begin{equation}\label{eq:75}
\Omega_{\text{SW}}{\rm h}^2=2.65\times 10^{-6} \Upsilon(\tau_{SW}) \left(\dfrac{\beta}{H_{\star}} \right) ^{-1} v_{w} \left(\dfrac{\kappa_{v} \alpha}{1+\alpha}\right)^2 \left(\dfrac{g_*}{100}\right)^{-\frac{1}{3}}\left(\dfrac{f}{f_{\text{SW}}}\right)^{3} \left[\dfrac{7}{4+3 \left(\dfrac{f}{f_{\text{SW}}}\right)^{2}}\right]^{\frac{7}{2}}\,\,,
\end{equation}
where $T_{\star}$ is the temperature just after the end of GW production and $H_{\star}$ is the Hubble rate at that time. For this analysis, we consider the approximation that $T_{\star} \approx T_n$. 
The present day peak frequency $f_{\text{SW}}$ for the sound wave contribution is
\begin{equation}\label{eq:77}
f_{\text{SW}}=1.9\times10^{-5}\hspace{1mm} \text{Hz} \left( \dfrac{1}{v_{w}}\right)\left(\dfrac{\beta}{H_{\star}} \right) \left(\dfrac{T_n}{100 \hspace{1mm} \text{GeV}} \right) \left(\dfrac{g_*}{100}\right)^{\frac{1}{6}}\,\,.
\end{equation}
It has been shown in recent studies~\cite{Guo:2020grp, Hindmarsh:2020hop} that sound waves have a finite lifetime, which causes $\Omega_{\text{sw}}h^2$ to be suppressed. The multiplication factor $\Upsilon(\tau_{SW})$ is considered in equation~\ref{eq:75} to take this suppression into account.
\beq
\Upsilon(\tau_{SW}) = 1 - \frac{1}{\sqrt{1 + 2 \tau_{\text{sw}} H_{\ast}}}.
\label{eq:upsilon}
\eeq
Here, the lifetime $\tau_{\text{sw}}$ is considered as the time scale when the turbulence develops, approximately given by~\cite{Pen:2015qta,Hindmarsh:2017gnf}:
\begin{eqnarray}
\tau_{\text{sw}} \sim \frac{R_{\ast}}{\bar{U}_f} ,
\end{eqnarray}
where $R_{\ast}$ is the mean bubble separation, and  $\bar{U}_f$ is
the root-mean-squared (RMS) fluid velocity. $R_{\ast}$ is related to the duration of the phase transition parameter, $\beta$, through the relation
$R_{\ast} = (8\pi)^{1/3} v_w /\beta$~\cite{Hindmarsh:2019phv, Guo:2020grp}. On the other hand, from hydrodynamic analyses, references~\cite{Hindmarsh:2019phv,Weir:2017wfa} has shown that the RMS fluid velocity is given by, $\bar{U}_f = \sqrt{3 \kappa_v \alpha/4}$. 
At $\tau_{\text{sw}} \rightarrow \infty$, $\Upsilon(\tau_{SW})$ approaches the asymptotic value $1$.

In contrast, the contribution $\Omega_{\text{turb}}h^2$ from the MHD turbulence in equation~\ref{TotalGW} can be modeled by~\cite{Caprini:2015zlo}
\beq
\label{GWturb}
\Omega_{\text{turb}}h^2 =3.35\times 10^{-4} \left(\frac{H_*}{\beta}\right)\left(\frac{\kappa_\text{turb} \alpha}{1+\alpha}\right)^{\frac{3}{2}}
\left(\frac{100}{g_s}\right)^{\frac{1}{3}}v_w\frac{(f/f_\text{turb})^3}{[1+(f/f_\text{turb})]^\frac{11}{3}(1+8\pi f/h_{\star})},
\eeq
where turbulence produced GW spectrum present day peak frequency $f_{\text{turb}}$ is given by,
\beq
\label{peakfreqturb}
f_{\text{turb}}= 2.7 \times 10^{-2}~\text{mHz}\frac{1}{v_w}\left(\frac{\beta}{H_*}\right)\left(\frac{T_*}{100~\text{GeV}}\right)\left(\frac{g_s}{100}\right)^{\frac{1}{6}},
\eeq
with the parameter
\begin{equation}\label{eq:h-star}
h_{*}=16.5\times10^{-6}\hspace{1mm} \text{Hz} \left(\dfrac{T_n}{100 \hspace{1mm} \text{GeV}} \right) \left(\dfrac{g_*}{100}\right)^{\frac{1}{6}}\,\,.
\end{equation}

Before ending our discussion on the framework of GW production from FOPT in the early Universe, a few words are necessary on EWBG.
For the early Universe to undergo a successful EWBG via an SFOEWPT, one of the parameters discussed in this subsection, the bubble wall velocity ($v_w$), needs to play an important role.
From the preceding GW calculation, it is clear that a larger $v_w$ is required for a stronger GW production, but to achieve the observed matter anti-matter asymmetry a small subsonic $v_w$ is needed. Consequently, a significantly large $v_w$, which can generate detectable GW signals, is detrimental to the process of producing the observed baryon asymmetry. 
However, recent studies recognise that $v_w$ may not be the velocity that enters EWBG calculation. This is due to the non-trivial plasma velocity profile surrounding the bubble wall~\cite{No:2011fi}. This will require an in-depth analysis of the effect of particle transport near the wall and we leave it for a future study. For the current work, we assume that expanding bubbles reach a relativistic terminal velocity in the plasma, i.e., $v_w \simeq 1$.

\section{Theoretical and experimental constraints}
\label{constraints}
Having set up the framework for our study, we discuss various theoretical and phenomenological constraints on the model parameters used in our analysis.
The theoretical constraints coming from the EW vacuum stability at zero temperature, perturbativity, partial wave unitarity and the phenomenological constraints  from the flavor sector, the neutrino oscillation data and the electroweak precision observables including the `$\rho$' parameter are well discussed in the literature~\cite{Yang:2021buu}.
 For the sake of completeness of this work, we discuss those in the appendix~\ref{theoreticalconstraints} and~\ref{Phenomenologicalconstraints}.
 Other phenomenological constraints from the DM experiments and those coming from the neutral and charged Higgs boson searches at the LHC are discussed below.
Such a discussion is useful to determine the targeted region of parameter space for our work.
\subsection{DM constraints}
\label{subsec:DMconstraints}
Besides neutrino masses, another objective of the model used in this paper is to explain the observed DM relic density of the Universe. In Section~\ref{darkmatter} we provided the analytical formulae necessary to calculate the DM abundance, $\Omega_{S} h^2$, and th$\sigma_{\rm DD}^{\rm SI}$) within this model. Here, we discuss how we study the DM phenomenology of the model numerically. First, we generated the model files using {\tt FeynRules}~\cite{Alloul:2013bka} and then feed the necessary files into {\micromegas}{\tt(v4.3)}~\cite{Belanger:2006is} and perform a scan over the model parameters relevant for the dark sector, as listed in equation~\ref{eq:DMpar}. The package {\micromegas} evaluate $\Omega_{S} h^2$ and  $\sigma_{\rm DD}^{\rm SI}$ for each benchmark point. 

It is well known that the Planck collaboration measured the DM relic abundance to be $0.120 \pm 0.001$, which we take into cognizance in our DM analysis. 
On the other hand, the most stringent bounds on $\sigma_{\rm DD}^{\rm SI}$ are provided by the latest XENON-1T~\cite{Aprile:2018dbl} and PANDA-4T~\cite{PandaX-4T:2021bab} results. While imposing these upper limits on $\sigma_{\rm DD}^{\rm SI}$ we take into account a
scale factor ($\xi_{_{\text DM}}$), which is the ratio of the computed relic density for the DM particle of the model and the observed DM relic density of the Universe, i.e., $\xi_{_{\text DM}} = \frac{\Omega_S h^2}{0.120}$. $\xi_{_{\text DM}}$ alleviates the strong bounds on $\sigma_{\rm DD}^{\rm SI}$ for a significant fraction of our scanned points. More details about how the above DM relic density and DMDD constraints are incorporated in our numerical analysis are discussed in Section~\ref{DMpar}.

The indirect search of DM followed from the experiments such as Fermi-LAT and MAGIC also put a constraint on the individual thermal average annihilation cross-section of DM into SM charge pairs as $\xi_{\rm DM}^2 \langle \sigma v \rangle_{S S \to \psi \overline{\psi}}$ where $\psi:=\{\mu^-,\tau^-,b,W^-\}$~\cite{MAGIC:2016xys}. When the density of DM is under-abundant, the scale factor $\xi_{\rm DM}$ further reduces the effective indirect search cross-section and alleviates bounds on the corresponding annihilation process. Like the direct search, same portal coupling, $\lambda_{S H}$ is also involved in the indirect search cross-section. The direct search of DM puts a strong constraint on the portal coupling, $\lambda_{SH}$, which leads to a small indirect search cross-section well below the existing upper bound of the corresponding process. Therefore in our discussion of available DM parameter space in Section~\ref{DMpar}, we ignore the indirect search constraints, which are automatically satisfied once direct search constraints take into account.    

\subsection{LHC constraints}
\label{subsec:LHCconstraints}
Finally, we discuss the phenomenology of the Higgs sector at the LHC in this subsection.
The production of the triplet scalars at the LHC via $s$-channel exchange of $Z/\gamma^*$ and $W^\pm$ lead to various final states of the SM particles from their prompt decays. The wide array of phenomenological consequences of type-II seesaw at the LHC has been thoroughly studied in the literature~\cite{Huitu:1996su,Chakrabarti:1998qy,Chun:2003ej,Akeroyd:2005gt,Garayoa:2007fw,Kadastik:2007yd,Akeroyd:2007zv,FileviezPerez:2008jbu,delAguila:2008cj,Akeroyd:2009hb,Melfo:2011nx,Aoki:2011pz,Akeroyd:2011zza,Chiang:2012dk,Chun:2012jw,Akeroyd:2012nd,Chun:2012zu,Dev:2013ff,Banerjee:2013hxa,delAguila:2013mia,Chun:2013vma,Kanemura:2013vxa,Kanemura:2014goa,Kanemura:2014ipa,kang:2014jia,Han:2015hba,Han:2015sca,Das:2016bir,Babu:2016rcr,Mitra:2016wpr,Cai:2017mow,Ghosh:2017pxl,Crivellin:2018ahj,Du:2018eaw,Dev:2018kpa,Antusch:2018svb,Aboubrahim:2018tpf,deMelo:2019asm,Primulando:2019evb,Padhan:2019jlc,Chun:2019hce,Ashanujjaman:2021txz, Mandal:2022zmy,Dutta:2014dba}.
Numerous collider searches have been carried out by the CMS and ATLAS collaborations to hunt for the same at the LHC~\cite{ATLAS:2012hi,Chatrchyan:2012ya,ATLAS:2014kca,Khachatryan:2014sta,CMS:2016cpz,CMS:2017pet,Aaboud:2017qph,CMS:2017fhs,Aaboud:2018qcu,Aad:2021lzu}. However, there has not been any evidence of an excess over the SM background expectations so far. Thus, these searches strongly constrain the parameter space relevant to the type-II seesaw model.
 \begin{figure}[t]
\begin{center}
\includegraphics[height=5.6cm,width=0.46\linewidth]{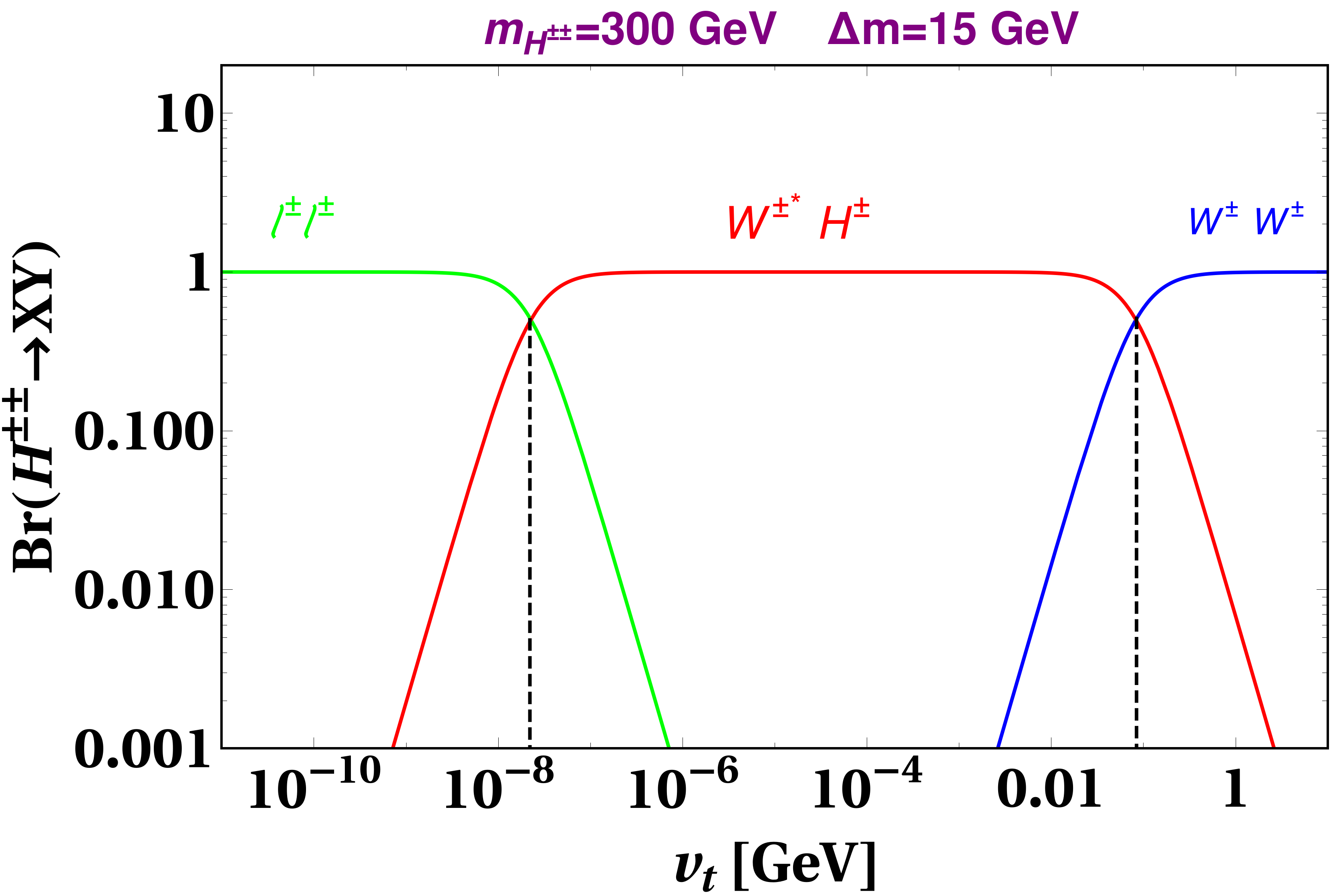}
\hskip 10pt
\includegraphics[height=5.6cm,width=0.46\linewidth]{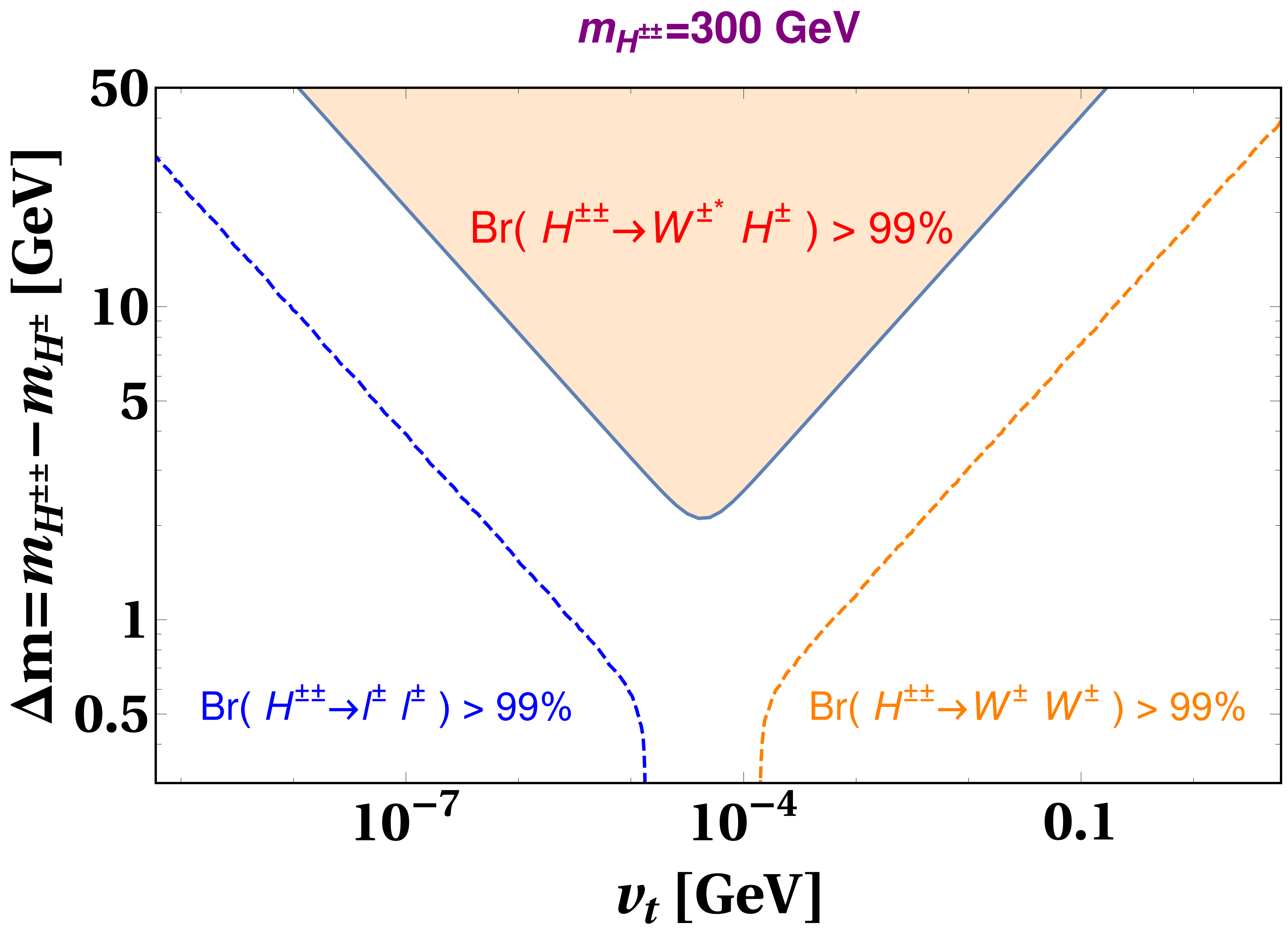}
\hskip 10pt
\caption{[Left] Branching ratios of doubly charged scalar, $H^{\pm\pm}$  as a function of $v_t$ for $m_{H^{\pm\pm}}=300$ GeV and $\Delta m =15$ GeV. [Right] Phase diagram for $H^{\pm\pm}$ decays to different modes is shown in $\Delta m-v_t$ plane by different contour lines: the leptonic  decay mode by blue dotted line, the gauge boson decay mode by orange dotted line and the cascade decay mode by the blue line. Each contour line corresponds to 
 $99 \% $ of the branching ratio for the corresponding decay channel. The mass of $H^{\pm\pm}$ is kept fixed as $m_{H^{\pm\pm}}=300$ GeV. The cascade dominated region ($Br(H^{\pm\pm} \to {W^{\pm}}^* H^{\pm} ) > 99\%$), which we are interested in our discussion is shown by the shaded orange region.}
\label{fig:BRs}
\end{center}
\vspace{-0.5cm}
\end{figure}

The most tantalizing collider signature of a type-II seesaw motivated model is the production and decay of $H^{\pm \pm}$. It is well known that $H^{\pm \pm}$ can decay to $\ell^{\pm} \ell^{\pm}$ \, ($W^{\pm} W^{\pm})$ for low (relatively large) values of $v_t$, when the triplet scalars are degenerate in mass. If one introduces a mass-splitting between the triplet scalars, additionally, the $H^{\pm \pm} \rightarrow H^{\pm} W^{\pm}$ decay channel opens up for $\Delta m > 0$ scenarios. However, as we have seen in Section~\ref{Sec:EWPO}, the EW precision observables limit the mass-splitting, $\Delta m$, to be less than 40 GeV. Hence, this restricts $H^{\pm \pm}$ to decay to either $H^{\pm} \ell^{\pm} \nu$ and $H^{\pm} q \bar{q'}$ via an off-shell $W^{\pm *}$. For $\Delta m \lesssim \mathcal{O}(2)$ GeV, when quarks cease to remain as free particles, then $H^{\pm \pm}$ decay to $H^{\pm} \pi^{\pm}$ has to be considered also. Similar behavior can be observed for decays of $H^{\pm}$ as well, again for $\Delta m > 0$, leading to a cascade decay of $H^{\pm\pm}$ in turn. In fact, for $v_t \sim 10^{-3} - 10^{-6}$ GeV, the $H^{\pm \pm}(H^{\pm}) \rightarrow H^{\pm} (H^0/A^0) W^{\pm *}$ decay becomes the dominant decay channel for $\Delta m \gtrsim 10$ GeV. For a detailed discussion on the various decay channels of $H^{\pm \pm}(H^{\pm})$ we refer interested readers to references~\cite{FileviezPerez:2008jbu,Ghosh:2018drw}. To illustrate the above points we show BFs of $H^{\pm\pm}$ to $\ell^{\pm} \ell^{\pm}$ \, ($W^{\pm} W^{\pm})$, and the cascade channel ($H^{\pm}W^{\pm*}$) for $m_{H^{\pm\pm}} = 300$ GeV with $\Delta m = 15$ GeV in the left panel of figure~\ref{fig:BRs}. In contrast, we display the Br$(H^{\pm\pm} \rightarrow H^{\pm}W^{\pm*}) > 99 \%$ region by the shaded area in the right panel of figure~\ref{fig:BRs} as a function of $v_t$ as well as $\Delta m$. 

In the $\Delta m =0$ case, the ATLAS collaboration places the most stringent bound of $m_{H^{\pm\pm}} \gtrsim 870$ GeV, assuming Br($H^{\pm\pm} \rightarrow \mu^{\pm} \mu^{\pm}$) to be 100$\%$~\cite{Aaboud:2017qph}, and using 36.1 fb$^{-1}$ data at $\sqrt{s}=13$ TeV. On the other hand, ATLAS again excludes $m_{H^{\pm\pm}} \gtrsim 350$ GeV in the $W^{\pm \pm}$ final state with 139 fb$^{-1}$ of data~\cite{Aad:2021lzu}. The first bound is applicable in the small $v_t$ regime, while the second is for relatively large $v_t$ values. It is clear from the discussion in the previous paragraph that the above limits are not applicable to the whole $\Delta m - m_{H^{\pm\pm}}$ plane for the range of $v_t$, where the cascade decay have a significant BF.  

Let us first consider $\Delta m< 0$ scenarios. For a large mass-splitting ($\mathcal{O}(10)$ GeV) and a moderate $v_t$, cascade decays of  $H^0/A^0 \to H^\pm W^{\mp *}$ and $H^\pm \to H^{\pm \pm} W^{\mp *}$, dominate over other decays. In this regime, $H^{\pm \pm}$ is the lightest triplet scalar, and decays to either $\ell^{\pm} \ell^{\pm}$ or $W^{\pm} W^{\pm}$ depending on the choice of $v_t$. Hence, the effective pair production cross-section of $H^{\pm \pm}$ is augmented by cascade decays of $H^0, A^0$ and $H^{\pm}$. So, one can expect the bounds on $m_{H^{\pm\pm}}$ to be strong for $\Delta m<$ 0 scenarios compared to the $\Delta m=0$ case. In a recent analysis, reference~\cite{Ashanujjaman:2021txz} has shown by recasting various ATLAS and CMS analyses that for $\Delta m = -10$ ($-30$)~GeV the present exclusion limit for $m_{H^{\pm\pm}}$ is 1115 (1076) GeV with $v_t \sim 10^{-5} - 10^{-6}$ GeV. In the same paper, the authors have improved the bound on $m_{H^{\pm\pm}}$ for $\Delta m =0$ to 955 GeV (420 GeV) for small (relatively large) $v_t$.

On the other hand, in $\Delta m> 0$ scenarios, $H^{\pm \pm}$ is the heaviest triplet member, and again for a mass-splitting of $\mathcal{O}(10)$ GeV and a moderate $v_t$, $H^{\pm \pm}$ will mostly undergo the cascade decay, $H^{\pm \pm} \rightarrow  W^{\pm *} H^{\pm} \rightarrow  W^{\pm *} W^{\pm *} \,  H^0/A^0$. 
$H^0$ and $A^0$ mostly decay to invisible neutrinos for $v_t <$ $10^{-4}$ GeV. $H^0$ can also decay to $b \bar{b}$ or $\tau^+ \tau^-$ if its mixing angle with the SM Higgs is sufficiently large. The EW precision bound of $\Delta m < 40$ GeV ensures that the leptons and jets arising from $H^{\pm \pm}$ cascade decay will be soft and may not cross the $p_T^{\ell/j}$ thresholds required in canonical $\ell^{\pm} \ell^{\pm}$ and $W^{\pm} W^{\pm}$ searches for $H^{\pm \pm}$. Reference~\cite{Ashanujjaman:2021txz} shows that even with 3 ab$^{-1}$ of integrated luminosity data, the LHC will be unable to constrain $m_{H^{\pm\pm}} \gtrsim$ 200 GeV for $\Delta m =$ 10 GeV (30 GeV), with  $v_t$ around $10^{-3}$ GeV to $10^{-5}$ GeV ($10^{-3}$ GeV to $10^{-6}$ GeV). This scenario is akin to compressed supersymmetry spectra and a dedicated search strategy is required to develop to probe this region~\cite{Baer:2014kya,Dutta:2012xe,Dutta:2014jda,Ajaib:2015yma,Dutta:2017nqv}. 

In addition, the measurement of the properties of the observed 125 GeV Higgs boson can also restrict some of our model parameters. As the observed Higgs boson's couplings to the SM particles are found to be almost SM-like, it will not allow a significant mixing between the neutral $CP$-even components of the triplet and the SM doublet. In order to satisfy the Higgs signal strength bounds, the mixing angle $\sin\theta_t$, defined in equation~\ref{thetat}, should be below 0.1~\cite{ATLAS:2016neq, CMS:2018uag}. The observed Higgs can decay to a pair of DM scalars if $m_S < m_{h^0}/2$. Hence, we impose the limit Br($h^0 \rightarrow$ invisible) $<$ 0.15 on our model parameters~\cite{CMS:2018yfx}.

Furthermore, the decay width of the SM Higgs into a photon pair can be affected by the presence of $H^{\pm}$ and $H^{\pm \pm}$ in the loops. $\Gamma(h^0 \rightarrow \gamma \gamma)$ can both be enhanced or reduced compared to the SM value depending on the sign of quartic couplings $\lambda_1$ and $\lambda_4$. 
Since the partial decay width of the SM into photons is tiny, the total decay width of $h^0$ barely changes.
The signal strength parameter of 
$h^0 \rightarrow \gamma \gamma$ channel is defined as,
\begin{equation}
\label{mugammagamma}
\mu_{\gamma \gamma} = \frac{\Gamma^{\text NP}[h^0 \rightarrow \gamma \gamma ]}{\Gamma^{\text SM}[h^0 \rightarrow \gamma \gamma ]}
\end{equation}
where $\Gamma^{\text{SM (NP)}}[h^0 \rightarrow \gamma \gamma ]$ denotes the decay rate without (with) the inclusion of new physics. We consider that the production cross-section of $h^0$ will remain the same when the effect of NP is included since $\cos \theta_t \rightarrow 1$. For more details on $\Gamma(h^0 \rightarrow \gamma \gamma)$ within the type-II seesaw model we refer the reader to references~\cite{Arhrib:2011vc, Gunion:1989we}. The current limits on $\mu_{\gamma \gamma}$ from ATLAS and CMS are $1.04^{+0.10}_{-0.09}$~\cite{ATLAS:2022tnm} and $1.12{\pm 0.09}$~\cite{CMS:2021kom}, respectively.
%
\section{Choice of parameter space}
\label{sec:motivatedregion}

\subsection{Implications from Higgs searches}
\label{htogammagamma}

As discussed in subsection~\ref{subsec:LHCconstraints}, LHC direct searches for $H^{\pm \pm}$ will not be able to limit the triplet-like scalar masses even at 200 GeV for a moderate $v_t$ and relatively large $\Delta m$ ($> 10$ GeV)~\cite{Ashanujjaman:2021txz}.
Depending on the value of $v_t$ and mixing with $h^0$, in this scenario, $H^0/A^0$  either decays into $\nu \nu/ b \bar{b}/\tau^+ \tau^-$ or $h^0h^0, ZZ/h^0Z$. 
Henceforth, in this work, we focus on 10 GeV $<\Delta m <$ 40 GeV and $10^{-3}$ GeV $<v_t<$ $10^{-6}$ GeV region of parameter space of the type-II seesaw part of the model. On the other hand, the rich scalar sector of this model can generate stochastic GW signals from FOPT in the early Universe. As the LHC fails to constrain this region from the direct searches of triplet-like scalars, we try to probe this region of parameter space from GW experiments.

For $\Delta m > 0$, $\lambda_4$ needs to be $-$ve (see equation~\ref{trplmassdiff}). Thus, we consider $\lambda_4 < 0$ region of parameter space in our analysis. For successful FOPT to take place in the Higgs sector, the two most important quartic couplings are $\lambda_1$ and $\lambda_4$. They are responsible for mixing the doublet and the triplet fields (see quation~\ref{Vtree}).
We vary the absolute value of these couplings from 0 to 4. Other quartic couplings of the triplet-sector $\lambda_2$ and $\lambda_3$ do not significantly affect FOPT. Thus, we set them to fixed values for the rest of our study. 
Since we focus on light triplet-like scalars, we require $m_{H^{\pm\pm}}$ within 200 GeV to 600 GeV, and for that purpose, we vary $M_{\Delta}$ from 100 GeV to 800 GeV. 
In addition, for the above choice of our $v_t$ and $m_{H^{\pm\pm}}$ values, our benchmark points pass the flavour constraints discussed in subsection~\ref{subsec:Flavourconstraints} as well.   

The precision measurements of $h^0$ couplings at the LHC restrict the mixing between the doublet-like and the triplet-like $CP$-even neutral scalars to small values. 
Therefore, we limit our scan to tiny $\sin\theta_t$ values to allow minimum mixing and maintain the SM-like nature of $h^0$. However, loop-induced $h^0$ decays, such as $h^0 \rightarrow \gamma \gamma$, can deviate significantly from the observed limits even at small $\sin\theta_t$ values. We discuss the variation of the signal strength, $\mu_{\gamma \gamma}$ of the $h^0 \rightarrow \gamma \gamma$ channel in the above-discussed region of parameter space below.

 \begin{figure}[t]
\begin{center}
\includegraphics[height=7.0cm,width=0.56\linewidth]{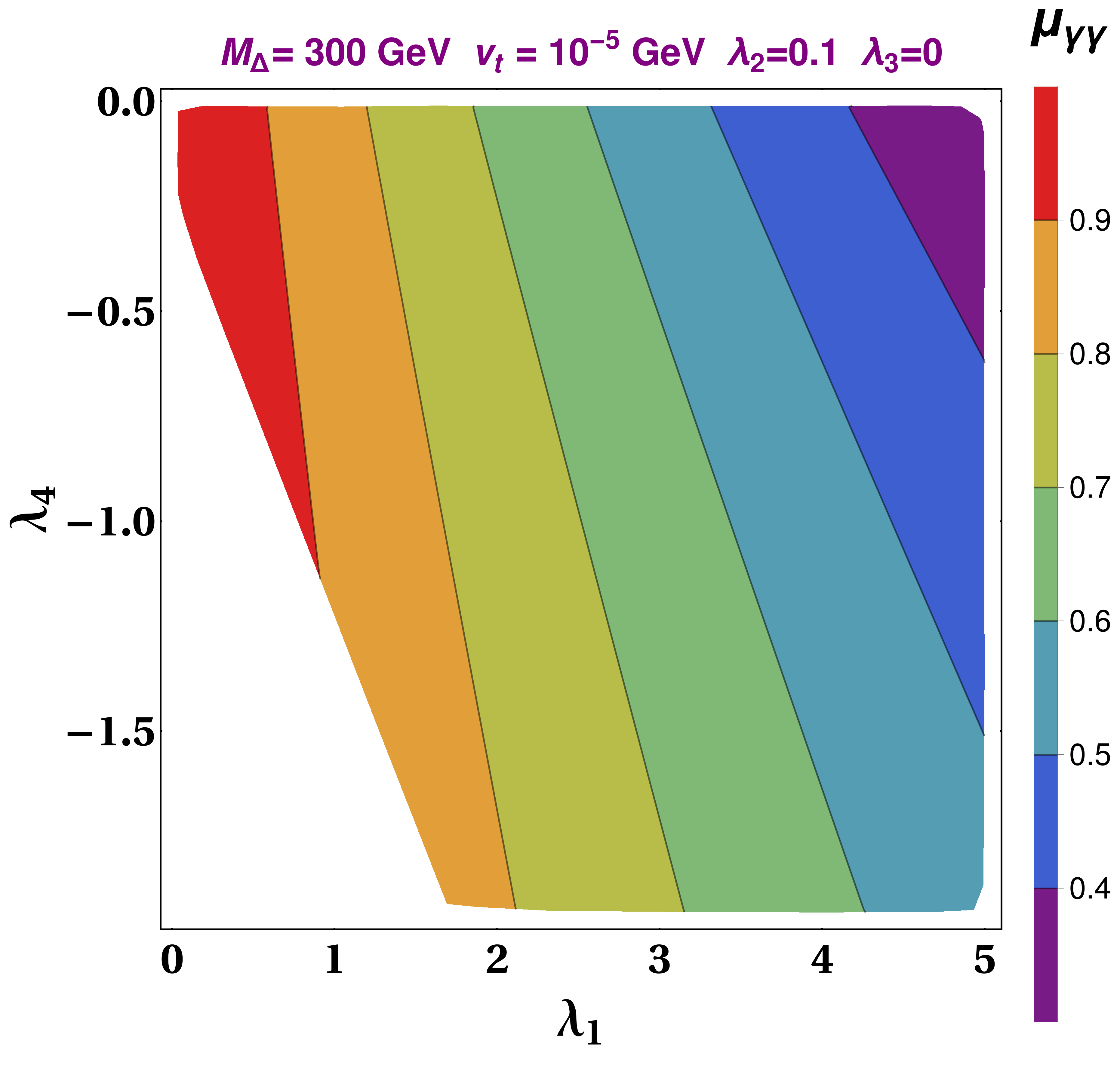}
\hskip 10pt
\caption{Variation of the signal strength $\mu_{\gamma \gamma}$ in the $\lambda_1-\lambda_4$ plane. The other  two parameters, $M_{\Delta}$ and $v_t$, of the triplet sector that can alter the signal strength are fixed at   300~GeV and $10^{-5}$~GeV, respectively.}
\label{fig:mugammagamma}
\end{center}
\vspace{-0.5cm}
\end{figure}
 At the one-loop level, the top quark, along with the charged weak gauge bosons and the charged triplet-like Higgs states contribute to the decay of $h^0 \rightarrow \gamma \gamma$. 
 The deviation of the signal strength of this decay channel $\mu_{\gamma \gamma}$, defined in equation~\ref{mugammagamma}, from 1 can become a signature of new physics beyond the SM. We present that the variation of $\mu_{\gamma \gamma}$ for $\lambda_1 >0$, $\lambda_4 < 0$, $M_{\Delta} = 300$ GeV and $v_t = 10^{-4}$ GeV in figure~\ref{fig:mugammagamma}.
The variation of $\lambda_1$ and $\lambda_4$ are presented in the $x$-axis and $y$-axis, respectively, whereas the variation of $\mu_{\gamma \gamma}$ is indicated by palette colors.
The figure also takes into consideration the theoretical constraints. Due to constraints from vacuum stability that are discussed in section~\ref{vacuumstabilityconstraints},\footnote{Particularly from the inequality relations: $(\lambda_{1}+\lambda_{4})+2\sqrt{\lambda_H(\lambda_2+\lambda_3)} \geq 0$ and $(2\lambda_{1}+\lambda_{4})+4\sqrt{\lambda_H(\lambda_2+\frac{\lambda_3}{2})} \geq 0$} the white empty section of figure~\ref{fig:mugammagamma} is excluded. We note that due to a {destructive interference} between the SM and BSM contributions to $\mu_{\gamma \gamma}$, it is mostly less than 1 for $\lambda_4 <0$. The deviation of $\mu_{\gamma \gamma}$ from 1 increases with the increase of the absolute values of $\lambda_1$ and $\lambda_4$. Thus the precision study of $\mu_{\gamma \gamma}$ at the LHC and other proposed collider experiments can exclude a significant amount of the parameter space of our interest in this paper. 
We shall discuss this issue further in section~\ref{results}.

\subsection{Implications from DM searches}
\label{DMimplications}

In contrast to the Higgs sector discussed above, in the dark sector, we vary $m_S$, $\lambda_{SH}$, and $\lambda_{S\Delta}$ within a certain range to satisfy all DM constraints. $\lambda_{SH}$ coupling has a significant impact on FOPT as it mixes the singlet $h_s$ field with the $h_d$-field (see equation~\ref{Vtree}). Thus, the possibility of a strong FOPT is expected to increase with a large $\lambda_{SH}$. However, as we have already discussed in section~\ref{darkmatter}, increasing the same coupling increases $\sigma_{\rm DD}^{\rm SI}$ as well. Hence, large $\lambda_{SH}$ values are ruled out by DMDD limits. 
We shall discuss this issue in more detail in section~\ref{FOPTduetoDM}. Another dark sector parameter, $\mu_3$, also impacts both DM and the FOPT. This coupling not only can produce the necessary barrier for FOPT but also contributes to DM semi-annihilation processes. We vary this parameter in our study satisfying the constraints $\mu_3/m_S \lesssim 2 \sqrt{\lambda_S}$ from the stability of the global minimum, which is discussed in subsection~\ref{vacuumstabilityconstraints}.

To summarize, the ranges of input parameters of the model we consider for the scan are listed in table~\ref{tab:ranges}. The other parameters are fixed at $\lambda_H = 0.129$, $\lambda_2 = 0.1$, $\lambda_3 = 0$, $\lambda_{S} = 0.1$.
\begin{table}[t]
\begin{center}
\centering{
{\tiny\fontsize{8.3}{7.8}\selectfont{
\begin{tabular}{|c|c|c|c|c|c|c|c|c|c|}
\hline
\makecell{Varying \\ parameters}  & $\lambda_1$ & $\lambda_4$ & \makecell{$M_{\Delta}$  \\ (GeV)}& \makecell{$v_t$ \\ (GeV)}&  \makecell{$m_S$ \\ (GeV)} &
\makecell{$|\lambda_{SH}|$ \\}
 & \makecell{$|\lambda_{S\Delta}|$ \\ } & \makecell{$\mu_3$ \\ (GeV)} & \makecell{$|\lambda_S|$  \\} \\
\hline
 & & & & & & & & &  \\
Ranges  & 0 -- 4 & $-4$ -- 0 &
100 -- 800 & $10^{-6}$ -- $10^{-3}$& 0 -- 1000 & $<4$ & $<6$ & $<1000$ & $<6$ \\
 & & & & & & & & &  \\
\hline
\end{tabular}
}}}
\caption{Ranges of various model  parameters employed for
scanning the present model parameter space.}
\label{tab:ranges}
\end{center}
\vspace{-0.5cm}
\end{table}
%


\section{Results}
\label{results}
In this section, we investigate the DM phenomenology, the thermal history regarding an FOEWPT, the associated production of stochastic GW and the interplay among the DM, GW and the collider physics in this model.
In the subsection~\ref{DMpar}, we analyse the dependencies of the DM phenomenology in terms of both DM relic density and DMDDSI cross-section on the triplet-sector scalar masses. We study the variation of various model parameters of the dark sector in the estimation of relic density and DMDDSI cross-section.
Implications of the various model parameters on FOEWPT  have been studied in two parts. The impact of the triplet-sector parameters on the FOEWPT, the production of GW and the interplay with LHC have been studied in the subsection~\ref{EWBGregion}. Two benchmark scenarios are presented to illustrate
the effect of the triplet-sector parameters on the FOEWPT. 
In the subsection~\ref{FOPTduetoDM}, we study the possibilities of FOEWPT  by altering the shape of the SM Higgs potential mainly from the dark sector.
The connection between the DM observables and an FOPT has been discussed. A benchmark scenario that exhibits a two-step phase transition in the early Universe has been presented to illustrate such a scenario.
\subsection{DM Parameter space}
\label{DMpar}
In this section, we discuss the DM parameter space in terms of model parameters which is consistent with DM relic density observed by the PLANCK experiment and the upper bound on DM-nucleon spin-independent cross-section provided by XENON-1T and PANDA-4T experiments. Before going to the details of the parameter space scan, we attempt to comprehend the variation of DM relic density with DM mass, $m_S$ $(\equiv m_{\rm DM})$ and other relevant
parameters of the model such as  scalar portal couplings $\lambda_{S H}$, $\lambda_{S \Delta}$ and trilinear coupling $\mu_3$.
The abundance of DM also significantly depends on the masses of triplet scalar, i.e.,  $m_{H^0},~m_{A^0},~m_{H^\pm}$ and $m_{H^{\pm\pm}}$ which can be expressed as the independent parameters as mentioned in equation~\ref{tripar}.
It is difficult to establish the significance of all the relevant parameters that vary simultaneously. Therefore, we fix the triplet sector parameters (in equation~\ref{tripar}), which simplifies the setup and allows us to investigate DM phenomenology. 
For DM discussion, we consider one benchmark point from the cascade region of the triplet sector (as discussed in section~\ref{DMimplications}) which is given by:
\bea
\{M_\Delta=367.7 ~{\rm GeV},~v_t=3.3\times 10^{-4}~ {\rm GeV},~\lambda_1=3.92,~\lambda_2=0.1,~\lambda_3=0,~\lambda_4=-0.989\} \nonumber \\
\eea
 which corresponds to the physical masses and mixing angle followed from equations~\ref{MCPE}, \ref{MCPO}, \ref{MSC} and \ref{MDC} as:
 \bea
 \{m_{H^0}= 367.7,~m_{A^0}=367.7,~m_{H^\pm}=387.5,~m_{H^{\pm\pm}}=406.4~ {\rm GeV}, \nonumber \\ ~\mu=0.00104,~\sin\theta_t=-1.03 \times 10^{-6}~ \} 
 \eea
It is worth mentioning here that one can also consider
another set of parameters as mentioned above, for which the underlying physics remains the same. The only thing that will change is the allowed region of DM mass depending on the triplet scalar mass.  Therefore the phenomenology of DM depends on the following independent parameters in the dark sector along with the above-fixed parameters
\bea
\{m_S,~\mu_3/m_S=1,~\lambda_{S H},~\lambda_{S \Delta} \}.
\eea
 Note that for DM  discussion we consider here $\mu_3/m_S=1$ which is responsible for the semi-annihilation process, $S~S \to S ~h^0$. The larger values of $\mu_3/m_S$ lead to a larger semi-annihilation contribution to DM abundance. However, a stable EW vacuum sets an upper bound on the trilinear coupling, $\mu_3 \leq 2 m_S$ as discussed earlier.   

\begin{figure}[t]
\begin{center}
\includegraphics[height=5.4cm,width=0.46\linewidth]{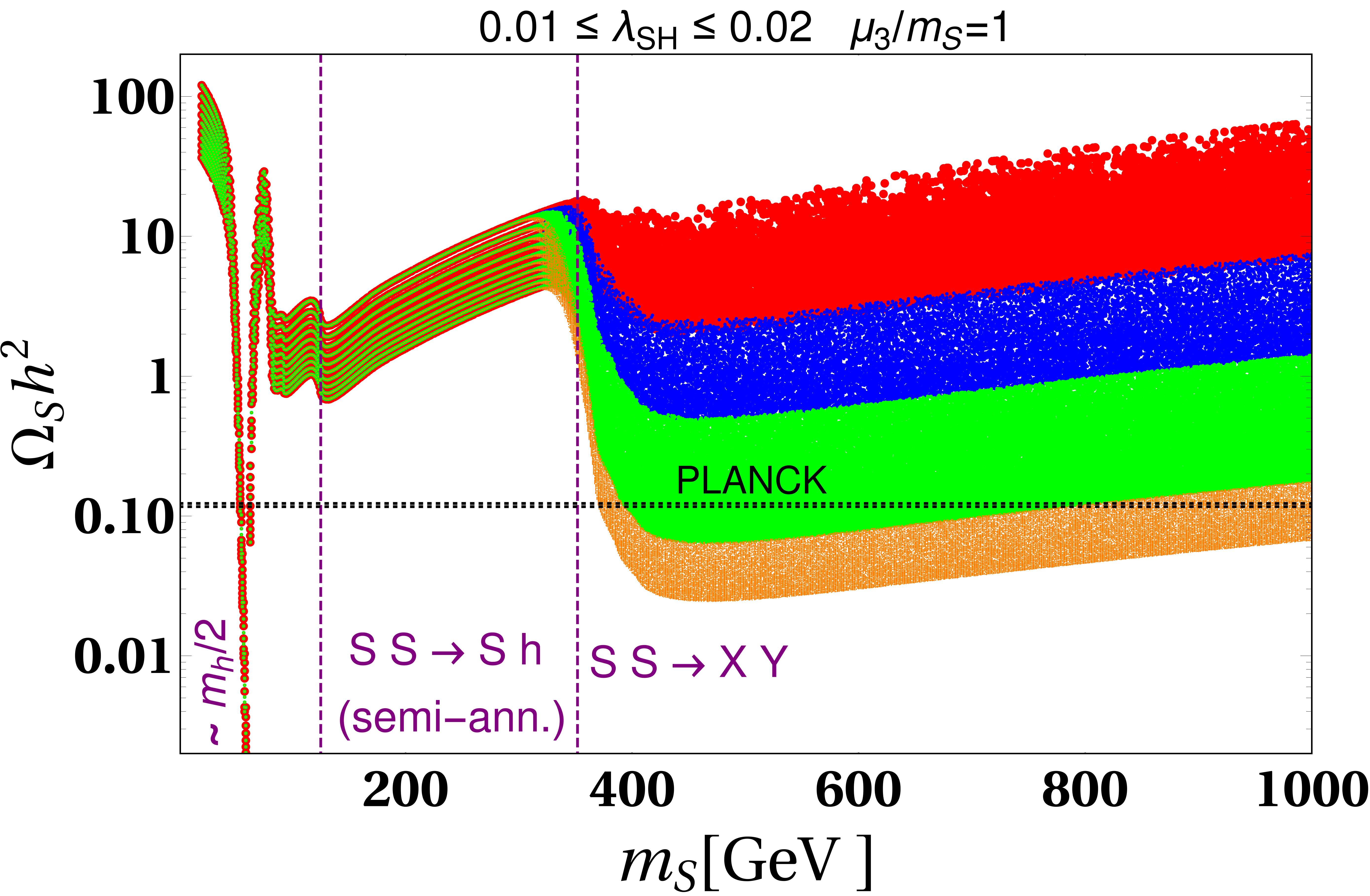}
\hskip 5pt
\includegraphics[height=5.4cm,width=0.46\linewidth]{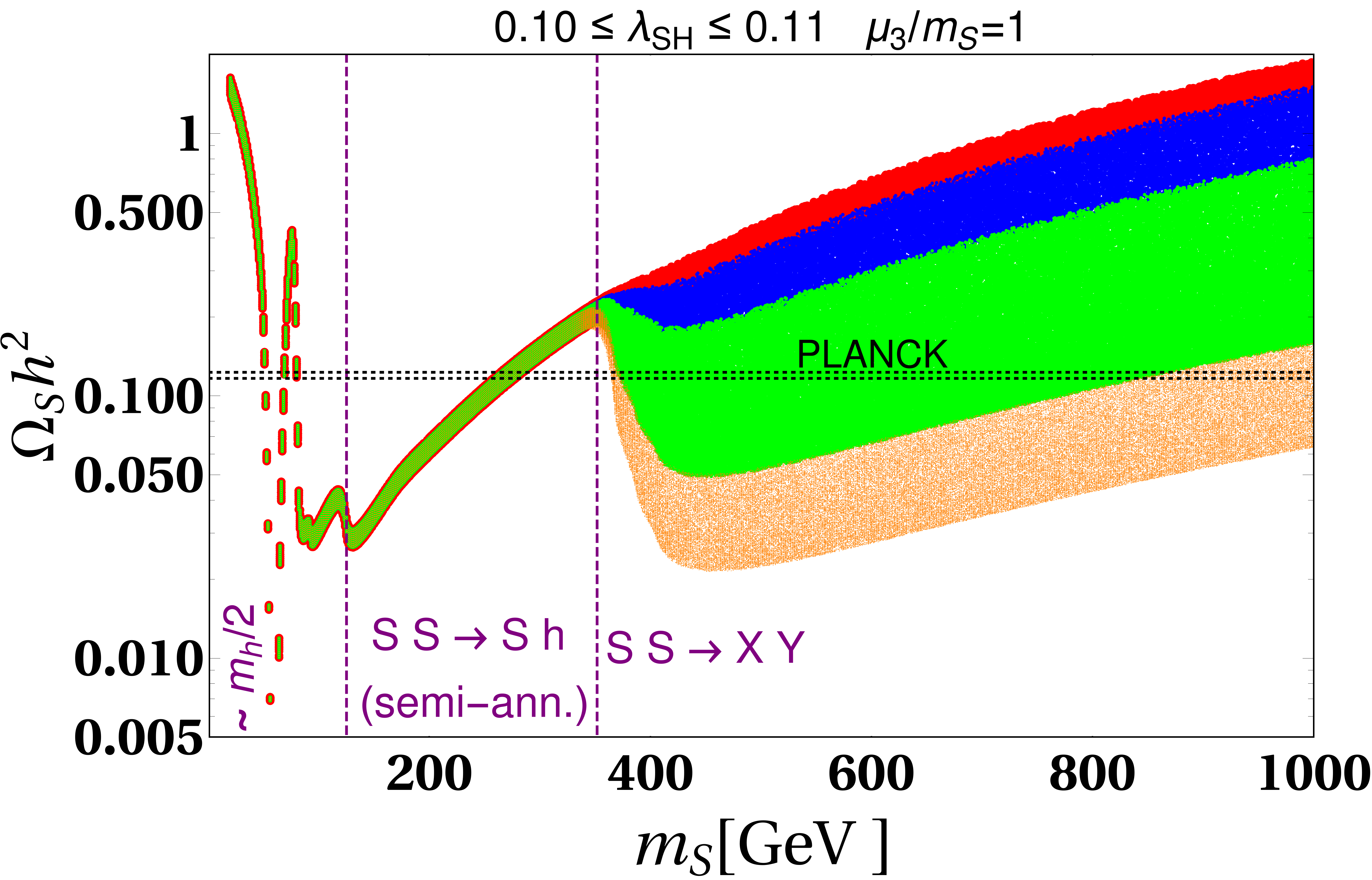}
\hskip 5pt
\caption{Variation of relic density as a function of DM mass for different ranges $\lambda_{S \Delta}$ : $0.01 \leq \lambda_{S \Delta} \leq 0.04$ (red), $0.04 < \lambda_{S \Delta} \leq 0.1$ (blue), $0.1 < \lambda_{S \Delta} \leq 0.3$ (green) and $0.3 < \lambda_{S \Delta} \leq 0.5$ (orange). We illustrate two different choices of $\lambda_{S H}:$ $\{0.01-0.02\}$ (left) and $\{0.10-0.11\}$ (right).}
\label{fig:dmpar1}
\end{center}
\end{figure}
%

 In Fig.\ref{fig:dmpar1}, we show the variation of DM abundance ($\Omega_S h^2$) as a function of $m_S$ for two different choices of $\lambda_{S H}$:  $0.01 \leq \lambda_{S H} \leq 0.02$ (left) and $0.10 \leq \lambda_{S H} \leq 0.11$ (right).
 Different colored patches indicate different ranges of DM portal coupling with triplet, $\lambda_{S \Delta}$ : $0.01 \leq \lambda_{S \Delta} \leq 0.04$ (red), $0.04 < \lambda_{S \Delta} \leq 0.1$ (blue), $0.1 < \lambda_{S \Delta} \leq 0.3$ (green) and $0.3 < \lambda_{S \Delta} \leq 0.5$ (orange) are considered for each plot. The black dotted horizontal lines indicate the observed DM relic density from the PLANCK data~\cite{Planck:2018vyg}, i.e., $\Omega_{\text{DM}} h^2 = 0.120$. 
With the increase in DM mass, different number-changing processes open up after crossing some threshold values resulting in a  drop in relic density.
 
 $\bullet$ {\textbf{$ m_S < m_{h^0}$}:} When DM mass, $ m_S < m_{h^0}$, the DM abundance  is mostly dominated by the $S ~S \to {\rm SM~SM}$ number changing process which is mediated by both the CP even Higgses, $h^0$ and $H^0$. And the density of DM varies as:
 \bea
 \Omega_{S}h^2 \propto \frac{1}{\langle \sigma v \rangle _{S S \to {\rm SM~SM}}}.
 \eea
 where for the small $\sin\theta_t$ limit, $\langle \sigma v \rangle _{S S \to {\rm SM~SM}} \propto {\lambda_{S H}^2}/{m_S^2}$ and almost independent of $\lambda_{S \Delta}$ which is shown from the above Figs.\ref{fig:dmpar1}. The annihilation contribution will get suppressed with the increase of DM mass and hence relic density increases.  For a fixed DM mass, with an increase of $\lambda_{\phi H}$, relic density decreases as it is shown in the right panel figure. A sharp drop in DM density due to the SM Higgs, ${h^0}$ resonance near $m_S \sim m_{h^0}/2$. Beyond the Higgs($h^0$) pole, $m_{S} > m_W,m_Z$, DM annihilates into a pair gauge final states resulting in a large enhancement in $\langle \sigma v \rangle$, which leads to less density.

 $\bullet$ {\textbf{$m_{h^0} \leq m_S < m_{H^0}$}:} A new annihilation channel, $S ~S \to S~ h^0$ (semi-annihilation)starts contributing when the DM mass becomes heavier than Higgs mass ($m_{S} > m_{h^0}$) thanks to the trilinear coupling, $\mu_3$ present in this scenario. Therefore the relic density of DM in this region is followed as:
 \bea
 \Omega_{S}h^2 \propto \frac{1}{\langle \sigma v \rangle _{S S \to {\rm SM~SM}}+\langle \sigma v \rangle _{S S \to S h^0}}.
 \eea
The second term in the above equation depends on $\mu_3$ and $\lambda_{S H}$ whereas the first term only depends on $\lambda_{S H}$. The semi-annihilation contributions dominate over the standard annihilation to SM for a larger value of $\mu_3$ which will help to evade the direct search bound on $\lambda_{S H}$. With the increase of $\lambda_{S H}$, both standard annihilation and semi-annihilation contributions are significantly enhanced and hence relic density decreases as it is depicted from the left and right panel of Fig.\ref{fig:dmpar1}. Note here that the effect of semi-annihilation becomes pronounced where the propagator suppression is minimal near $m_S \gtrsim m_{h^0}$. There is no variation of relic density with  $\lambda_{S \Delta}$ because both the contributions are almost insensitive to the change of $\lambda_{S \Delta}$ for the small $\sin\theta_t$ limit. There is no drop in relic density near $m_{S} \sim m_{H^0}/2$  due to the second heavy Higgs, $H^0$. This is because of the heavy Higgs mediated diagrams, $S~S \to {\rm SM~SM}$, are strongly suppressed by small $\sin\theta_t$ and $v_t$.

$\bullet$ {\textbf{$m_S \gtrsim m_{H^0}$}:} The 
new annihilation processes start to contribute to DM relic when DM mass is larger than the masses of the triplet states i.e. $2 m_S > m_X+m_Y$ and the relic density of DM turns out to be
 \bea
 \Omega_{S}h^2 \propto \frac{1}{\langle \sigma v \rangle _{S S \to {\rm SM~SM}}+\langle \sigma v \rangle _{S S \to S h^0}+\langle \sigma v \rangle _{S S \to X~Y}}~, ~\{X,Y\}:=\{H^0,A^0,H^\pm,H^{\pm\pm}\}. \nonumber \\
 \eea
 The Higgs mediated s and t channel diagrams for $S S \to X Y$ process are suppressed by a small mixing angle, $\sin\theta_t$ and small $v_t$. But the four-point contact diagrams for the $S S \to X Y$ process enhance the cross-section significantly with the increase of $\lambda_{S\Delta}$. Therefore the effect of new annihilation to DM relic density becomes more prominent with the increase of $\lambda_{S\Delta}$. Since this coupling is almost insensitive to $\sigma_{\rm DD}^{\rm SI}$ for the small mixing angle, one can play with it to satisfy observed DM density. For the smaller values of $\lambda_{S H}$ and $\mu_3(=m_S)$, the most dominant contribution to DM relic is coming from the new annihilation channels, $S ~S \to X~Y$ whereas the other two contributions, $S~S \to {\rm SM,SM}$ and $S~S \to S ~h^0$  are sub-dominate. With the increase of $\lambda_{S \Delta}$, $\langle \sigma v \rangle _{S~S \to X,Y}$ increases and hence relic density decreases which are shown from the left panel of Fig.\ref{fig:dmpar1} for fixed choices of $\lambda_{S H}: \{0.01-0.02\}$ and  $\mu_3/m_S=1$. In the right panel of Fig.\ref{fig:dmpar1}, we considered comparably large values of $\lambda_{S H}: \{0.10-0.11\}$ which further increases the $S~S \to {\rm SM~SM}$ and $S~S \to S ~h^0$ contributions and hence total effective annihilation cross-section increases. As a result, relic density decreases compared with the left panel figure. Note that a new DM semi-annihilation, $S S \to S H^0$ is kinematically allowed here but  it has almost negligible contribution to relic density as the cross-section is suppressed by the small  mixing angle, $\sin\theta_t$ and small $v_t$.   

\begin{figure}[t]
\begin{center}
\includegraphics[height=5.4cm,width=0.46\linewidth]{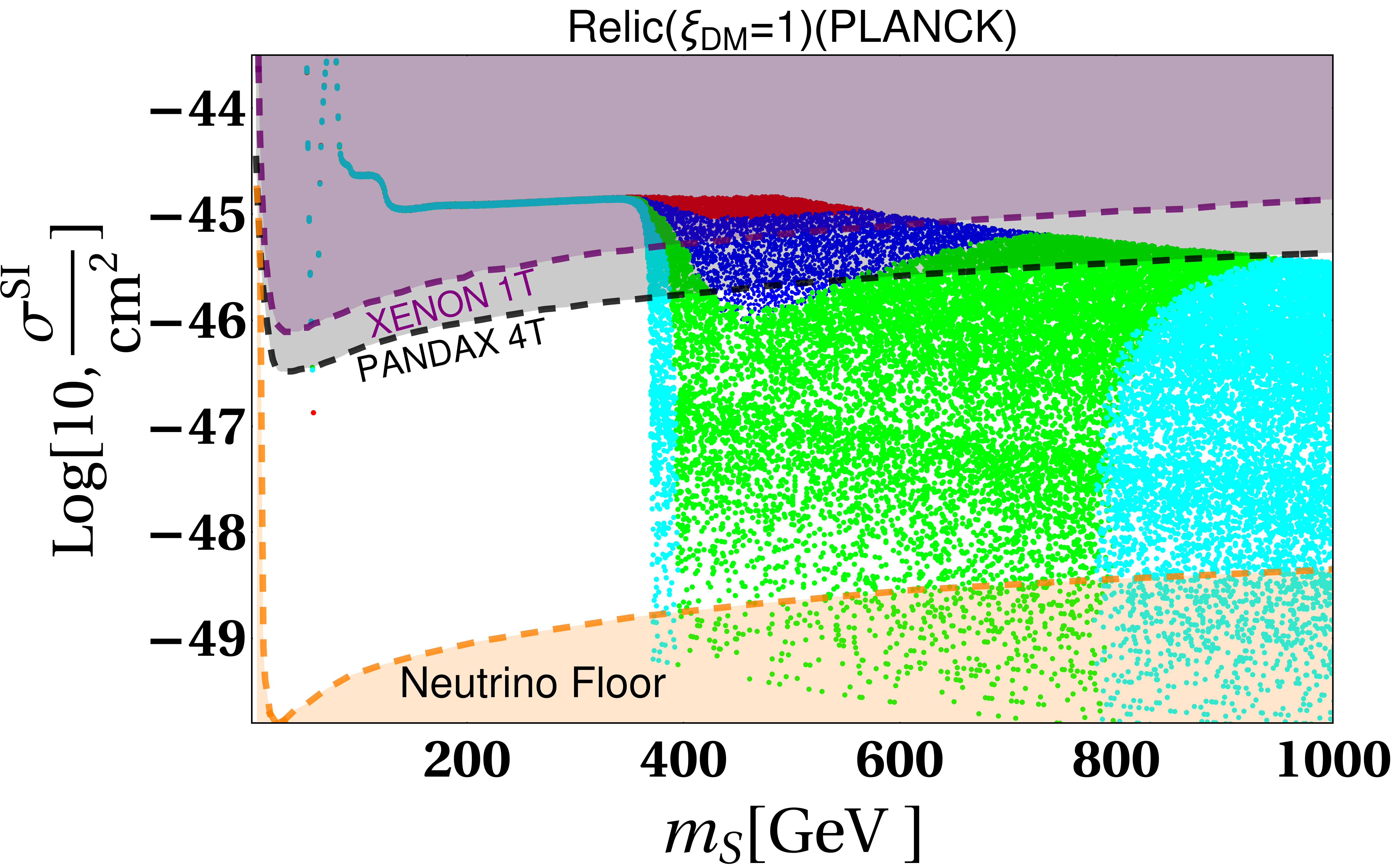}
\hskip 5pt
\includegraphics[height=5.4cm,width=0.46\linewidth]{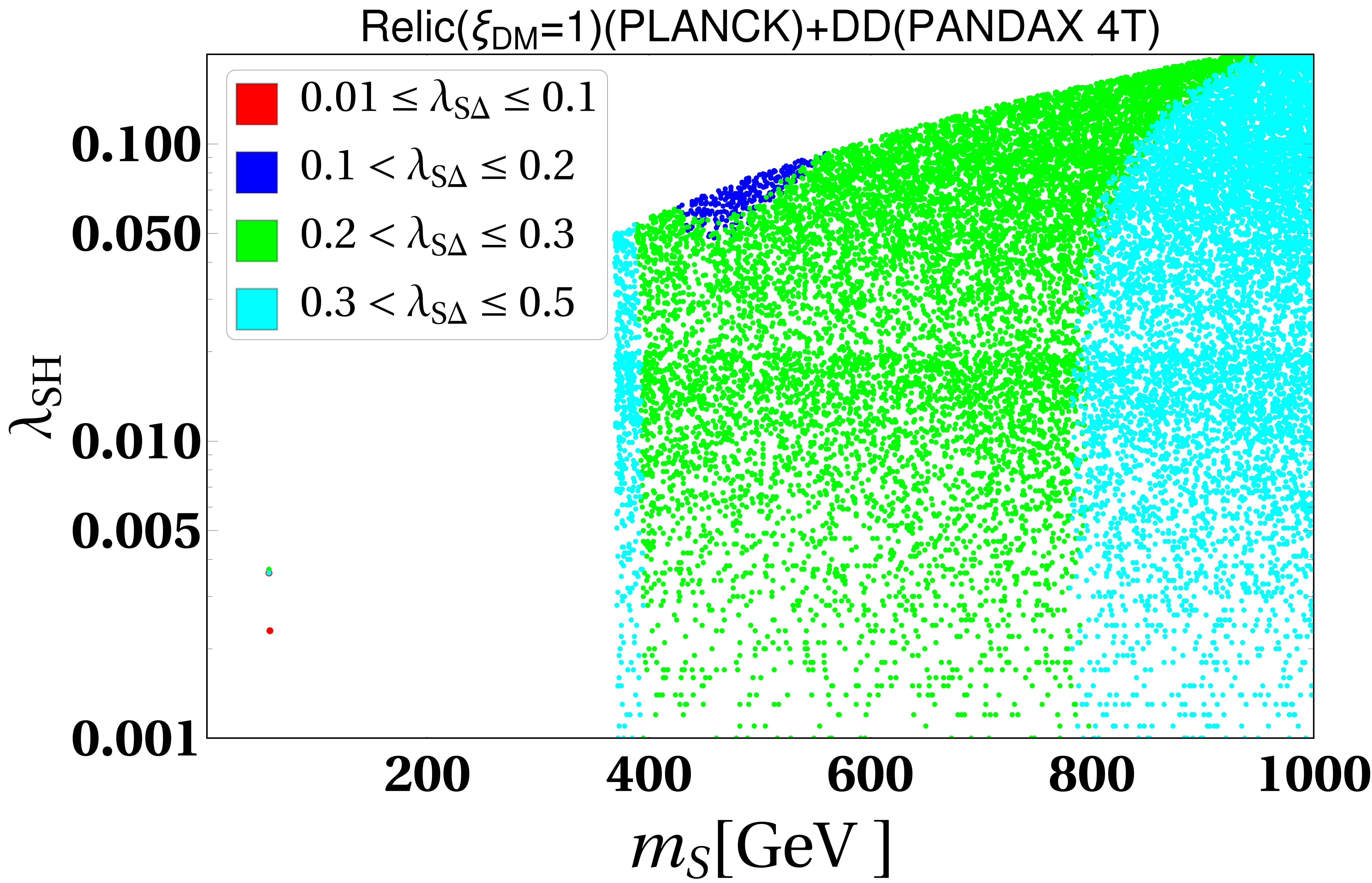}
\hskip 5pt
\caption{[Left]SI spin-independent DM-nucleon scattering cross-section, ($\sigma^{\rm SI}_{\rm DD}$) for relic density allowed(PLANCK) parameter space  as a function of DM mass($m_S$). For comparison, the current upper bound on the SI DM-nucleon cross-section from XENON 1T~\cite{Aprile:2018dbl}  and PANDAX 4T~\cite{PandaX-4T:2021bab} data are shown in the same plane. The  Neutrino Floor is represented by a shaded orange region. [Right] Relic(PLANCK)$+$DD(PANDAX-4T) allowed parameter space is shown in $m_{S}-\lambda_{S H}$ plane for different ranges of $\lambda_{S \Delta}$ shown by different color patches. Note that the relic density of DM, $S$ considered here $100\%$ ($\xi_{\rm DM}=1$) of observed DM density.}
\label{fig:dmpar2}
\end{center}
\end{figure}
%
We shall now move to the parameter space scan. To get the allowed region of parameter space, we perform a three-dimensional random scan over the following parameters keeping other parameters fixed as discussed above.
\bea
\{m_{S}:~ \{30-1000 \} {\rm GeV},~~\lambda_{S H}: \{0.001-0.3 \},~~\lambda_{S \Delta}: \{0.01-0.5 \} \}~.
\eea
In the left panel of Fig.\ref{fig:dmpar2}, we show the relic density allowed
parameter space (with $\xi_{\rm DM}=1$) emerging from the random scan in the plane of $\sigma_{\rm DD}^{\rm SI}$ versus DM mass, $m_S$. As discussed in section \ref{darkmatter}, for a small mixing limit($\sin\theta_t \to 0$), the DD cross-section is proportional to $\lambda_{S H}^2$ and almost independent of  $\lambda_{S \Delta}$. Therefore with an increase of $\lambda_{S H}$, the DD cross-section increases. The latest upper bounds on $\sigma_{\rm DD}^{\rm SI}$ against DM mass from  XENON 1T (purple dashed line)  and PANDAX 4T (black dashed line) are shown in the same plane.  The parameter space above the dashed line can be disfavoured from the non-observation of the DM signal of the corresponding direct search experiments. We find that DM mass below the mass of the lightest triplet state ($m_S< m_{H^0}$), the DM-SM Higgs portal coupling, $\lambda_{S H}$ is required comparably large to satisfy observed relic density which corresponds to large DD cross-section lies above the experimental upper limit. As a result, most of the region($m_{S}< m_{H^0}$) is excluded from non-observation of DM signal apart from Higgs poles, $m_{S}\sim m_{h^0}/2$. Once the triplet final states open up, the DM portal coupling with a triplet, $\lambda_{S \Delta}$ takes up the major role in controlling DM relic density. With the help of $\lambda_{S \Delta}$ which is insensitive to DD-cross-section, the pressure on $\lambda_{S H}$ can be reduced to produce correct relic density which can evade the direct search bound. This phenomenon can be observed from the right panel of Fig.\ref{fig:dmpar2}. As a result, the region above the DM mass, $m_{S} > m_{H^0}$ is allowed from both relic and direct search constraints with the help of large $\lambda_{S \Delta} (\gtrsim 0.1)$.

\subsection{Implications of the triplet sector parameters on SFOEWPT}
\label{EWBGregion}
\begin{figure}[t]
\begin{center}
\includegraphics[height=5.6cm,width=0.46\linewidth]{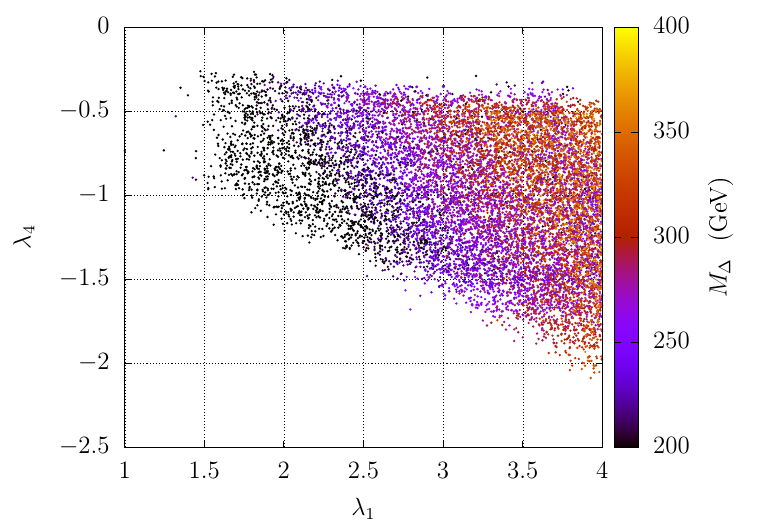}
\hskip 10pt
\includegraphics[height=5.6cm,width=0.46\linewidth]{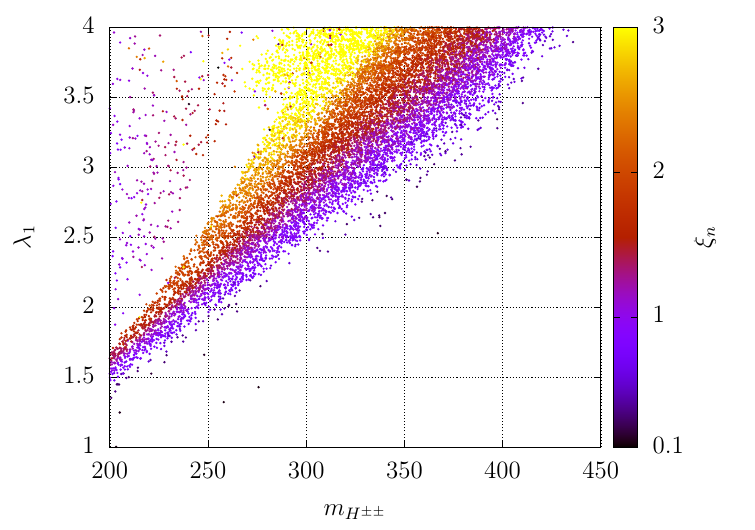}
\hskip 10pt
\includegraphics[height=5.6cm,width=0.46\linewidth]{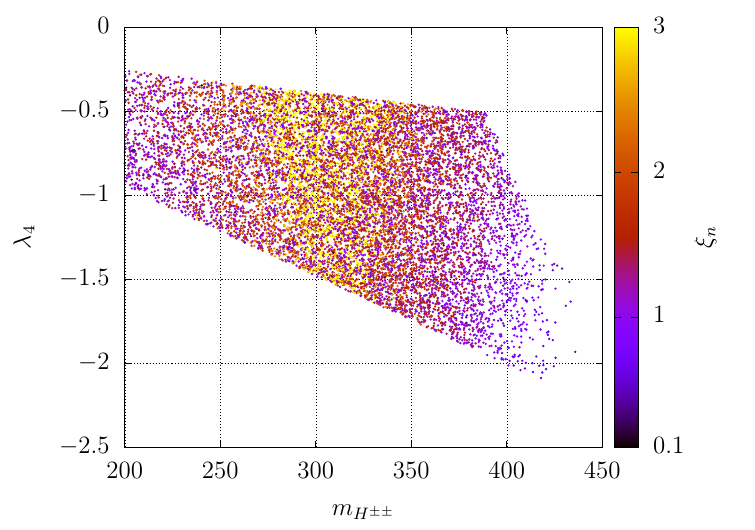}
\hskip 10pt
\includegraphics[height=5.6cm,width=0.46\linewidth]{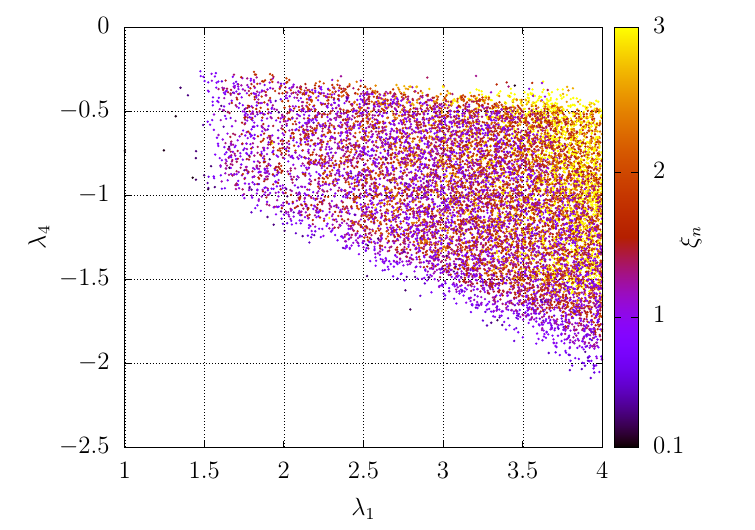}
\hskip 10pt
\caption{Scattered points that exhibits FOPT along the $h_d-$field direction in the $\lambda_1-\lambda_4$ plane with $M_{\Delta}$ (strength of the phase transition, $\xi_n$) indicated by the palette
in the top left (bottom right) plot.
The variation of the same points in the $m_{H^{\pm\pm}}-\lambda_1$ ($m_{H^{\pm\pm}}-\lambda_4$) plane with the palette indicating the strength of the phase transition $\xi_n$ is shown in the top right (bottom left) plot.
}
\label{fig:FOPTallowedregion}
\end{center}
\vspace{-0.5cm}
\end{figure}

We have already discussed how the discovery of a 125 GeV Higgs mass in the SM demonstrates that electroweak breaking occurs through a smooth cross-over transition. 
However, it can be an FOPT in the presence of additional scalars, going Beyond-the SM. 
A further benefit of FOEWPT
is that it can supply one of the necessary conditions for the explanations of BAU.  In this section, we illustrate the possibility of FOPT in the electroweak sector in our choice of parameter region of the current model. To explain the observed matter-antimatter asymmetry via the mechanism of EWBG, the FOPT in the electroweak sector needs to satisfy another additional condition, i.e.,
\begin{equation}
   \xi_n = \frac{v_n}{T_n} > 1.
\end{equation}
$v_n$ is the $\vev$ in the electroweak minimum at the nucleation temperature $T_n$. The strength of the FOPT is quantified via $\xi_n$. This criterion is required to prevent the wash-out of the generated baryon asymmetry after the EWPT.
We present four plots of the FOPT-allowed scan results in figure~\ref{fig:FOPTallowedregion} to illustrate the correlations among the model parameters of the triplet-sector in connection with the FOPT and its strength, $\xi_n$. 
Here, We fix the various couplings of dark sectors to small values to ensure that their impact on the FOPT is minimal and we vary $\lambda_1$, $\lambda_4$, $v_{t}$ and $M_{\Delta}$ only in the range as described in table~\ref{tab:ranges}. Thus, here we want to study the impact of the triplet-sector parameters on the FOPT in the $h_d$-field direction. In our scan results $v_n \sim v_h$, since the triplet $\vev$ always remains very small.

%

%
\begin{figure}[h]
\begin{center}
\includegraphics[height=5.6cm,width=0.46\linewidth]{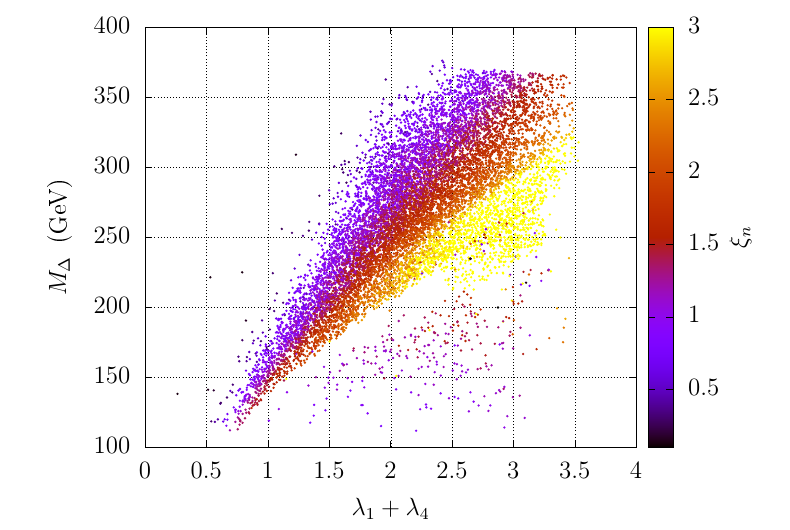}
\hskip 10pt
\includegraphics[height=5.6cm,width=0.46\linewidth]{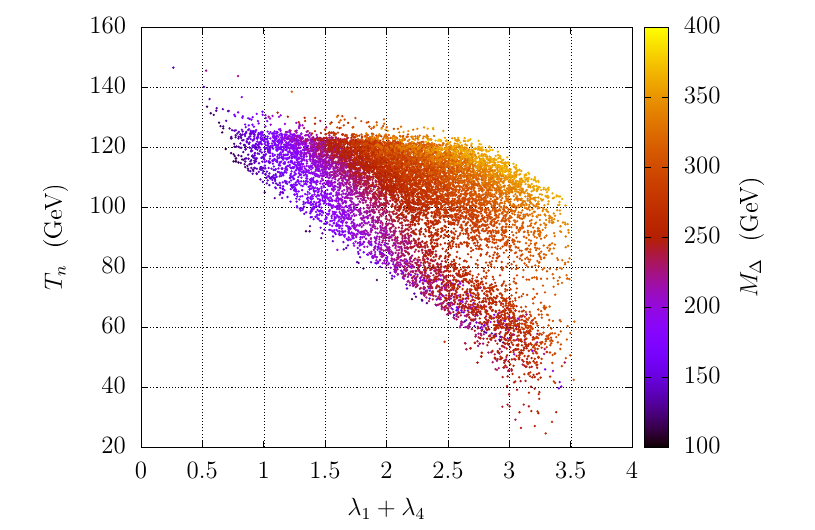}
\hskip 10pt
\caption{Left: Scatter points in the $(\lambda_1+\lambda_4)-M_{\Delta}$ plane that satisfies FOPT along the $h_d-$field direction with the palettes indicating the strength of the phase transition, $\xi_n$. Right: Similar to the scatter spots on the left plot but in the $(\lambda_1+\lambda_4)-T_n$ plane with showing the variation of $M_{\Delta}$ via the palette.}
\label{fig:FOPTregioneffcoupl}
\end{center}
\vspace{-0.5cm}
\end{figure}
%

The plot on top, left of figure~\ref{fig:FOPTallowedregion} shows the FOPT allowed region of $\lambda_1$ and $\lambda_4$ with the $M_{\Delta}$ being indicated by the palette-color.
It shows that FOPT prefers relatively larger $\lambda_1$. We have already discussed that in our choice of parameter region ($\Delta m > 0$), $\lambda_4$ has to be $-$ve and from the theoretical constraints (particularly constraints from the stability) that we have discussed in section~\ref{theoreticalconstraints}, the absolute value of $\lambda_4$ cannot be too large compared with $\lambda_1$.
As a result, the permitted range of $\lambda_4$ rises with larger $\lambda_1$. Color variation reveals that $M_{\Delta}$ needs to be on the smaller side at the relatively smaller values of $\lambda_1$. From equation~\ref{MDC}, $m_{H{^\pm\pm}} \sim M_{\Delta}$ at smaller $\lambda_4$. Thus, FOPT demands relatively light triplet-like scalars at smaller quartic coupling $\lambda_1$. At the relatively larger triplet-like scalar masses (larger $M_{\Delta}$), FOPT demands larger $\lambda_1$. On the other hand, for a given $\lambda_1$, the decrease of the absolute value of $\lambda_4$ demands increasing $M_{\Delta}$ for FOPT.

The shape of the SM Higgs potential gets modification from the quartic terms that are proportional with $\lambda_1$ and $\lambda_4$ in equation~\ref{Vtree}. Thus, the phase transition pattern in the Higgs sector strongly depends on these quartic couplings $\lambda_1$ and $\lambda_4$.
In the plot on top, right and bottom, left of figure~\ref{fig:FOPTallowedregion}, we present the variation of $m_{H^{\pm\pm}}$ with $\lambda_1$ and $\lambda_4$, respectively, from our scanned points that satisfy FOPT. The palette-color indicates the associated strength of the FOPT, $\xi_n$, values. For a fixed $m_{H^{\pm\pm}}$, color variation shows that the strength of the FOPT increases with $\lambda_1$. Thus, larger $\lambda_1$ is favoured for an SFOEWPT, which is required for EWBG. In addition, for a fixed value of the quartic coupling $\lambda_1$, lighter $m_{H^{\pm\pm}}$ is preferred for an SFOEWPT. Thus, larger quartic coupling $\lambda_1$ and lower triplet-like scalars increase the strength of the FOPT. The other plot indicates that for a given $m_{H^{\pm\pm}}$, the strength of the FOPT increases with the decrease of the absolute value of $\lambda_4$. The dependence of $\lambda_1$ and $\lambda_4$ on $\xi_n$ can clearly be understood from the bottom and the right plot of figure~\ref{fig:FOPTallowedregion}. It shows that the increase of $\lambda_1$ and the decrease of the absolute value of $\lambda_4$ increases the strength of the phase transition. 

The effective dependence of these quartic couplings on the phase transition pattern can be quantified in terms of the sum of these two quartic couplings, i.e., $\lambda_1 + \lambda_4$.
We present the variation of $\lambda_1 + \lambda_4$ with $M_{\Delta}$ with $\xi_n$ being indicated by the palette-color in the left plot of figure~\ref{fig:FOPTregioneffcoupl}. The variation of colors indicates that the strength of FOPT increases with the increase of the effective quartic coupling ($\lambda_1 + \lambda_4$) and decrease of $M_{\Delta}$ (or, the triplet-like scalar masses).

Another very important quantity of the FOPT is the temperature at which the PT starts to occur, i.e., the nucleation temperature ($T_n$). We have discussed in section~\ref{GW_section} that the production of GW peak amplitude and the peak frequency from an FOPT depend on $T_n$. Thus, studying the variations of $T_n$ with the various model parameters is important to identify the region of parameter space for a strong GW signal from FOPT. Keeping this in mind, We show the variation of $T_n$ with respect to the effective quartic coupling ($\lambda_1+\lambda_4$) in the right plot of figure~\ref{fig:FOPTregioneffcoupl}. The variation of $M_{\Delta}$ is presented in palette color. It can be seen that at relatively small effective quartic coupling, $T_n$ is relatively large. Also, low effective quartic coupling demands relatively low $M_{\Delta}$. On the other hand, at relatively larger effective quartic coupling, $T_n$ can go down to even below 50 GeV at moderate values of $M_{\Delta}$.

Finally, after studying all these plots in figures~\ref{fig:FOPTallowedregion} and~\ref{fig:FOPTregioneffcoupl}, we can point out that stronger FOEWPT demands relatively larger effective quartic couplings in the potential and relatively low triplet-like scalars in the parameter space. However, various Higgs searches at the LHC can constrain this parameter space. In the following subsection, we discuss the connection between the strength of the FOPT and the production of GW signals and the interplay between the detection of GW and the LHC searches in this region of parameter space. 
\begin{figure}[t!]
\begin{center}
\includegraphics[height=5.6cm,width=0.46\linewidth]{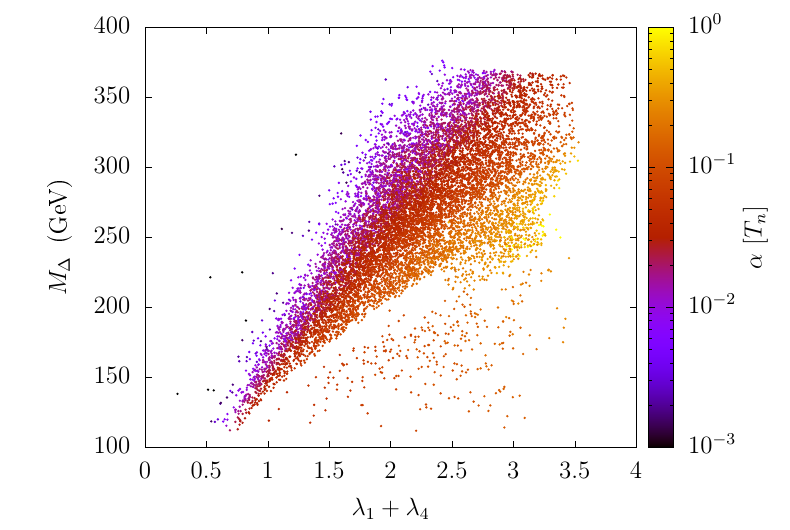}
\hskip 10pt
\includegraphics[height=5.6cm,width=0.46\linewidth]{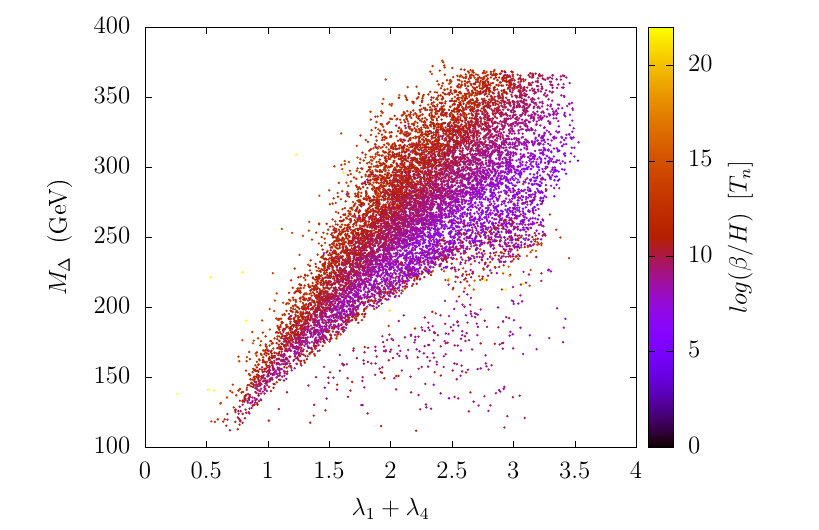}
\caption{Scatter points in the $(\lambda_1+\lambda_4)-M_{\Delta}$ plane that exhibits FOPT along the $h_d-$field direction. The variation of $\alpha$  ($log(\beta/H_n)$) is shown in the left (right) plot via the palette.}
\label{fig:alphabeta}
\end{center}
\vspace{-0.5cm}
\end{figure}
%
\subsubsection{Production of GW and the interplay with LHC}
\label{GWresults}
In section~\ref{GW_section}, we have discussed the possibility of producing stochastic GW background from the cosmological FOPT in the early Universe. This can be detected in the various future proposed GW space/ground-based detectors. The important portal parameters that control the GW intensity are $\alpha$, $\beta/H_n$, $T_n$ and $v_w$. We set $v_w = 1$ for this work. we have already discussed the variation of $T_n$ in the the bottom, right plot of figure~\ref{fig:FOPTallowedregion}.
In this subsection, we discuss the variation of the other two main portal parameters $\alpha$ and $\beta/H_n$, which control the GW spectrum.

In figure~\ref{fig:alphabeta}, we present two plots to show the variation of $\alpha$ (top, left) and $\beta/H_n$ (top, right) at $T= T_n$ in colors in the $(\lambda_1+\lambda_4) - M_{\Delta}$ plane. $\alpha$ ($\beta/H_n$) increases (decreases) with increasing ($\lambda_1+\lambda_4$) and decreasing $M_{\Delta}$. These variations have a direct connection with $\xi_n$. Stronger FOPT corresponds to larger $\alpha$ and lower $\beta/H_n$ at $T = T_n$. 
From the discussion in section~\ref{GW_section}, it can be found that the magnitude of the peak of the GW intensity is proportional to $\alpha$ and inversely proportional to $\beta/H_n$ for fixed $v_w$ and $T_n$ (see equations~\ref{eq:75} and~\ref{GWturb}).
These dependencies can be understood from their physical definitions.
A larger $\alpha$ corresponds to the more energy transfer from the plasma to the form of GW and a smaller $\beta/H_n$ implies a longer phase transition period. Thus larger $\alpha$ and small $\beta/H_n$ enhance the GW intensity.
Therefore, one can expect an increase in the GW intensity with $\xi_n$. As a result, the parameter space with larger effective quartic coupling and smaller triplet-like scalars would produce stronger GW spectrum intensity. In addition to this, as we discuss in section~\ref{GW_section}, it can be found that the value of $\beta/H_n$ is crucial to set the peak frequency range (see equations~\ref{eq:77} and~\ref{peakfreqturb}) which is also important for the detection purpose at the various future proposed GW experiments.
%
\begin{figure}[t!]
\begin{center}
\includegraphics[height=5.6cm,width=0.46\linewidth]{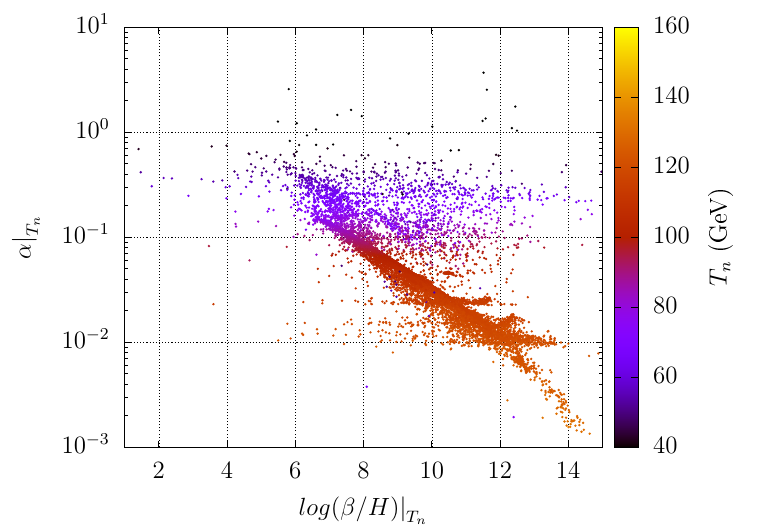}
\hskip 10pt
\includegraphics[height=5.6cm,width=0.46\linewidth]{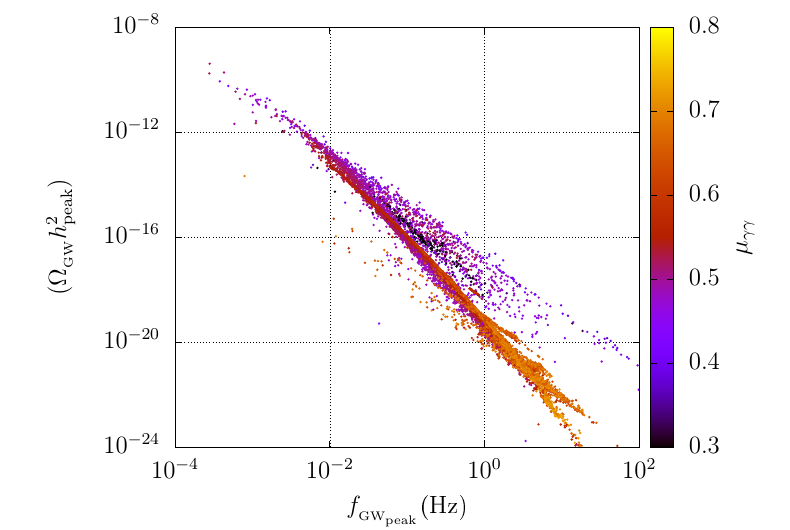}
\caption{Left: Parameter points of the scan in the $log(\beta/H_n)-\alpha$ plane with $T_n$ indicating
via the palette. Right: Scatter plots of GW peak amplitude ($\Omega_{\text{GW}} h^2$) vs the peak frequency ($f_{\text{peak}}$) in the unit of Hz considering only the contribution coming from the sound waves mechanism. Variation of the signal strength, $\mu_{\gamma\gamma}$, is indicated
via the palette.}
\label{fig:gwscanplots}
\end{center}
\vspace{-0.5cm}
\end{figure}
%

Therefore, in the left plot of figure~\ref{fig:gwscanplots}, we present the variation of $\alpha$ and $log(\beta/H_n)$ at $T= T_n$. $T_n$ is presented in palette color. The color variation shows that relatively lower $T_n$ corresponds to higher $\alpha$. On the other hand, $\beta/H_n$ measures the inverse duration of the phase transition. In some situations, the phase transition takes longer to start, corresponding to relatively lower $T_n$ and a larger gap between $T_c$ and $T_n$. These scenarios correspond to lower $\beta/H_n$. On the contrary, even at relatively low temperatures, the phase transition can happen quickly, corresponding to higher $\beta/H_n$. In the plot, the purple points (lower $T_n$) is across the wide range of $\beta/H_n$.

Variations of these portal parameters $\alpha$, $\beta/H_n$, $T_n$ correspond to a wide range of GW peak amplitude and peak frequency.
We have already discussed in section~\ref{GW_section} that the sound wave contribution mainly dominates the GW peak amplitude and the peak frequency (see equations~\ref{eq:77} and \ref{eq:75}).
In the right plot of figure~\ref{fig:gwscanplots}, we present the peak amplitude ($\Omega_{\text{GW}} h^2$) vs the peak frequency ($f_{\text{peak}}$) in the unit of Hz considering only the contribution coming from the sound waves mechanism.
The future proposed GW experiments have different sensitive regions in the intensity amplitude and the frequency range. Thus, studying the variation of the peak frequency and the peak amplitude of the produced GW is crucial for detecting the spectrum in future proposed experiments. The projected sensitivity of those experiments is presented in figures~\ref{fig:GW-freq-plot} and~\ref{GWplotbp3}.

We have already discussed that the regions of parameter space with lower triplet-like scalar masses and correspondingly higher effective quartic couplings can generate larger GW intensity signals.  In section~\ref{htogammagamma}, we show the variation of $\mu_{\gamma \gamma}$ with the various parameters of the model like $\lambda_1, \lambda_4$. It is expected that at relatively larger effective quartic couplings and light charged triplet-like scalars, $\mu_{\gamma \gamma}$ deviates significantly from 1. Therefore, we present the variation of 
$\mu_{\gamma \gamma}$ in the palette-color in the bottom, right plot of figure~\ref{fig:gwscanplots}. The plot shows that the $\mu_{\gamma \gamma}$  significantly diverges from 1 for larger GW peak amplitudes. 
We find that most of the region of parameter space of FOPT, which can be probed via the GW experiments, $\mu_{\gamma \gamma}$ is more than $3\sigma$ away from the current ATLAS and CMS limits, whereas few points lie within the $3\sigma$ limit from the latest observations. 
Therefore, most of the region of parameter space of this scenario that lies within the sensitivity of various GW detectors are already excluded from the latest precision study of SM-like Higgs boson, particularly $h^0 \rightarrow \gamma \gamma$ decays channel.
The region that lies within the $3\sigma-$limit entire region is expected to be tested from the precision study of SM-like Higgs boson in the HL-LHC and/or future colliders. This scenario has an interesting interplay between LHC physics and various GW detectors regarding the possibility of FOPT in the present model. The possibilities of detecting GW at the future proposed GW detectors due to FOPT from the current model would be severely limited in the absence of new physics at the HL-LHC. 

In the above discussions, we talk about the phase transition that always happens in the $h_d$-field direction. We have only discussed how different model parameters from the triplet sector of this model affect the 
phase transition quantities and the correlations
 the production of stochastic GW signals and the LHC searches.
We find that such correlations excluded most of the regions of the triplet sector that can produce GW as a result of an FOEWPT.
On the other hand, the model parameters from the dark sector, particularly $\lambda_{SH}$ and $m_S$, can play a crucial role in altering the SM Higgs potential in favour of FOEWPT.
 Before looking at the influence of the dark sector parameters on the FOPT in the next section, we first provide two benchmark points from the above scan results to illustrate the effect of the triplet-sector parameters on FOEWPT where the dark sector couplings are small. 
 The implications of the dark sector parameters on the FOPT and the relationships between the generated GW and other DM experimental, observational constraints are then discussed in section~\ref{FOPTduetoDM}.
\begin{table}[t!]
\renewcommand{\arraystretch}{1.14}
\begin{center}
{\tiny\fontsize{8.2}{7.7}\selectfont{
\begin{tabular}{|c|@{\hspace{0.06cm}}c@{\hspace{0.06cm}}|@{\hspace{0.06cm}}c@{\hspace{0.06cm}}|@{\hspace{0.06cm}}c@{\hspace{0.06cm}}|c|@{\hspace{0.06cm}}c@{\hspace{0.06cm}}|c@{\hspace{0.06cm}}|c|@{\hspace{0.06cm}}c@{\hspace{0.06cm}}|c|@{\hspace{0.06cm}}c@{\hspace{0.06cm}}|}
\hline
\Tstrut
Input/Observables & \makecell{BP1} & \makecell{BP2} \\
\hline
\Tstrut
$\lambda_1$   &  3.92 &3.59 \\
$\lambda_4$  & $-0.989$ & $-0.56$ \\
$M_{\Delta}$     & 367.7 & 366.0 \\
$v_{t}$~(GeV)   & $3.3 \times 10^{-4}$ & $6.1 \times 10^{-4}$  \\
$\mu_3$~(GeV)  & $-540.9$ & 441.1  \\
$\lambda_{SH}$    & 0.0053 & 0.0013   \\
$\lambda_{ST}$     & 0.913 & 0.02 \\
$m_{_{\text DM}}$   & 432.2 & 420.0  \\[0.05cm]
\hline
$m_{H^{++}}$~(GeV)   & 406.4 & 388.3  \\
$m_{H^{+}}$~(GeV)    & 387.5 & 377.3\\
$m_{H^{0} \sim A^0}$~(GeV)      & 367.7 & 365.9 \\
$m_{h}$~(GeV)    & 125  & 125 \\ 
$\sin\theta_t$~(GeV)      & $-1.03 \times 10^{-6}$ & $5.69 \times 10^{-7}$  \\[0.05cm] \hline
$\Omega h^2$ & 0.008 & 0.120  \\
$\xi_{\text{DM}} \, \sigma_{\rm DD}^{\rm SI}$~(cm$^2$)    & $(9.17)9.31\times 10^{-50}$ &  $8.81 (8.99)\times 10^{-50}$    \\[0.15cm] \hline
SNR (LISA) & 16.7 & $<< 1$ \\
[0.07cm]\hline
\end{tabular}
}}
\caption{Set-I benchmark scenarios with relatively larger (smaller) portal couplings between the triplet (singlet) and the SM Higgs doublet that exhibit an FOEWPT. Shown are the various model input parameters, relevant masses, mixing, DM relic density, DMDDSI cross-section and the signal-to-noise ratio of the produced GW as a result of an FOEWPT in the LISA experiment. The details of the EWPT of these benchmark scenarios are presented in tables~\ref{phasetransitiontable_set1} and~\ref{tab:Table_GW_data_set1}.}
\label{tab:BP_set1}
\end{center}
\end{table}
%
 %
\subsubsection{Benchmark scenarios (Set-I)}
\label{sbenchmarkpoints}
In this subsection, we present two benchmark points, BP1 and BP2, in table \ref{tab:BP_set1} to illustrate the effect of the triplet-sector parameters on the FOEWPT. 
In the triplet-sector, $v_{t}$ is around $10^{-4}$ GeV, $H^{\pm \pm}$ around 400 GeV and for BP1 (BP2) $\Delta m$ is around 20 GeV (10 GeV).
In the dark sectors, $m_s > m_{H^0,A^0,H^{\pm},H^{\pm\pm}}$ and the DM annihilation cross-section was large at the early Universe as a result of the opening of the various annihilation channels of the triple-sector.
Since $\lambda_{S\Delta}$ is smaller in BP2 than in BP1, the DM relic density is lower in BP1. 
Lower values of $\lambda_{SH}$ in BP2 result in a substantially low $\sigma_{\rm DD}^{\rm SI}$ compared to BP1. 
However, relic density scaling sets $\sigma_{\rm DD}^{\rm SI}$ value at the same range. For both BP1 and BP2, $\sigma_{\rm DD}^{\rm SI}$ is below the neutrino floor. Therefore, it is difficult to probe from the DMDD experiments. However, a significant amount of parameter space left will be probed (corresponds to relatively larger $\lambda_{SH}$) from the future DMDD experiments. 
%
\begin{table}[t!]
\renewcommand{\arraystretch}{1.3}
\small{\begin{tabular}{|c||c|c|c|c|c|c|}
\hline
 BM No &\multicolumn{2}{c|}{$T_i$ (GeV)} & ${\{h_d, h_t, h_s\}}_{\text{false}}$ & $\xrightarrow[\text{type}]{\text{Transition}}$ & ${\{h_d, h_t, h_s\}}_{\text{true}}$ (GeV) & $\gamma_{_{\rm{EW}}}$ \\
\hline 
\hline
\multirow{2}{*}{BP1} & $T_c$ & 76.9 & \{0, 0, 0\} & FO & \{230.1, 0, 0\} & \\   
\cline{2-6} 
 & $T_n$ & 58.1 & \{0, 0, 0\} & ,, & \{241.3, 0, 0\} & 4.15\\ 
\hline
\hline
\multirow{2}{*}{BP2} & $T_c$ & 115.0 & \{0, 0, 0\} & ,, & \{120.9, 0, 0\} & \\   
\cline{2-6} 
 & $T_n$ & 114.4 & \{0, 0, 0\} & ,, & \{130.4, 0, 0\} & 1.14\\ 
\hline
\end{tabular}}
   \caption{Phase transition characteristics  of the benchmark scenarios BP1 and BP2 presented in table~\ref{tab:BP_set1}. Values of $T_c$, $T_n$, the corresponding field values at the false and true phases and the strength of the phase transition, ($\gamma_{_{\rm{EW}}}$), along the $SU(2)$-direction are presented for each benchmark scenario. `FO' corresponds that the phase transition is first-order type.}
  \label{phasetransitiontable_set1}
\end{table}
%
%
\begin{figure}[t!]
\begin{center}
\includegraphics[height=5.9cm,width=0.49\linewidth]{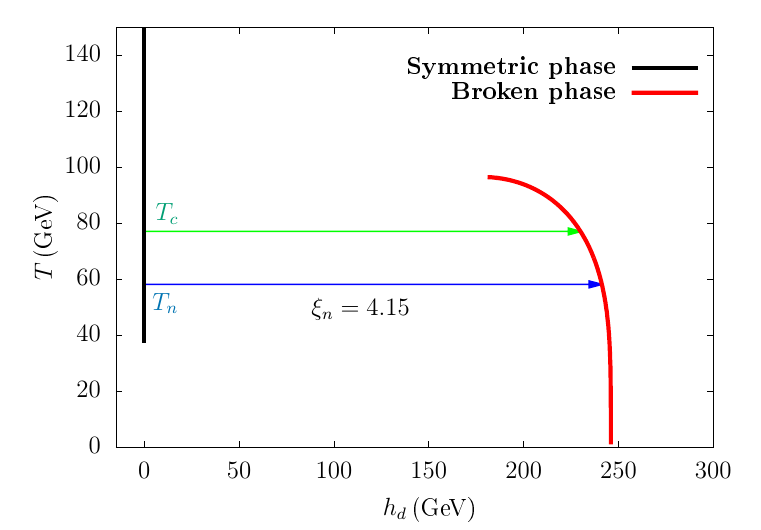}
\hskip 5pt
\includegraphics[height=5.9cm,width=0.49\linewidth]{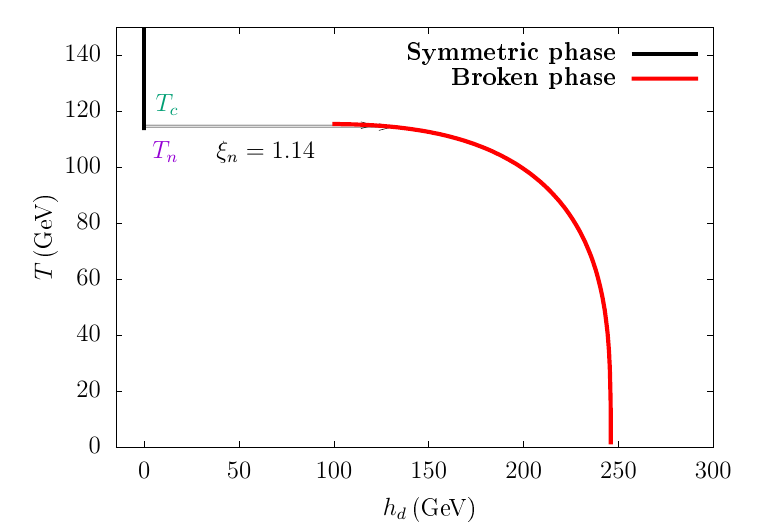}
\caption{Phase flows in benchmark scenarios BP1 and BP2. Individual color corresponds to a specific
minimum of the potential (phase) while each line indicates the evolution of the phase
in the $h_d-$field direction as a function of temperature.
The arrows reflect the direction of transition from a false vacuum to a true vacuum, as calculated at $T_c$ and $T_n$ along the $h_d$-field. 
See the text for the details.}
\label{phasediagram_set1}
\end{center}
\vspace{-0.5cm}
\end{figure}
%

 The values of $T_c$ and $T_n$, the strength of the phase transition $\xi_n$ and the field values of the phases at $T_c$ and $T_n$ are presented in Table~\ref{phasetransitiontable_set1}. The calculation of $T_c$ and $T_n$ indicates that these EWPTs are a one-step transition. In BP1, large supercooling is possible since $T_n$ and $T_c$ have a significant difference. This makes the strength of the transition relatively large. On the other hand, for BP2, the $T_n$ and $T_c$ are very close to each other. Also, $T_n$ in BP1 is much smaller than in BP2.

We present the 
evolution of the symmetric and broken phases with temperature in figure~\ref{phasediagram_set1} for these two benchmark points. Each line (red and black color) shows the field values at a particular minimum as a function
of temperature. The black line represents the symmetric phase, whereas the red line indicates the broken phase. 
In addition, the green arrow indicates that an FOPT may happen in the direction of the arrow since at that temperature and the two phases connected by the arrow are degenerate.
The blue color arrow shows the transition direction at the nucleation temperature along which phase transition actually starts.
In contrast to BP2, where $T_c$ and $T_n$ are pretty near, BP1 exhibits a significant separation between them, resulting in a stronger phase transition than BP2. 
The evolution of the broken phase for both benchmark points indicates that the $h_d$ value (red line) finally evolves to 246 GeV at $T=0$. Thus, the Universe finally evolves to the correct EW minima, i.e., $v = \sqrt{v_d^2 + v_{t}^2} = 246$ GeV. Note that the contribution of the triplet $\vev$ to $v$ is minimal as $v_t \sim 10^{-4}$ GeV at $T= 0$.

\begin{table}[t]
\renewcommand{\arraystretch}{1.25}
   \centering
   \setlength{\extrarowheight}{5.5pt}
\small{\begin{tabular}{|c|c|c|c|c|c|}
\hline   
 \multirow{1}{*}{BM No.}& \multirow{1}{*}{$T_n$ (GeV)} & \multirow{1}{*}{$\alpha$} & \multirow{1}{*}{$\beta/H_n$} \\
\hline 
  BP1 & 58.1  & $0.325$ & $33.58$\\
   \hline
  BP2 & 114.4  & 0.021 & 26689.5\\
\hline
\end{tabular}}
   \caption{Values of the parameters $T_n$, $\alpha$ and $\beta/H_n$ (that control the GW intensity) for the benchmark points BP1 and BP2, presented in
   table~\ref{tab:BP_set1}.}
   \label{tab:Table_GW_data_set1}
   \vspace{-0.3cm}
\end{table}
%
%
\begin{figure}[t!]
\begin{center}
\includegraphics[height=6.6cm,width=0.6\linewidth]{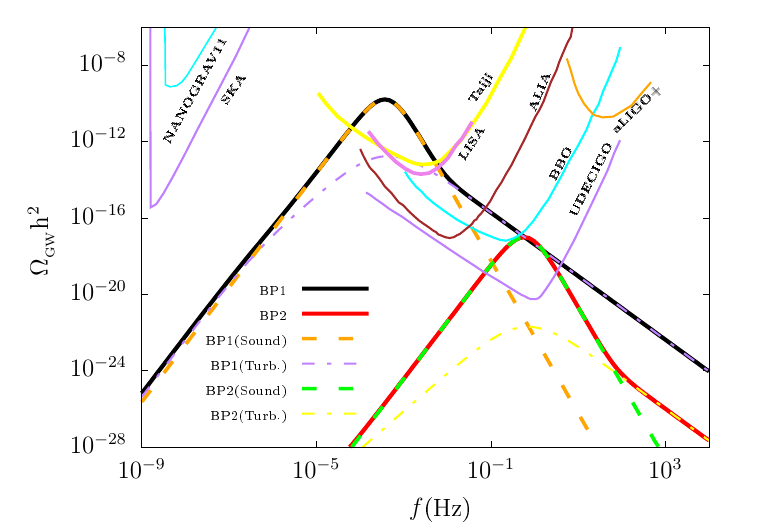}
\caption{GW energy density spectrum with respect to frequency for the benchmark scenarios BP1 and BP2 illustrated
against the experimental sensitivity curves of the future proposed GW detectors such as LISA,  SKA, TianQin, Taiji, aLigo+,
BBO, NANOGrav, and U-DECIGO. 
The solid lines indicate the overall GW energy density produced for the specific benchmark scenario, while the broken lines in various colours reflect the individual contributions from sound waves and turbulence. See the text for details.}
\label{fig:GW-freq-plot}
\end{center}
\vspace{-0.5cm}
\end{figure}
%
Various key parameter values ($\alpha$, $\beta$) pertaining to the GW spectra generated from the FOPTs for these two benchmark scenarios (BP1 and BP2) are presented in table~\ref{tab:Table_GW_data_set1}. The corresponding
GW (frequency) spectra are presented in figure~\ref{fig:GW-freq-plot} using equations~\ref{TotalGW} to \ref{eq:h-star}.
For each phase transition process, the contributions from sound waves and turbulence are shown in different colors with broken lines. 
The GW peak amplitude and the peak frequency are mostly dominated by sound contributions.
These GW spectra are further compared with the projected sensitivity of some ground- and space-based GW detectors, viz., LISA~\cite{LISA:2017pwj}, ALIA~\cite{Gong:2014mca}, Taiji~\cite{Hu:2017mde}, TianQin~\cite{TianQin:2015yph}, aLigo+~\cite{Harry:2010zz}, SKA~\cite{Carilli:2004nx, Janssen:2014dka, Weltman:2018zrl}, NANOGrav~\cite{McLaughlin:2013ira},   Big Bang Observer (BBO)~\cite{Corbin:2005ny} and  Ultimate(U)-DECIGO~\cite{Kudoh:2005as} in Fig.~\ref{fig:GW-freq-plot}. Note that, a significant amount of supercooling happens in BP1 and the corresponding nucleation temperature is relatively low compared with BP2. This also enhances the $\alpha$ value for BP1 compared with BP2. Also, the nucleation takes a much longer time for BP1 compared with BP2. The $\beta$ parameter, which indicates the inverse of the duration of the phase transition, is smaller in BP1 than in BP2. Thus, as the FOPT happens much stronger in BP1, the GW spectrum peak is expected to be much larger in BP1 than in BP2. This behaviour can be observed in Fig.~\ref{fig:GW-freq-plot}. Note that the GW intensity obtained for BP1
lies within the sensitivities of LISA, Taiji, ALIA, BBO and UDECIGO, while BP2 lies only within
the sensitivity of UDECIGO and marginally touches the BBO sensitivity.

The quantity known as the signal-to-noise ratio (SNR) is used to measure the detection of the GW signal in different experiments. It is defined
as~\cite{Caprini:2015zlo}
\beq\small{
\text{SNR} = \sqrt{\delta\times \mathcal{T}\int_{f_{min}}^{f_{max}}df\bigg[\frac{h^2\Omega_{\text{GW}}(f)}{h^2\Omega_{\text{exp}}(f)}\bigg]^2},}
\eeq
where $\delta$ corresponds to the number of independent channels for
cross-correlated detectors (to determine the stochastic origin of GW). It is 2 for BBO, U-DECIGO and 1 for LISA. The duration of the experimental mission in years is defined by $\mathcal{T}$.  In this work, we consider $\mathcal{T}=5$. The effective power spectral density of the experiment's strain noise is indicated by $\Omega_{\text{exp}}(f) \, h^2$. For the detection prospects of these experiments, the SNR values need to cross the threshold value which depends on the configuration details of the experiment. Such as, the recommended threshold number is around 50 for a four-link LISA setup, however, a six-link design allows for a much lower value of around 10~\cite{Caprini:2015zlo}. 
In this work, we calculate the SNR value only for LISA for these two benchmark points. For BP1, it is 16.7 and for BP2, the SNR value is way below 1. It is expected that the SNR value of BP2 will be substantially lower because its GW signal power spectrum does not fall within the LISA sensitivity curves. However, other experiments like U-DECIGO can detect this type of benchmark scenario.

Note that in both the benchmark scenarios, $\mu_{\gamma \gamma}$ is more than the  $3\sigma$ away from the latest LHC limits. Thus, as we discussed in the description of the right plot of figure~\ref{fig:gwscanplots}, FOEWPT preferred region of parameter space of the triplet sector is mostly excluded from the precision study of the SM Higgs boson di-photon decay modes. 
Since the FOEWPT is crucial for the EWBG, we investigate its viability in the current model by changing the parameters of the dark sector. 
In the next section, we will examine such possibilities and determine the consequences of the dark sector parameters on the FOEWPT as well as the establishing connections between the produced GW and other DM experimental constraints.

\subsection{FOPT from the dark sector}
\label{FOPTduetoDM}
In the previous discussions of this article, we set the dark sector portal couplings $\lambda_{SH}$ and $\lambda_{S\Delta}$ at smaller values and the phase transition along the $h_d-$field direction is mostly controlled by the triplet sector parameters of the model, i.e., $\lambda_1$, $\lambda_4$ and $M_{\Delta}$. However, larger DM Higgs portal coupling, particularly $\lambda_{SH}$ as the $h_d-$field couples with the $h_s-$field through this portal coupling in the dark sector, can alter the Higgs potential in such a way that FOEWPT occurs.
In addition to this, another intriguing possibility is the FOPT in the dark sector. 
In this case, $h_s-$field changes from zero $\vev$ at high temperature to non-zero $\vev$ at the electroweak temperature scale, and then it evolves to zero $\vev$ at $T=0$.
Moreover, a two-step FOPT in the scalar potential is even another intriguing possibility. In this category, the $h_s-$field acquires non-zero $\vev$ at a certain temperature in the early Universe through an FOPT and subsequently, relatively at a lower temperature another phase transition occurs where the $h_d-$field develops a non-zero $\vev$ whereas the $\vev$ of the $h_s-$field goes from a non-zero value to a zero value.
It is crucial to remember that among these three forms of phase transitions, the one-step phase transition along the $h_s-$direction is capable of generating GW signal but is ineffective for the electroweak baryogenesis. 

\begin{figure}[t!]
\begin{center}
\includegraphics[height=5.6cm,width=0.46\linewidth]{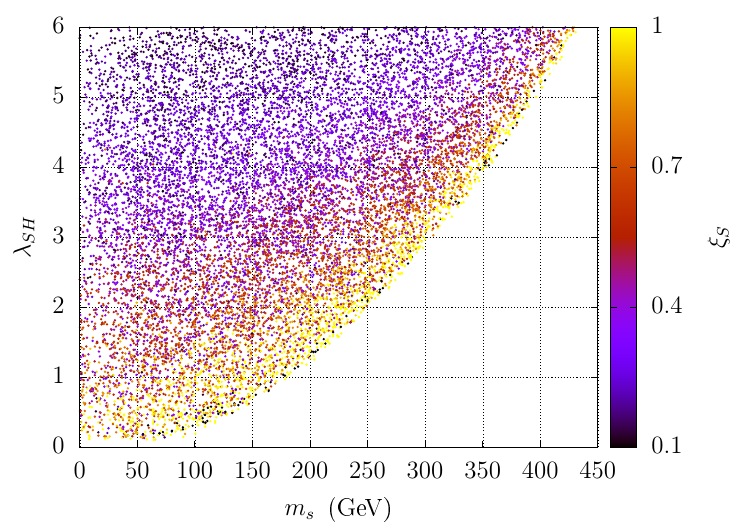}
\hskip 10pt
\includegraphics[height=5.6cm,width=0.46\linewidth]{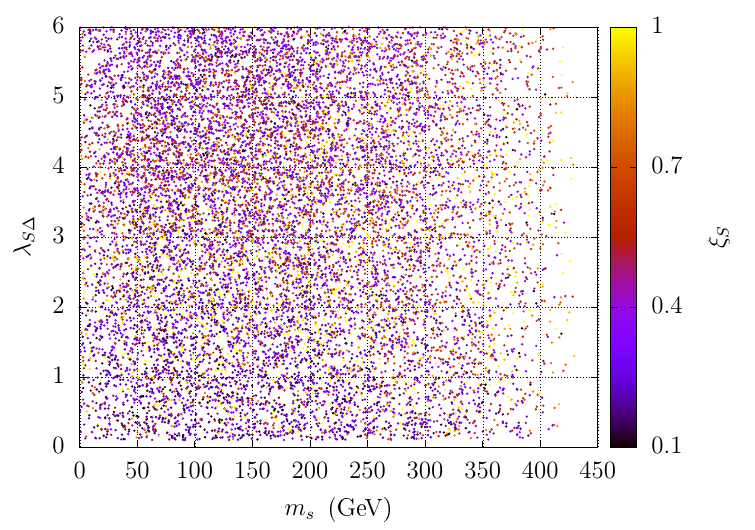}
\hskip 10pt
\caption{Scatter plots in the $m_S-\lambda_{SH}$ ($m_S-\lambda_{S\Delta}$) plane in the left (right) plot that exhibits FOPT along the $h_s-$field direction. The variation of the strength of the phase transition $\xi_S$ is shown via the palette.}
\label{fig:dmpt_strength}
\end{center}
\vspace{-0.5cm}
\end{figure}

To study the phase transition along the $h_s-$field direction, we first fix the triplet sector parameters the same as the benchmark scenario BP1 and vary the dark sector parameters $m_S$, $\lambda_{SH}$, $\lambda_{S\Delta}$, $\lambda_{S}$ and $\mu_3$. The DM phenomenology due to the variation of these DM input parameters for the same triplet-sector input parameters has already been discussed in section~\ref{DMpar}. In figure~\ref{fig:dmpt_strength}, we present the variation of $m_s$ and $\lambda_{SH}$ ($\lambda_{S\Delta}$) in the left (right) plot that exhibits FOPT along the $h_s-$field direction. The strength of the phase transition along the $h_s-$field direction, i.e., $\xi_s = \frac{<h_s>}{T_n}$ where $<h_s>$ is the $\vev$ of the $h_s$ field at the broken phase at $T = T_n$, is presented via the palette color.

In the left plot, the variation of $m_s$ and $\lambda_{SH}$ reveals that for the phase transition along the $h_s-$field direction, $\lambda_{SH}$ must be large for larger $m_s$ and lower for smaller $m_s$. The region of parameter space with $m_s \gtrsim 425$~GeV is almost disfavoured for FOPT for $\lambda_{SH} \lesssim 6.0$.
 On the other hand, the right plot indicates that the phase transition does not have a preferential bias over the choice of $\lambda_{S\Delta}$. 
Equation~\ref{DMmassT0} elucidates such connection between $m_s$ and $\lambda_{SH}$.
The phase transition along the singlet direction generally prefers relatively low bare mass squared term ($\mu_s^2$) in the potential~\cite{Ghorbani:2018yfr}.
Therefore, the parameter space that corresponds to an FOPT along the $h_s$ direction, the maximum possible value of $m_s$ is mainly controlled by the term $\lambda_{SH} v_d^2$ and it is almost independent of $\lambda_{S\Delta}$ as $v_t << 1$. The color variation in the left plot reveals that the strength of the phase transition is maximum at the outer edge, corresponding to the parameter space consisting of very small bare mass $\mu_s$.

The variations of these parameters will also have effects on the observables of the dark sector, such as the $\Omega_{S}h^2$ and $\sigma_{\rm DD}^{\rm SI}$, and make these scenarios more sensitive to the various DM experiments. 
It is therefore anticipated that the production of GW as a result of this type of FOPT in the dark sector will be connected with DM observables.
In the following subsection, we discuss the relationship between the DM phenomenology and FOPT along the $h_s-$field direction.
Note that, in the previous discussion, we found a similar type of correlation between the FOEWPT-induced GW intensity and the precision measurement of the signal strength, $\mu_{\gamma \gamma}$ of the $\hsm \rightarrow \gamma \gamma$ channel at the LHC.

%
\subsubsection{Correlations among the DM observables and an FOPT}
\label{DMPTcorrelation}
In this section, we now turn our attention to the link between the DM observables and FOPT along the $h_s-$field direction.  There is an effect on the various DM observables from the parameter space that exhibits such a phase transition.

\begin{figure}[t!]
\begin{center}
\includegraphics[height=5.6cm,width=0.46\linewidth]{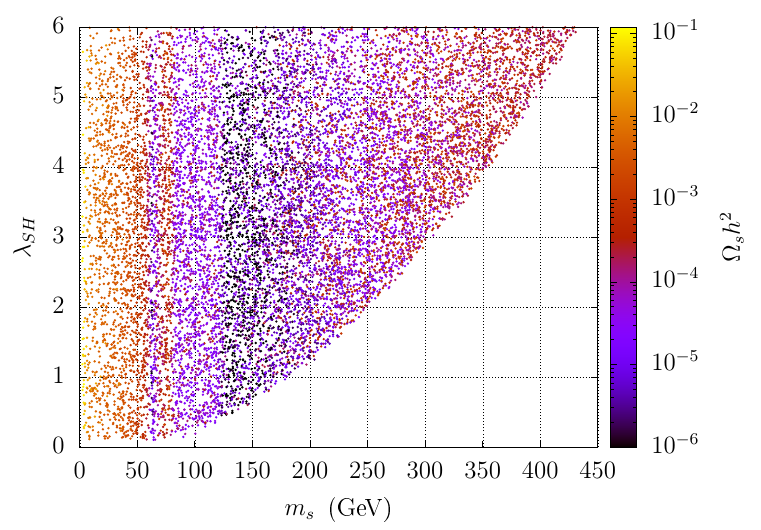}
\hskip 10pt
\includegraphics[height=5.6cm,width=0.46\linewidth]{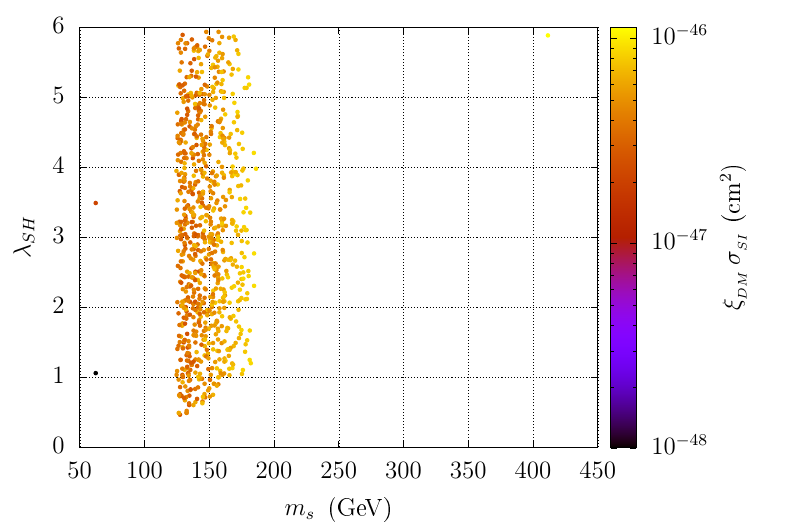}
\hskip 10pt
\caption{Left: Scatter points that exhibits FOPT along the $h_s-$field direction and $\Omega_S h^2 < 0.120$ in the $m_S-\lambda_{SH}$ plane. Right: Similar to the scatter spots on the left figure, but comply with the latest DMDDSI experimental limits. The variation of $\Omega_S h^2$ ($\sigma_{\rm DD}^{\rm SI}$) is shown via the palette in the left (right) plot.}
\label{fig:dmpt_correlation}
\end{center}
\vspace{-0.5cm}
\end{figure}

The left plot figure~\ref{fig:dmpt_correlation} is otherwise similar with the left plot of figure~\ref{fig:dmpt_strength} but here the points satisfy the Planck experimental constraints on the relic density, i.e., $\Omega_s h^2 \leq 0.120$.
 The relic density of the singlet scalar field is indicated by the palette color in the plot. Note that, as we described earlier $m_s < \hsm$, $\lambda_{SH}$ mostly controls the $\Omega_s h^2$ whereas for $m_s > \hsm$, the $\mu_3$ proportional semi-annihilation process starts to contribute to DM density. $\lambda_{S\Delta}$ starts to contribute significantly when $m_s$ is larger than the triplet-like scalar masses which is in this case around $\sim$ 400 GeV.

The $CP$-even scalars $h^0$ and $H^0$ contribute $\sigma_{\rm DD}^{\rm SI}$.
It has been discussed in section~\ref{darkmatter} that for smaller values of $v_t$ and $\sin\theta_t$, $\sigma_{\rm DD}^{\rm SI}$ mostly depends on the SM-like Higgs boson medicated process and it increases with $\lambda_{S H}$.
Such enhancement of $\sigma_{\rm DD}^{\rm SI}$ with $\lambda_{SH}$ excludes a significant amount of parameter space from the latest constraints from XENON-1T and PANDA-4T. 
On the other hand, if the DM mass is larger than the triple-like scalars then larger $\lambda_{S\Delta}$ decreases the relic density without affecting the $\sigma_{\rm DD}^{\rm SI}$. With this, at a relatively larger $\lambda_{S\Delta}$ value, the DM could be underabundant. The small scaling factor due to the low relic density actually reduces the effective $\sigma_{\rm DD}^{\rm SI}$ value. Such scenarios allow the parameter space with larger $\lambda_{SH}$ which is otherwise excluded from the latest constraints.
%

In the right plot of figure~\ref{fig:dmpt_correlation}, we present the points of the left plot that pass the latest $\sigma_{\rm DD}^{\rm SI}$ constraint from the XENON-1T and PANDA-4T experiment. The effective $\sigma_{\rm DD}^{\rm SI}$ is presented by the palette color.
Note that due to the requirement of larger $\lambda_{SH}$ for the FOPT a significant amount of the parameter space is excluded from the $\sigma_{\rm DD}^{\rm SI}$ constraints. Note that for $m_s > h^0$, the $\mu_3$ dependent semi-annihilation process starts to contribute and because of that the relic density drops significantly which helps the effective $\sigma_{\rm DD}^{\rm SI}$ rates to remain within the exclusion limit result of a significant downward scaling. Because of that, we find some parameter space for $m_s$ after the SM-like Higgs Boson threshold value. Therefore, it is important to note that the term $\mu_3$ has an important role in the dark sector and the FOPT sector. As it generates the semi-annihilation processes for the DM annihilation, also it can alter the potential shape and generate a barrier between the minima which is necessary for an FOPT.
In the scenario for $m_s$ larger than the triplet-like scalars, the term proportional to $\lambda_{S\Delta}$ starts to contribute to the DM annihilation, which can drastically reduce the relic density.
However, in this scenario, as the triplet-like scalar masses are relatively large $\sim 400$~GeV, this effect does not start to occur significantly in the FOPT preferred allowed parameter space as $m_s$ is bounded up to $\sim425$~GeV for $\lambda_{SH} \lesssim 6.0$. 

The discussion of this subsection points out the strong correlations between the FOPT-favoured parameter space and the DM observables.
Such correlations rule out a significant amount of parameter space of having an FOPT along the $h_s-$field direction.
In this case,  as we consider the triplet-like scalars a bit heavier, approximately $\sim$400~GeV, the contribution from the $\lambda_{S\Delta} h_d^2 h_s^2$ interaction term in the potential on the relic density estimation significantly start to contribute for comparatively higher $m_s$. 
In the scenario with relatively smaller triplet-like scalars, the same interaction becomes significant at relatively small $m_s$ that can open up much more allowed parameter points satisfying the latest DMDDSI cross-section limits from the XENON-1T and PANDA-4T experiments due to downward scaling of the underabundant DM. In the next subsection, we choose a benchmark scenario from that category where the triplet-like scalars are relatively low, around $\sim$200~GeV, and exhibit FOEWPT in the early Universe and discuss the such correlation between the FOPT, production of GW and the DM observables is also discussed.
\subsubsection{Benchmark scenario (Set-II)}
\label{BPset2}
\begin{table}[t!]
\renewcommand{\arraystretch}{1.14}
\begin{center}
{\tiny\fontsize{8.2}{7.7}\selectfont{
\begin{tabular}{|c|@{\hspace{0.06cm}}c@{\hspace{0.06cm}}|@{\hspace{0.06cm}}c@{\hspace{0.06cm}}|@{\hspace{0.06cm}}c@{\hspace{0.06cm}}|c|@{\hspace{0.06cm}}c@{\hspace{0.06cm}}|c@{\hspace{0.06cm}}|c|@{\hspace{0.06cm}}c@{\hspace{0.06cm}}|c|@{\hspace{0.06cm}}c@{\hspace{0.06cm}}|}
\hline
\Tstrut
Input/Observables & \makecell{BP3} \\
\hline
\Tstrut
$\lambda_1$  & 0.17 \\
$\lambda_4$   & $-0.30$ \\
$M_{\Delta}$   & 193.2 \\
$v_{t}$~(GeV)   & $4.5 \times 10^{-4}$  \\
$\mu_3$~(GeV)   & -100.8  \\
$\lambda_{SH}$     & 1.2   \\
$\lambda_{ST}$     & 6.9 \\
$\lambda_{ST}$      & 2.17 \\
$m_{_{\text DM}}$  & 252.2 \\[0.05cm]
\hline
$m_{H^{++}}$~(GeV)  & 215.3  \\
$m_{H^{+}}$~(GeV)    & 204.5\\
$m_{H^{0} \sim A^0}$~(GeV)      & 193.2 \\
$m_{h}$~(GeV)    & 125 \\ 
$\sin\theta_t$~(GeV)      & $8.0 \times 10^{-6}$  \\[0.05cm] \hline
$\Omega h^2$ & $4.3 \times 10^{-5}$  \\
$\xi_{\text{{DM}}} \, \sigma_{\rm DD}^{\rm SI}$~(cm$^2$)  &  $7.1\times 10^{-47}$    \\[0.15cm] \hline
SNR (LISA)& $<< 1$ \\
[0.07cm]\hline
\end{tabular}
}}
\caption{Set-II benchmark scenario with relatively smaller (larger) portal couplings between the triplet (singlet) and the SM Higgs doublet that exhibits a two-step EWPT in the early Universe. Similar to table~\ref{tab:BP_set1}, various model input parameters, masses, mixing, DM relic density, DMDDSI cross-section and the signal-to-noise ratio of the produced stochastic GW signal in the LISA experiment are shown. The details of the phase transition are shown in tables~\ref{phasetransitiontableBP3} and~\ref{tab:Table_GW_databp3}.}
\label{tab:BPs}
\end{center}
\end{table}
%
In this section, we present a benchmark point (BP3) exhibiting a two-step FOPT in the early Universe. We consider a relatively light triplet-like scalar of around 200 GeV and much smaller $\lambda_1$ and $|\lambda_4|$ values such that $\mu_{\gamma\gamma}$ is within the 1$\sigma$ (2$\sigma$)-limit of ATLAS (CMS) latest results. 
Scenarios like this demand larger quartic couplings (particularly $\lambda_{SH}$) in the dark sector and smaller DM mass for an FOPT in the early Universe.
In the discussion in section~\ref{EWBGregion}, we consider very low quartic couplings of the dark sector, i.e., much smaller $\lambda_{SH}$ and $\lambda_{S\Delta}$. This can be seen in BP1 and BP2 in table~\ref{tab:BPs}.
Because of such consideration, we do not find points with smaller quartic coupling points in the triplet-sector, i.e., parameter points with smaller $\lambda_1$ and $|\lambda_4|$, in section~\ref{EWBGregion} that can generate an FOEWPT in the early Universe.
 
Compared to the scenarios discussed in section~\ref{EWBGregion}, in the BP3  $\lambda_{SH}$ is relatively large$\sim1.2$, which facilitates successful FOPT in the scalar potential. 
However, $\sigma_{\rm DD}^{\rm SI}$ is expected to be much larger as a result of a large coupling. To comply with the latest experimental constraints from XENON-1T and PANDA-4T we consider $m_{s}$ larger than the triplet-like scalar masses and much larger $\lambda_{S\Delta}$ $\sim 6.9$, such that the 
 $S~ S^* ~\to {\rm X~~Y}; ~~~\{\rm X,Y\}=\{H,A^0,H^\pm, H^{\pm\pm}\}$ annihilation processes open up and contribute significantly to the annihilation cross-section so that DM becomes underabundant. 
 Such a lower relic density corresponds to a much smaller scaling fraction $\xi_{\text{DM}}$ which
 helps the computed $\sigma_{\rm DD}^{\rm SI}$ rate to comply with the latest experimental stringent limits. 
 
Note that the value of $m_S$ is also crucial for the FOPT and cannot be too large.
Due to the comparatively smaller triplet-like scalar masses $\sim$ 200 GeV in this benchmark point, we can keep $m_S$ within the range of $\sim 250$ GeV and also take the benefit of the downward relic density scaling factor on the DMDDSI experimental constraints.
This aids in keeping the DMDDSI cross-section of BP3 below the most recent constraints from the XENON-1T and PANDA-4T experiments.
\begin{table}[t]
\renewcommand{\arraystretch}{1.4}
\small{\begin{tabular}{|c||c|c|c|c|c|c|}
\hline
 BM No &\multicolumn{2}{c|}{$T_i$ (GeV)} & ${\{h_d, h_t, h_s\}}_{\text{false}}$ & $\xrightarrow[\text{type}]{\text{Transition}}$ & ${\{h_d, h_t, h_s\}}_{\text{true}}$ (GeV) & $\gamma_{_{\rm{EW}}}$ \\
\hline 
\hline
\multirow{4}{*}{BP3} & \multirow{2}{*}{$T_c$} & 163.2 & \{0, 0, 0\} & FO &\{0, 0, 14.8\} & \\
\cline{3-6}
 &  & 145.7 & \{0, 0, 63.6\} & ,, & \{78.9, 0, 0\} &\\ 
\cline{2-6} 
 & \multirow{2}{*}{$T_n$} & 163.1 & \{0, 0, 0\} & ,, & \{0, 0, 15.1\} &\\
 \cline{3-6} 
 &  & 116.3 & \{0, 0, 86.7\} & ,, & \{197.3, 0, 0\} & 1.7 \\ 
  \hline
\end{tabular}}
   \caption{Phase transition characteristics  of the benchmark point BP3 presented in table~\ref{tab:BPs}. Values of $T_c$, $T_n$, the corresponding field values at the false and true phases and the strength of the phase transition, ($\gamma_{_{\rm{EW}}}$), along the $SU(2)$-direction are presented. `FO' denotes to FOPT.
   }
  \label{phasetransitiontableBP3}
\end{table}
%
%
\begin{figure}[t!]
\begin{center}
\includegraphics[height=5.9cm,width=0.49\linewidth]{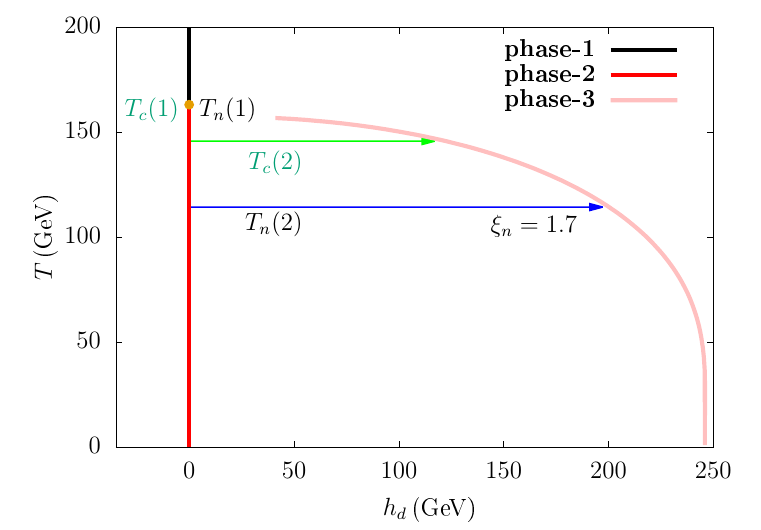}
\hskip 5pt
\includegraphics[height=5.9cm,width=0.49\linewidth]{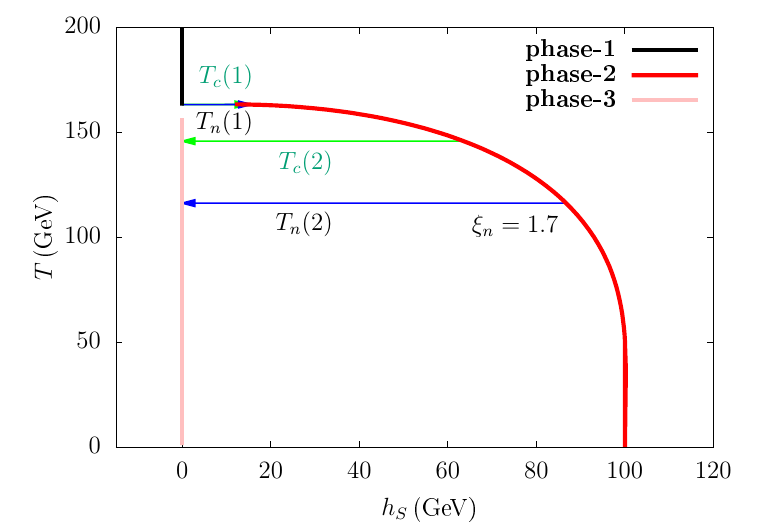}
\caption{Phase flows in benchmark point BP3. Individual color corresponds to a specific
minimum of the potential (phase) while the each line indicates the evolution of the phase
in the $h_d-$field direction in the left and $h_s-$field direction in the right as a function of temperature.
The arrows reflect the direction of transition from a false vacuum to a true vacuum, as calculated at $T_c$ and $T_n$ along the $h_d$-field, whereas an orange bullet indicates that the corresponding field value does not change much throughout that transition.
See the text for the details.}
\label{phasediagrambp3}
\end{center}
\vspace{-0.5cm}
\end{figure}
%
In contrast with BP1 and BP2, the phase transition pattern is quite different in BP3.
The values of $T_c$ and $T_n$ for the BP3 benchmark scenario are presented in table~\ref{phasetransitiontableBP3} and it suggests that the transition is a two-step FOPT. The evolution of the field and the various phases with temperature are presented in figure~\ref{phasediagrambp3}. 
Note that three types of phase exist. `phase-I' denotes the trivial phase, i.e., all the field values are always zero.
On the other hand, `phase-II' corresponds to the minimum where only the $h_s-$ field acquires a non-zero $\vev$ and other field $\vevs$ remain zero. `phase-III' corresponds to that minimum of the potential where only $h_d-$ field obtains a non-zero $\vev$ and all other field $\vevs$ remain at zero.
In the right and left plots of figure~\ref{phasediagrambp3}, the trivial minimum $\{0 ,0, 0\}$ at the high temperature is denoted by the black line (phase-I) where the broken phases in the direction of $h_s$ and $h_d$ are presented in the red color (phase-II) and magenta color (phase-III).
These lines correspond to the field values of a particular minimum as a function of temperature. The arrows in green color correspond to the temperature and field values of two phases where they degenerate and a possible FOPT can happen along the direction of the arrow. The phase at the end of the arrow indicates that it has a deeper minimum than the other phase for $T < T_c$.
The transitions finally start to occur when a bubble of the true minimum finally successfully nucleates, i.e., at $T = T_n$.
The arrows in blue color represent the direction of the FOPT at $T = T_n$ in the field space.
The small circle in yellow color represents that an FOPT happens in the scalar potential at that temperature but its field values do not change. 
\begin{table}[t]
\renewcommand{\arraystretch}{1.25}
   \centering
   \setlength{\extrarowheight}{5.5pt}
\small{\begin{tabular}{|c|c|c|c|c|c|}
\hline   
 \multirow{1}{*}{BM No.}& \multirow{1}{*}{$T_n$ (GeV)} & \multirow{1}{*}{$\alpha$} & \multirow{1}{*}{$\beta/H_n$} \\
\hline 
    \multirow{2}{*}{BP3} & \multirow{1}{*}{163.1}  & \multirow{1}{*}{$2.3 \times 10^{-4}$} & \multirow{1}{*}{$4.8 \times10^6$}\\
       & \multirow{1}{*}{116.3}  & \multirow{1}{*}{$0.027$} & \multirow{1}{*}{312.7}\\
\hline
\end{tabular}}
   \caption{Values of the parameters $T_n$, $\alpha$ and $\beta/H_n$ (that control the GW intensity) for the benchmark point BP3, shown in
   table~\ref{tab:BPs}.}
   \label{tab:Table_GW_databp3}
   \vspace{-0.3cm}
\end{table}

Note that, first, a broken phase $\{ 0, 0, 14.8\}$ appears only along the $h_s-$field direction at $T_c = 163.2$~GeV with a possibility of an FOPT.  The successful nucleation occurs at $T_n = 163.1$~GeV.
It has been shown in the right plot of figure~\ref{phasediagrambp3}. The small difference between $T_c$ and $T_n$ makes the two arrows overlap each other.
In the left plot of figure~\ref{phasediagrambp3}, a yellow circle has been shown at $T = 163.1$~GeV as during this FOPT the $h_d-$field does not acquire any non-zero $\vev$, i.e., EW symmetry remains unbroken. Thus, the first FOPT occurs from $\{h_d =0, h_t =0, h_s=0\} \rightarrow \{h_d =0, h_t =0, h_s \neq 0\}$ phase direction. 
Subsequently, later stage, another minimum form (phase-III), designated by the notation $\{h_d \neq 0, h_t =0, h_s=0\}$. 
At $T= 145.2$~GeV phase-II ($\{ 0, 0, 63.6\}$) and phase-III ($\{ 78.9, 0, 0\}$) become degenerate and for $T < 145.2$~GeV phase-III becomes the global minimum of the scalar potential. The corresponding nucleation occurs at $T = 116.3$~GeV. 
During this transition both $h_d-$ and $h_s-$ fields undergo a change, with the former acquiring a non-zero $\vev$ from zero and the latter developing a zero $\vev$ from a non-zero $\vev$ value. Note that, at $T = 0$, $h_d$ field in the phase-III evolves to the value of 246 GeV and $h_s$ remains at zero in this phase. During these phase transitions, the triplet field always remains very small. 
Therefore, the system finally evolves to the correct EW minimum. Note that in this benchmark scenario, the minimum phase-II is still present at T= 0 where $<h_s> \sim 100$~GeV and all other fields $\vevs$ are almost zero. But, the global potential minimum of the potential is the phase-III, not the phase-II. 

\begin{figure}[t!]
\begin{center}
\includegraphics[height=5.9cm,width=0.49\linewidth]{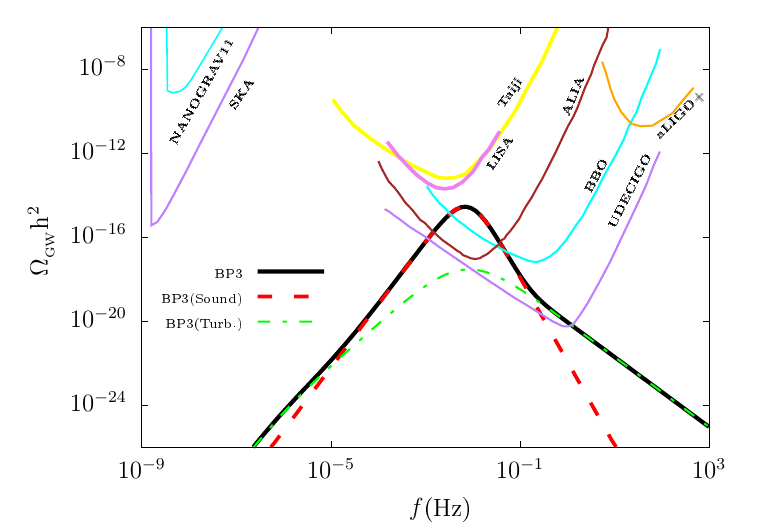}
\caption{GW energy density spectrum with respect to frequency for the benchmark point BP3 illustrated
against the experimental sensitivity curves of the future proposed GW detectors such as LISA,  SKA, TianQin, Taiji, aLigo+,
BBO, NANOGrav, and U-DECIGO. The black line denotes the total GW energy density whereas the red (green) broken line represents the contributions from sound waves (turbulence). Note that in this benchmark scenario the contribution from the first phase transition along the $h_s-$field direction does not appear in these plots as its contribution is too small.}
\label{GWplotbp3}
\end{center}
\vspace{-0.5cm}
\end{figure}

The values of the key parameters ($\alpha$, $\beta$), which are related to the computation of GW spectra generated from the FOPTs for the benchmark scenarios BP3, are presented in table~\ref{tab:Table_GW_databp3}. The
GW frequency spectrum of BP3 is presented in figure~\ref{GWplotbp3} using equations~\ref{TotalGW} to \ref{eq:h-star}.
Note that the first FOPT along the $h_s-$field direction happens very quickly (much larger $\beta/H_n$ value) and also the phase transition is not strong enough which corresponds to a much lower $\alpha$ value. Thus, the contributions of the first FOPT in the $h_s-$field direction to the total GW spectrum are very small compared to the ones coming from the second FOPT which happens along the $h_d$ and $h_s$ field directions simultaneously. Therefore, a spectral peak due to the first phase transition in this two-step phase transition does not appear in the plot.
%

Similar to figure~\ref{fig:GW-freq-plot}, here also each phase transition process, the contributions from sound waves and turbulence are shown in different colors with broken lines.
The GW peak amplitude and the peak frequency are mostly dominated by sound contributions.
This GW spectrum is further compared with the projected sensitivity of some ground- and space-based GW detectors, viz., LISA, Taiji, TianQin, aLigo+, SKA, NANOGrav, Big Bang Observer (BBO) and  Ultimate(U)-DECIGO in Fig.~\ref{GWplotbp3}.  
The GW intensity obtained for BP3
lies within the sensitivities of ALIA, BBO and UDECIGO. As expected, the SNR for LISA is very low for detection. Although the experiments like ALIA, BBO and UDECIGO can detect such a GW signal.

Therefore, the FOPT along the singlet-direction permissible parameter space we address in this section is either excluded from the most recent limits from the various DMDD experiments or will be investigated in the near future. The parameter space that is still allowed corresponds to a significantly underabundant DM.
It is projected that the HL-LHC will be unable to limit a substantial portion of the parameter space.
Focusing on this model, the observation of any signal at future DM and GW experiments but nothing at the HL-LHC would indicate that DM is severely underabundant.

\section{Summary and conclusion}
\label{subsec:summary}
In spite of many successes of the SM, it cannot explain the smallness of neutrino masses and the observed baryon asymmetry of the Universe and does not have a DM candidate.
In this article, we examine the $Z_3-$symmetric singlet scalar extended type-II seesaw model as a potential contender for simultaneously providing solutions to these three deficiencies of the SM. 
The type-II seesaw mechanism generates small neutrino masses from the $\vev$ of the scalar triplet.
The complex scalar field, $S$, of the model can be a
stable WIMP-like DM particle under the so-called freeze-out mechanism.
In addition, new interactions among the SM Higgs doublet, the complex triplet, and the complex singlet can induce an FOEWPT in the early Universe,
which is necessary for the solution of the BAU problem via the EWBG mechanism.
Such an FOEWPT can generate stochastic GW that can be detected in future proposed GW detectors.
In this work, we investigate the correlation among the production of GW, DM phenomenology, and BSM LHC searches within this model.

In this scenario, the scalar DM can annihilate mainly in three ways to freeze out in the early Universe and those processes are given by: DM annihilation$ \hspace{0.3cm}  (a)~ S~ S^{*} ~\to {\rm SM~~SM}$, (b) $S~ S^{*} ~\to {\rm X~~Y}; ~~~\{\rm X,Y\}= \{H^0,A^0,H^\pm, H^{\pm\pm}\}$ and DM semi-annihilation (c) $~ S~ S ~\to  S~U; ~~\{U\}=\{\hsm,H^0\} $. The coupling of the scalar DM with the SM Higgs doublet, $\lambda_{S H}(S^{*}S)(H^\dagger H)$, and the triplet, $\lambda_{S \Delta}(S^{*} S){\rm Tr}[\Delta^\dagger \Delta]$, control the DM-annihilation processes (a) and (b), respectively. In contrast, the DM semi-annihilation processes are initiated via the cubic term $\mu_3(S^3 + {S^*}^3)$ in the scalar potential.
Such semi-annihilation processes exist because of the $Z_3-$discrete symmetry of the model that settles the singlet as a DM particle. 
At a fixed value of $m_s$, the DMDDSI cross-section mostly depends on the $\lambda_{SH}$ portal coupling, which has an upper bound coming from the latest DMDD experiments.
When $m_s < \mhsm$, the same portal coupling $\lambda_{SH}$ predominantly controls the DM relic density.
Such dependencies on $\lambda_{SH}$ significantly restrict the DM parameter space for $m_s < \mhsm$ and only the resonance region around $\mhsm$ is still allowed after complying with the DM relic density and the latest DMDD constraints.
 The cubic interaction based on $\mu_3$ expands the parameter space for $m_s \gtrsim \mhsm$.
In addition to this, the interaction between the DM and the triplet sector through the portal coupling $\lambda_{S\Delta}$ opens up a wide range of DM allowed region of parameter space if $m_s \gtrsim m_{H^0,A^0,H^\pm, H^{\pm\pm}}$. 
Therefore, the allowed region of DM parameter space depends on the triplet-like scalar masses, which are constrained by the most recent direct searches for heavy neutral and charged scalars at the LHC.
Such phenomena show some correlations between DM phenomenology and the physics at LHC.

It has been pointed out in the literature that the direct searches at the LHC will be unable to restrict $H^{\pm\pm}$ mass even at 200 GeV with 3 ab$^{-1}$ data for moderate $v_t$ and reasonably large $\Delta m$ ($> 10$~GeV) values, which corresponds to $\lambda_4 < 0$. So, we focus on this part of the parameter space in order to investigate the feasibility of FOPT in the scalar potential at finite temperatures.
We find that the strength of an SFOEWPT along the $h_d-$field direction increases with larger $(\lambda_1+\lambda_4)$ and smaller $m_{H^{\pm\pm}}$ values.
For $\lambda_1 \lesssim 4.0$, $m_{H^{\pm\pm}}$ greater than 420~GeV is disfavored since the triplet becomes decoupled from the SM Higgs potential, rendering an SFOEWPT impossible. 

Furthermore, we have thoroughly studied the stochastic GW spectra that will carry the imprints of FOPT from new physics beyond the SM.
Relatively light triplet-like
scalars and large effective quartic couplings $(\lambda_1+\lambda_4)$ can generate GW with large enough intensity to fall within the projected sensitivity of some ground- and space-based GW detectors. 
However, we also show that the decay rate of $\hsm \rightarrow \gamma \gamma$ is significantly affected by the SFOEWPT-favored region of the parameter space, deviating significantly from the SM-predicted value for large effective quartic couplings and light triplet-like scalars.
The correlation between the produced GW spectra and the signal strength $\mu_{\gamma \gamma}$ of that decay channel has been pointed out.
LHC precision study of SM Higgs di-photon decay channel has already excluded most of the parameter space that can generate stochastic GW signal. Any further narrowing of $\mu_{\gamma \gamma}$ error bars at the HL-LHC will completely exclude the possibility of FOEWPT induced by large interactions between the SM Higgs doublet and $SU(2)$ triplet.

The interaction between the complex scalar and the SM Higgs doublet can be another source of FOPT in this model.
The singlet scalar is employed in this context to serve dual roles as a DM particle and to offer a favorable condition for the electroweak phase transition to be strongly first-order. 
Such modification in the scalar potential can also generate the possibility of a two-step FOPT, with the first transition happening in the singlet direction and the second in the $h_d-$field direction.
The scenario of an FOPT along the singlet direction prefers relatively large $\lambda_{SH}$. 
By imposing such a criteria for the FOPT, we have shown that a significant amount of the parameter space is already excluded from the latest constraints on the DMDDSI cross-section from various DM experiments.
Our analysis further shows that this model can generate an FOPT in the $h_s-$field direction and escape the DMDD constraints when the complex scalar DM is significantly underabundant.
Such a low relic density allows the effective $\sigma_{\rm DD}^{\rm SI}$ rates to stay below the exclusion limit owing to a considerable scaling down.
Such a situation comes into the picture only when for $m_s \gtrsim \hsm$, where the $\mu_3$ dependent DM semi-annihilation process can significantly contribute to a drop in the relic density.
Scaling down is also conceivable for $m_s \gtrsim m_{H^0,A^0,H^\pm, H^{\pm\pm}}$ by increasing $\lambda_{S\Delta}$. On the other hand, significantly large values of $m_s$ are disfavoured for an FOPT. 
Thus, the $\mu_3$ term in the potential plays a unique role since it still permits a specific region of parameter space that can comply with DMDDSI rates and also because it can generate a barrier in the tree-level potential in favour of FOPT along the $h_s-$field direction.

However, a more interesting case of a two-step phase transition from the singlet-doublet interaction is when the phase transition in the $h_d-$field direction is strongly first-order. This is what we need from the EWBG perspective. We explore this prospect by analysing a representative benchmark point where $\mu_{\gamma\gamma}$ remains very close to 1 since the various triplet sectors couplings are fixed at tiny values. We find that such an FOPT demands relatively large $\lambda_{SH}$ and as a consequence, future DMDD experiments will investigate the feasibility of generating GW as a result of an FOPT in this scenario.
Thus, focusing on this model, observing any signal at future DMDD and GW experiments but nothing at the HL-LHC will indicate that DM is severely underabundant. The absence of new physics signal at the HL-LHC and various DMDD experiments in future would severely limit the prospects of detecting GW at future GW experiments.

\section*{Acknowledgments}
PG would like to acknowledge Indian Association for the Cultivation of Science, Kolkata for the financial support. The work of TG and SR is supported by the funding available 
from the Department of Atomic Energy (DAE), Government of India for 
Harish-Chandra Research Institute (HRI). SR is also supported by the Infosys award for excellence in research. SR acknowledges the High-Performance Scientific Computing facility at the Regional Centre for Accelerator-based Particle Physics (RECAPP) and HRI.
\appendix
\label{appendix}
\section*{Appendix}
\section{Tree-level tadpole relations}
\label{tadpoleeq}
\vspace{0.1cm}
The mass parameters $\mu_H^2$, $\mu_{\Delta}^2$ and $\mu$ are determined via the 
minimization conditions (the tadpoles) of 
$V_0$ of equation~\ref{Vtree}
and are given by 
\begin{equation}
\label{eq:TreeLvlSymmBrkCond}
\footnotesize{
\begin{aligned}
\mu_H^2 &=
 \tfrac12 (\lambda_1 + \lambda_4) v_t^2
 + \lambda_H v_d^2
 - \sqrt{2} \mu v_t,
\\
\mu_{\Delta}^2 &=
 - (\lambda_2 + \lambda_3) v_t^2
 - \tfrac12 (\lambda_1 + \lambda_4) v_d^2
 + \frac{\mu v_d^2}{\sqrt{2} v_t}.
\\
\mu &= \frac{v_t}{\sqrt{2}v_d}\frac{-(\lambda_2 + \lambda_3)\sin2\theta_t v_t^2 - (\lambda_1 + \lambda_4) v_d v_t \cos2\theta_t + \lambda_H v_d^2 \sin2\theta_t}{v_d \sin\theta_t \cos\theta_t - 2 v_t \cos^2\theta_t + 2 v_t \sin^2\theta_t}.
\end{aligned}}
\end{equation}
\vspace{0.1cm}
\section{Set of UV-finite counterterm coefficients}
\label{CTcoeff}
The various coefficients of the counterterm potential, defined in equation~\ref{VCT} are given by,
\begin{subequations}
\small{\begin{align}
\delta_{\lambda_{S\Delta}}= - \frac{1}{4 v_d v_t} DV_{12}\, \, , \quad\quad \quad\quad \quad\quad \quad \quad\quad \quad\quad \quad\quad \quad 
\end{align}}
\end{subequations}
\vspace{-0.5cm}
\begin{subequations}
\small{\begin{align}
\delta_{\lambda_{H}}= - \frac{1}{8 v_d^3} DV_1 - \frac{1}{8 v_d^2} DV_{11}\, \, , \quad\quad \quad\quad \quad\quad \quad\quad\quad\quad
\end{align}}
\end{subequations}
\vspace{-0.5cm}
\begin{subequations}
\small{\begin{align}
\delta\mu_H^2= - \frac{3}{4 v_d} DV_1 + \frac{1}{4}DV_{11} + \frac{v_t}{4 v_d}DV_{12}\, \, , \quad\quad \quad\quad \quad\quad
\end{align}}
\end{subequations}
\vspace{-0.5cm}
\begin{subequations}
\small{\begin{align}
\delta\mu_{\Delta}^2= - \frac{3}{4 v_t} DV_2 + \frac{1}{4}DV_{22} + \frac{v_d}{4 v_t}DV_{12}\, \, , \quad\quad \quad\quad \quad\quad
\end{align}}
\end{subequations}
\vspace{-0.5cm}
\begin{subequations}
\small{\begin{align}
\delta_{\lambda_{2}}= \frac{1}{8 v_t^3} DV_2 - \frac{1}{8 v_t^2} DV_{22}\, \, , \quad\quad \quad\quad \quad\quad \quad\quad\quad \quad\quad 
\end{align}}
\end{subequations}
\vspace{-0.5cm}
\begin{subequations}
\small{\begin{align}
\delta_{\lambda_{SH}}= - \frac{1}{2 v_t^2} DV_{33}\, \,. \quad\quad \quad\quad \quad\quad \quad\quad\quad \quad\quad \quad \quad
\end{align}}
\end{subequations}
$DV_i = \frac{\partial V_\text{CW}}{\partial \phi_{_i}}$ and $DV_{ij} = \frac{\partial^2 V_\text{CW}}{\partial \phi_{_i}  \partial \phi_{_j}}$ where $\phi_{_i}, \phi_{_j} = \{h_d, h_t, h_s\}$. All the derivatives are taken at the true EW minima, i.e., $h_d = v_d$, $h_t = v_t$ and $h_s =0. $
\vspace{0.3cm}
\section{Daisy resummation improved thermal field-dependent masses}
\label{field-dependent-masses}
\vspace{0.1cm}
From the high-temperature expansion of the thermal one-loop 
potential $V_T$ (of equation~\ref{Vthermal} for bosons), the daisy coefficients can be found from the following relation:
\vspace{0.1cm}
\begin{equation}\small{
c_{ij}  =   \left.\frac{1}{T^2}\frac{\partial^2   V_T}{\partial  \phi_i  \partial  \phi_j}\right|_{T^2  \gg m^2}.}
\end{equation}
\vspace{0.1cm}
The eigenvalues of the thermally improved
mass-squared matrices for the Higgs and the gauge bosons, i.e., $M^2 (h_d, h_t, h_s, T) =$ eigenvalues$[m^2 (h_d, h_t, h_s) + \Pi (T^2)]$ where
$\Pi (T^2) = c_{ij} T^2$ and $c_{ij}$'s are the above described daisy 
coefficients.
The daisy resummation is particularly significant for an FOPT because it affects the all-important cubic term of the potential at a finite temperature. Only the longitudinal modes of the vectors and scalars get thermal mass corrections.
Due to gauge symmetry, thermal effects to the transverse modes are suppressed~\cite{Espinosa:1992kf}. With these in mind, the various daisy 
coefficients are as follows~\cite{Comelli:1996vm, Basler:2018cwe, Carrington:1991hz}:
\begin{subequations}
\label{eq:daisy-coeff}
\small{\begin{align}
c_{H} &= \frac{1}{16} ({g_1}^2+ 3 g_2^2) + \frac{1}{4} y_t^2 + \frac{1}{24}(6\lambda_1 + 3\lambda_4 + 12 \lambda_H + 12 \lambda_{SH}) ,\label{Eq:DaisyCoeffs_Hu}\\
c_{\Delta}  &= \frac{1}{4} ({g_1}^2+ 2 g_2^2) + \frac{1}{12}(2\lambda_1 + 8\lambda_2 + 6 \lambda_3 + \lambda_4 + \lambda_{S\Delta}),\label{Eq:DaisyCoeffs_Hd}\\
c_{S} &= \tfrac{1}{12} 4 \lambda_S + 2 \lambda_{SH} + 3 \lambda_{S\Delta}.\label{Eq:DaisyCoeffs_S}
\end{align}}
\end{subequations}
\vspace{0.1cm}
The field-dependent thermally mass-improved mass-squared matrices for the $2\times2$, symmetric matrix (${\cal M}_{H}^2$) for the $CP$-even scalars, 
in the basis $\{h_d, h_t\}$, is given by
\begin{subequations}
\label{cpevenmasssq}
\small{\begin{align}
M_{{H}_{11}}^2 = \tfrac12 (\lambda_1 + \lambda_4) h_t^2
 + 3 \lambda_H h_d^2 + \tfrac12 \lambda_{SH} h_s^2
 -\sqrt{2}\mu h_t  -\mu_{SH}^2 + c_{H} T^2,\label{Eq:DaisyCoeffs_Hu}\\
M_{{H}_{22}}^2 = \tfrac12 (\lambda_1 + \lambda_4) h_d^2
 + 3 (\lambda_2 +\lambda_3) h_t^2 + \tfrac12 \lambda_{ST} h_s^2  + \mu_{\Delta}^2 + c_{\Delta} T^2 , \quad\quad \quad\label{Eq:DaisyCoeffs_Hd}\\
M_{{H}_{12}}^2 =  M_{{H}_{21}}^2 = (\lambda_1 + \lambda_4) h_d h_t
 - \sqrt{2} \mu h_d .\label{Eq:DaisyCoeffs_S} \quad\quad \quad\quad \quad\quad \quad\quad
\end{align}}
\end{subequations}
\vspace{0.1cm}
The field-dependent thermally mass-improved mass-squared matrices for the $2\times2$, symmetric matrix (${\cal M}_{H}^2$) for the $CP$-odd scalars, 
in the basis $\{a_d, a_t\}$, is given by
\begin{subequations}
\label{cpoddmasssq}
\small{\begin{align}
M_{{A}_{11}}^2 = \tfrac12 (\lambda_1 + \lambda_4) h_t^2
 + \lambda_H h_d^2 + \tfrac12 \lambda_{SH} h_s^2 + c_{H} T^2
 + \sqrt{2}\mu h_t  -\mu_{SH}^2,\label{Eq:DaisyCoeffs_Hu}\\
M_{{A}_{22}}^2 = \tfrac12 (\lambda_1 + \lambda_4) h_d^2
 + (\lambda_2 +\lambda_3) h_t^2 + \tfrac12 \lambda_{ST} h_s^2  + \mu_{\Delta}^2 + c_{\Delta} T^2, \quad\quad \quad\label{Eq:DaisyCoeffs_Hd}\\
M_{{A}_{12}}^2 =  M_{{H}_{21}}^2 =  - \sqrt{2} \mu h_d .\label{Eq:DaisyCoeffs_S} \quad\quad \quad\quad \quad\quad \quad\quad \quad\quad \quad\quad \quad\quad
\end{align}}
\end{subequations}
\vspace{0.1cm}
The field-dependent thermally mass-improved mass-squared matrices for the $2\times2$, diagonal matrix (${\cal M}_{H}^2$) for thedark sector, 
in the basis $\{h_s,a_s\}$, is given by
\begin{subequations}
\label{DMmasssq}
\small{\begin{align}
M_{{DM}_{11}}^2 = 3 \lambda_S h_s^2 + \tfrac12 \lambda_{SH} h_d^2 + \tfrac12 \lambda_{S\Delta} h_t^2 + \frac{1}{\sqrt{2}} \mu_3 h_s  + \mu_{S}^2 + c_{S} T^2, \quad\quad \quad\label{Eq:DaisyCoeffs_Hu}\\
M_{{DM}_{22}}^2 = \lambda_S h_s^2 + \tfrac12 \lambda_{SH} h_d^2 + \tfrac12 \lambda_{S\Delta} h_t^2 - \frac{1}{\sqrt{2}} \mu_3 h_s + \mu_{S}^2 + c_{S} T^2 , \quad\quad \quad\label{Eq:DaisyCoeffs_Hd}\\
M_{{DM}_{12}}^2 =  M_{{H}_{21}}^2 =  0 .\label{Eq:DaisyCoeffs_S} \quad\quad \quad\quad \quad\quad \quad\quad \quad\quad \quad\quad \quad\quad \quad\quad \quad\quad
\end{align}}
\end{subequations}
\vspace{0.1cm}
The field-dependent thermally mass-improved mass-squared matrices for the $2\times2$, symmetric matrix (${\cal M}_{H{^\pm}}^2$) for the $CP$-even singly charged Higgs bosons, 
in the basis $\{h_d, h_t\}$, is given by
\begin{subequations}
\label{chargmasssq}
\small{\begin{align}
M_{{H{^\pm}}_{11}}^2 = \tfrac12 \lambda_1 h_t^2
 + \lambda_H h_d^2 + \tfrac12 \lambda_{SH} h_s^2 
  -\mu_{H}^2 + c_{H} T^2,\label{Eq:DaisyCoeffs_Hu} \quad\quad \quad\quad \quad\quad \quad\\
M_{{H{^\pm}}_{22}}^2 = \tfrac12 \lambda_1 h_d^2
 + (\lambda_2 +\lambda_3) h_t^2 + \frac{1}{4} \lambda_4 h_d^2 +\frac{1}{2} \lambda_{S\Delta} h_s^2  + \mu_{\Delta}^2 + c_{\Delta} T^2, \quad\label{Eq:DaisyCoeffs_Hd}\\
M_{{H{^\pm}}_{12}}^2 =  M_{{H}_{21}}^2 = \frac{1}{2\sqrt{2}}\lambda_4 h_d h_t
 - \mu h_d .\label{Eq:DaisyCoeffs_S} \quad\quad \quad\quad \quad\quad \quad\quad \quad
\end{align}}
\end{subequations}
\vspace{0.1cm}
The thermally mass-improved field-dependent mass-squared doubly-charged Higgs mass is given by
\begin{equation}
\label{Doubchargmasssq}
M_{{\Delta{^{\pm\pm}}}}^2 = \tfrac12 \lambda_1 h_d^2 + \lambda_2 h_t^2
 + \tfrac12 \lambda_{S\Delta} h_s^2 
  +\mu_{\Delta}^2 + c_{\Delta} T^2 \, ,\quad\quad \quad\quad 
\end{equation}
\vspace{0.1cm}
In the fermionic sector, we only consider the top quark degrees of freedom and its field-dependent mass is given by
\begin{equation}
\label{eq:FermionMasses}
\begin{gathered}
m_t  =  \tfrac{1}{\sqrt{2}}  y_t \Huzr \,  \quad.
\end{gathered}
\end{equation}
\vspace{0.1cm}
The field-dependent Daisy resummation improved thermal mass squared of the charged gauge boson (only the longitudinal modes get thermal mass corrections) field $W^{\pm}$ and the neutral electroweak gauge boson fields $W^3$ of $SU(2)$ and the $B$ of $U(1)$ group are given by,
\vspace{0.1cm}
\begin{equation}
\label{eq:gaugebosonmasses}
\small{
\begin{gathered}
m_{W^{^\pm}_{_T}}^2  =  \tfrac{1}{4} g_2^2 \left(h_d^2  +  2 h_t^2 \right), \quad\\
m_{W^{^\pm}_{_L}}^2  =  \tfrac{1}{4} g_2^2 \left(h_d^2  +  2 h_t^2 + \frac{5}{2} g_2^2 T^2 \right),\\
m_{W^{^3}_{_T}}^2  =  \tfrac{1}{4} g_2^2 \left(h_d^2  +  4 h_t^2 \right), \\
m_{B_{_T}}^2  =  \tfrac{1}{4} g_1^2 \left(h_d^2  +  4 h_t^2 \right), \\
m_{{W^{^3}_{_T}}{B_{_T}}}^2  =  - \tfrac{1}{4} g_1 g_2 \left(h_d^2  +  4 h_t^2 \right)\\
m_{W^{^3}_{_L}}^2  =  m_{W^{^3}_{_T}}^2 + \frac{5}{2} g_2^2 T^2, \\
m_{B_{_L}}^2  = m_{B_{_T}}^2 + \frac{17}{6} g_1^2 T^2, \\
m_{{W^{^3}_{_L}}{B_{_L}}}^2   = m_{{W^{^3}_{_T}}{B_{_T}}}^2 
\end{gathered}}
\end{equation}
The physical masses of the gauge bosons can be found after diagonalizing the above matrices. Note that the photon field also acquires non-zero mass at finite temperatures.  The zero-temperature field-dependent mass relations can be found by putting $T=0$ in the above mass relations.
\vspace{0.4cm}
\section{Feynman diagrams for DM annihilation}
\label{feynmandiag}
\begin{figure}[ht]
 \begin{center}
    \begin{tikzpicture}[line width=1.2 pt, scale=1.1]
        \draw[dashed] (-6,1)--(-5,0);
	\draw[dashed] (-6,-1)--(-5,0);
	\draw[dashed] (-5,0)--(-4,1);
	\draw[dashed] (-5,0)--(-4,-1);
	\node at (-6.2,1.1) {$S$};
	\node at (-6.2,-1.1) {$S^*$};
	\node at (-3.8,1.1) {$h^0$};
	\node at (-3.8,-1.1) {$h^0$};
	\draw[dashed] (-1.8,1.0)--(-0.8,0.5);
	\draw[dashed] (-1.8,-1.0)--(-0.8,-0.5);
	\draw[dashed] (-0.8,0.5)--(-0.8,-0.5);
	\draw[dashed] (-0.8,0.5)--(0.2,1.0);
	\draw[dashed] (-0.8,-0.5)--(0.2,-1.0);
	\node at (-2.0,1.1) {$S$};
	\node at (-2.0,-1.1) {$S^*$};
	\node at (-1.0,0.07) {$S$};
	\node at (0.4,1.1) {$h^0$};
	\node at (0.4,-1.1) {$h^0$};
	%
        \draw[dashed] (2.4,1)--(3.4,0);
	\draw[dashed] (2.4,-1)--(3.4,0);
	\draw[dashed] (3.4,0)--(4.4,0);
	\draw[solid] (4.4,0)--(5.4,1);
	\draw[solid] (4.4,0)--(5.4,-1);
	\node  at (2.2,-1) {$S$};
	\node at (2.2,1) {$S^*$};
	\node [above] at (3.9,0.05) {$h^0,H^0$};
	\node at (5.8,1.3){$f/W^+/Z/h^0$};
	\node at (5.8,-1.3) {$\overline{f}/W^-/Z/h^0$};
     \end{tikzpicture}
 \end{center}
\caption{Feynman diagrams for DM annihilate to the SM particles: $S ~S^* \to $ SM ~SM .}
\label{Feyn_diag1}
 \end{figure}
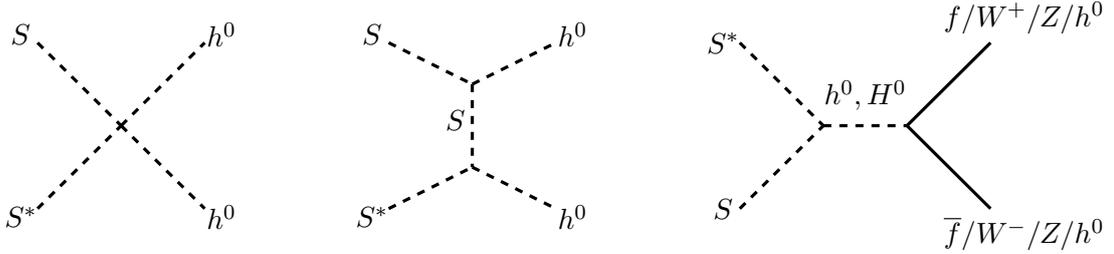

 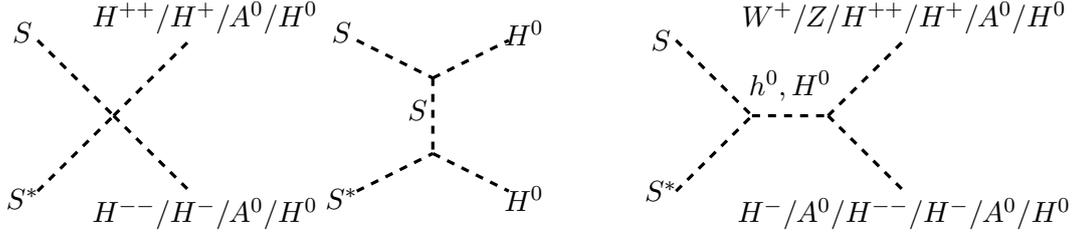
\begin{figure}[h]
 \begin{center}
    \begin{tikzpicture}[line width=1.2 pt, scale=1.0]
        \draw[dashed] (-6,1)--(-5,0);
	\draw[dashed] (-6,-1)--(-5,0);
	\draw[dashed] (-5,0)--(-4,1);
	\draw[dashed] (-5,0)--(-4,-1);
	\node at (-6.2,1.1) {$S$};
	\node at (-6.2,-1.1) {$S^*$};
	\node at (-3.8,1.3) {$H^{++}/H^+/A^0/H^0$};
	\node at (-3.8,-1.3) {$H^{--}/H^-/A^0/H^0$};
	\draw[dashed] (-1.8,1.0)--(-0.8,0.5);
	\draw[dashed] (-1.8,-1.0)--(-0.8,-0.5);
	\draw[dashed] (-0.8,0.5)--(-0.8,-0.5);
	\draw[dashed] (-0.8,0.5)--(0.2,1.0);
	\draw[dashed] (-0.8,-0.5)--(0.2,-1.0);
	\node at (-2.0,1.1) {$S$};
	\node at (-2.0,-1.1) {$S^*$};
	\node at (-1.0,0.07) {$S$};
	\node at (0.4,1.1) {$H^0$};
	\node at (0.4,-1.1) {$H^0$};
	%
        \draw[dashed] (2.4,1)--(3.4,0);
	\draw[dashed] (2.4,-1)--(3.4,0);
	\draw[dashed] (3.4,0)--(4.4,0);
	\draw[dashed] (4.4,0)--(5.4,1);
	\draw[dashed] (4.4,0)--(5.4,-1);
	\node  at (2.2,-1) {$S^*$};
	\node at (2.2,1) {$S$};
	\node [above] at (3.9,0.05) {$h^0,H^0$};
	\node at (5.4,1.3){$W^+/Z/H^{++}/H^+/A^0/H^0$};
	\node at (5.4,-1.3) {$H^-/A^0/H^{--}/H^-/A^0/H^0$};
     \end{tikzpicture}
 \end{center}
\caption{Feynman diagrams for DM annihilate to scalar triplet(s):$S~S^* \to W^+ H^-, Z A^0, X Y$. Here $\{X,Y\}=\{ H^{\pm\pm},H^\pm,A^0,H^0 \}$.}
\label{Feyn_diag2}
 \end{figure}


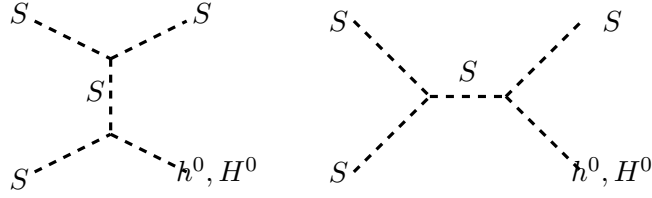
\begin{figure}[h]
 \begin{center}
    \begin{tikzpicture}[line width=1.2 pt, scale=1.0]
	\draw[dashed] (-1.8,1.0)--(-0.8,0.5);
	\draw[dashed] (-1.8,-1.0)--(-0.8,-0.5);
	\draw[dashed] (-0.8,0.5)--(-0.8,-0.5);
	\draw[dashed] (-0.8,0.5)--(0.2,1.0);
	\draw[dashed] (-0.8,-0.5)--(0.2,-1.0);
	\node at (-2.0,1.1) { $S$};
	\node at (-2.0,-1.1) {$S$};
	\node at (-1.0,0.07) {$S$};
	\node at (0.4,1.1) {$S$};
	\node at (0.6,-1.0) {$h^0,H^0$};
	%
        \draw[dashed] (2.4,1)--(3.4,0);
	\draw[dashed] (2.4,-1)--(3.4,0);
	\draw[dashed] (3.4,0)--(4.4,0);
	\draw[dashed] (4.4,0)--(5.4,1);
	\draw[dashed] (4.4,0)--(5.4,-1);
	\node  at (2.2,-1) {$S$};
	\node at (2.2,1) {$S$};
	\node [above] at (3.9,0.05) {$S$};
	\node at (5.8,1.0){$S$};
	\node at (5.8,-1.0) {$h^0,H^0$};
     \end{tikzpicture}
 \end{center}
\caption{Feynman diagrams for DM semi-annihilation processes: $S~S \to S~h^0, S ~H^0$.}
\label{Feyn_diag3}
 \end{figure}

\vspace{1.3cm}
\section{Theoretical constraints}
\label{theoreticalconstraints}
In this section, we highlight some of the key theoretical constraints on our model parameter space.
\subsection{Vacuum stability}
\label{vacuumstabilityconstraints}
The scalar potential of the model requires to have a stable vacuum. In other words, it means that at large field values the potential is bounded from below so that the scalar fields do not run away. At large field values, the quadratic and cubic terms in the scalar potential can be neglected in comparison to the quartic term to determine the stability conditions of the scalar potential. Using the copositivity conditions of vacuum stability, as outlined in references~\cite{Kannike:2012pe,Yang:2021buu}, we obtain the following relations for our model:
\begin{align}
 \lambda_H \geq 0, ~~\lambda_S \geq 0, \nonumber \\
 \lambda_{2}+\lambda_{3} \geq 0 ,~~ \lambda_2+\frac{\lambda_3}{2} \geq 0 , \nonumber \\
 \lambda_{1}+2\sqrt{\lambda_H(\lambda_2+\lambda_3)} \geq 0 ,~~(\lambda_{1}+\lambda_{4})+2\sqrt{\lambda_H(\lambda_2+\lambda_3)} \geq 0  , \nonumber \\
 (2\lambda_{1}+\lambda_{4})+4\sqrt{\lambda_H(\lambda_2+\frac{\lambda_3}{2})} \geq 0  ,~~~
 (\lambda_{2}+\lambda_{3})+\sqrt{(\lambda_{2}+\lambda_{3})(\lambda_2+\frac{\lambda_3}{2})}\geq 0 ,\nonumber \\   
 \lambda_{S H} +2 \sqrt{\lambda_S \lambda_H} \geq 0,~~
 \lambda_{S \Delta} + 2 \sqrt{\lambda_S (\lambda_2+\lambda_3)} \geq 0, \nonumber \\
 \lambda_{S \Delta} + 2 \sqrt{\lambda_S (\lambda_2+\frac{\lambda_3}{2})} \geq 0~.
\end{align}
It is also important to note that a stable global minimum sets an upper bound on the dark scalar cubic coupling, $\mu_3$ as $\mu_3/m_S \lesssim 2 \sqrt{\lambda_S}$~\cite{Belanger:2012zr,Hektor:2019ote}. For $\lambda_S=1$, this relation becomes $\mu_3 \lesssim 2 m_S$.  
\subsection{Perturbativity}
In order to ensure the validity of standard perturbation theory where the one-loop level correction to coupling should be smaller than the tree-level couplings. The quartic and the Yukawa couplings of the interaction Lagrangian should obey the following upper bounds~\cite{Lerner:2009xg,Yang:2021buu} as: 
\begin{align}
 |\lambda_H| \lesssim \frac{2\pi}{3},\nonumber \\
 |\lambda_1| \lesssim 4\pi,~|\lambda_2| \lesssim 2\pi, ~|\lambda_3| \lesssim 2 \sqrt{2}\pi,~|\lambda_4| \lesssim 4\pi,\nonumber \\
 |\lambda_1+\lambda_4| \lesssim 4\pi ,~|\lambda_2+\lambda_3| \lesssim \frac{2\pi}{3} ,~|\lambda_1+\frac{\lambda_4}{2}| \lesssim 4\pi ,~|\lambda_2+\frac{\lambda_3}{2}| \lesssim \pi , \nonumber \\
 |\lambda_S| \lesssim \pi,~|\lambda_{S H}| \lesssim 4\pi,~|\lambda_{S \Delta}| \lesssim 4\pi \nonumber \\
 {\rm and}~~~ |y_L| \lesssim \sqrt{4\pi} ~.
\end{align}

\subsection{Partial wave unitarity}
The quartic couplings of the scalar potential also be constrained from tree level unitarity of the theory, coming from all possible $2 \to 2$ scattering amplitudes which will form the $S$ matrix~\cite{Lee:1977eg,Arhrib:2011uy,Yang:2021buu}. The eigenvalues of the $S$ matrix are bounded from above as: 

\begin{align}
 |\lambda_H| \leq 4\pi,~~|\lambda_1| \leq 8\pi,~~|\lambda_2| \leq 4\pi, \nonumber \\
 |\lambda_{S H}| \leq 8\pi,~~|\lambda_{S \Delta}| \leq 8\pi, \nonumber \\
 |\lambda_2+\lambda_3| \leq 4 \pi,~~|\lambda_1+\lambda_4| \leq 8 \pi , \nonumber \\
 |\lambda_1-\frac{\lambda_4}{2}| \leq 8 \pi ,~~|\lambda_1+\frac{3\lambda_4}{2}| \leq 8 \pi , \nonumber \\ 
 |x_{1,2,3}| \leq 8 \pi,
\end{align}
where $x_{1,2,3}$ are the roots of the following polynomial equation:
\bea
&& 24 \lambda_1^2 \lambda_S +24 \lambda_1 \lambda_{4} \lambda_S -12 \lambda_1 \lambda_{S \Delta} \lambda_{S H}-192 \lambda_{2} \lambda_{H} \lambda_S +16 \lambda_{2}\lambda_{S H}^2-144 \lambda_{3} \lambda_{H} \lambda_S +12 \lambda_{3}\lambda_{S H}^2+6 \lambda_{4}^2 \lambda_S \nonumber \\
&&  -6 \lambda_{4} \lambda_{S \Delta} \lambda_{S H}+18 \lambda_{H} \lambda_{S \Delta}^2 + x \Big(-12 \lambda_1^2-12 \lambda_1 \lambda_{4}+96 \lambda_{2} \lambda_{H}+32 \lambda_{2} \lambda_S +72 \lambda_{3} \lambda_{H}+24 \lambda_{3} \lambda_S -3 \lambda_{4}^2 \nonumber \\ 
&& +24 \lambda_{H} \lambda_S -3 \lambda_{S \Delta} ^2-2\lambda_{S H}^2\Big)+x^2 (-16 \lambda_{2}-12 \lambda_{3}-12 \lambda_{H}-4 \lambda_S )  +2 x^3 =0
\eea
\section{Phenomenological constraints}
\label{Phenomenologicalconstraints}
Here, we turn our attention to the phenomenological constraints on our model parameter space.
\subsection{Neutrino Oscillation data}
Explaining the smallness of neutrino masses is one of the key motivations of the model under consideration. Hence, we briefly mention the constraints imposed on the model parameter space coming from neutrino oscillation data. Neutrino oscillation experiments provide bounds on neutrino sector parameters such as mass-squared differences and the mixing angles~\cite{Esteban:2020cvm}. The $3 \sigma$ allowed ranges for these observables depend slightly on the ordering of neutrino masses. Below we list the latest limits from a global fit~\cite{Esteban:2020cvm}, for the normal ordering (inverted ordering) of neutrino masses
\begin{align}
\Delta m^2_{21} 
	\in [6.82-8.04]\times 10^{-5}  {\text{eV}^2},
    \nonumber\\
	\Delta m^2_{3\ell}
	\in [2.431-2.598]~\Big([-2.583, -2.412]\Big)\times 10^{-3} {\text{eV}^2},
    \nonumber\\    
    \sin^2\theta_{12}\in [0.269-0.343],
	\nonumber\\
	\sin^2\theta_{23} \in [0.407-0.618]~\Big( [0.411-0.621] \Big),
    \nonumber\\
	\sin^2\theta_{13}\in [0.02034-0.02430]~\Big( [0.02053-0.02436] \Big),
    \nonumber\\
    \delta_{\text{CP}} \in [107-403]^{\circ}~\Big( [192-360]^{\circ} \Big).
	\label{obs_para}
\end{align}
Only one set of $3 \sigma$ allowed values are quoted above for $\Delta m^2_{21}$ and $\sin^2\theta_{12}$ since these two observables are insensitive to the mass ordering of neutrinos.
In addition, cosmological observations put an upper bound on the sum of neutrino masses, $ \sum_{i=1}^{3} m_i < 0.12$ eV~\cite{Planck:2018vyg}.

\subsection{Flavor constraints}
\label{subsec:Flavourconstraints}
The Yukawa interactions of the Higgs triplet with the SM lepton doublets, which generate the non-zero neutrino masses, can also induce charged lepton flavor violating (cLFV) processes. The cLFV processes can arise both at tree-level (e.g. $\ell_\alpha \to \ell_\beta \ell_\gamma \ell_\delta$) or loop-level (e.g. $\ell_\alpha \to \ell_\beta \gamma$), where $\alpha,\beta,\gamma, \delta$ are flavor indices. The bounds from the cLFV processes can be quite stringent, and the two channels that impose the strongest limits are muon decay to $3e$ and $e \gamma$~\cite{Dev:2021axj}. The branching fractions (BF) of these two important processes are given by~\cite{Kakizaki:2003jk,Akeroyd:2009nu,Dinh:2012bp,Ashanujjaman:2021txz}
\begin{eqnarray*}
&& {\rm Br}(\mu^- \to e^+e^-e^-) = \frac{|(y_L)^\dagger_{ee} (y_L)_{\mu e}|^2}{4 \, G_F^2 m_{H^{\pm \pm}}^4}~,
\\
&& {\rm Br}(\mu^- \to e^- \gamma) = \frac{\alpha|\sum_{i=1}^{3}(y_L)^\dagger_{e i} (y_L)_{\mu i}|^2}{192\pi G_F^2} \left(\frac{1}{m_{H^\pm}^2}+\frac{8}{m_{H^{\pm \pm}}^2} \right)^2~,
\end{eqnarray*}
where $(y_L)_{\alpha \beta} = \frac{\sqrt{2} \, m_{\alpha \beta}^\nu}{ v_t}$.
The experimental upper limits for ${\rm Br}(\mu^- \to e^+e^-e^-)$ and ${\rm Br}(\mu^- \to e^- \gamma)$ are $1.0\times 10^{-12}$~\cite{Bellgardt:1987du} and $4.2\times 10^{-13}$~\cite{TheMEG:2016wtm}, respectively.
Using them we can compute the following lower bounds on $v_t$ as a function of $m_{H^{\pm \pm}}$
\begin{equation*}
v_t \gtrsim 0.78\text{--}1.5 \, \big(0.69\text{--}2.3 \big) \times 10^{-9}  {\rm ~GeV} \times \bigg( \frac{1~ \rm TeV}{m_{H^{\pm \pm}}} \bigg)
\end{equation*}
for the normal (inverted) ordering of neutrino masses.
\subsection{Electroweak precision observables}
\label{Sec:EWPO}
Since the $CP$-even neutral scalar of the triplet develops an induced $\vev$ in this model, it can lead to a deviation of the oblique $T$ parameter from the SM prediction at the tree level itself. In addition, a mass-splitting, $\Delta m$, between the components of the triplet can further shift the oblique parameters $S, \, T$ and $U$ at one-loop level from the SM predictions~\cite{Chun:2012jw, Aoki:2012jj}. 
The tree-level masses of $W$ and $Z$ bosons within this model get altered, which leads to the following expression of the EW $\rho$ parameter
\bea
\rho=\frac{m_W^2}{m_Z^2 \cos^2\theta_W}=\frac{v_d^2+2v_t^2}{v_d^2+4v_t^2}\simeq 1-\frac{2v_t^2}{v_d^2},
\eea
where $v_d=\sqrt{v^2-2v_t^2}$ with $v=246$ GeV. $\rho$ is connected to the $T$ parameter at the tree level by  $T_{\text{tree}} \simeq - 2 v_t^2/v_d^2/\alpha = (1 - \rho)/\alpha$. It is needless to say that the above relations hold only under the assumption $v_t << v_d$. This assumption is a natural choice in type-II seesaw motivated models since the triplet $\vev$ generates neutrino masses. A tiny $v_t$ will allow us to fit neutrino masses and mixings with small Yukawa couplings satisfying all flavor constraints. The current measurement of the EW observables constrain the $\rho$-parameter as : $\rho=1.00038 \pm 0.00020$~\cite{ParticleDataGroup:2020ssz}, which translates into a bound $v_t \leq 2.6$ GeV at $3 \sigma$ level.

Now, assuming $v_t << v_d$, we incorporate the limits from the three EW oblique parameters $S, \, T$ and $U$ on $\Delta m$. In this paper, we exclusively consider $v_t << 1$ GeV, and in this regime, the correction to the $T$ parameter is dominated by 1-loop level contributions compared to the small tree level correction shown above. References~\cite{Chun:2012jw, Aoki:2012jj} show that oblique parameters constrain $\Delta m$ to be below 40 GeV for a large range of $m_{H^{\pm \pm}}$ values.


\begin{thebibliography}{10}
%
\bibitem{ATLAS:2012yve}
G.~Aad \textit{et al.} [ATLAS],
Phys. Lett. B \textbf{716} (2012), 1-29
doi:10.1016/j.physletb.2012.08.020
[arXiv:1207.7214 [hep-ex]].

\bibitem{CMS:2012qbp}
S.~Chatrchyan \textit{et al.} [CMS],
Phys. Lett. B \textbf{716} (2012), 30-61
doi:10.1016/j.physletb.2012.08.021
[arXiv:1207.7235 [hep-ex]].

\bibitem{Sakharov:1967dj}
A.~D.~Sakharov,
Pisma Zh. Eksp. Teor. Fiz. \textbf{5} (1967), 32-35
doi:10.1070/PU1991v034n05ABEH002497

\bibitem{Trodden:1998ym}
M.~Trodden,
Rev. Mod. Phys. \textbf{71} (1999), 1463-1500
doi:10.1103/RevModPhys.71.1463
[arXiv:hep-ph/9803479 [hep-ph]].

\bibitem{Anderson:1991zb}
G.~W.~Anderson and L.~J.~Hall,
Phys. Rev. D \textbf{45} (1992), 2685-2698
doi:10.1103/PhysRevD.45.2685

\bibitem{Huet:1995sh}
P.~Huet and A.~E.~Nelson,
Phys. Rev. D \textbf{53} (1996), 4578-4597
doi:10.1103/PhysRevD.53.4578
[arXiv:hep-ph/9506477 [hep-ph]].

\bibitem{Morrissey:2012db}
D.~E.~Morrissey and M.~J.~Ramsey-Musolf,
New J. Phys. \textbf{14} (2012), 125003
doi:10.1088/1367-2630/14/12/125003
[arXiv:1206.2942 [hep-ph]].

\bibitem{Hu:2001bc}
W.~Hu and S.~Dodelson,
Ann. Rev. Astron. Astrophys. \textbf{40} (2002), 171-216
doi:10.1146/annurev.astro.40.060401.093926
[arXiv:astro-ph/0110414 [astro-ph]].

\bibitem{WMAP:2006bqn}
D.~N.~Spergel \textit{et al.} [WMAP],
Astrophys. J. Suppl. \textbf{170} (2007), 377
doi:10.1086/513700
[arXiv:astro-ph/0603449 [astro-ph]].

\bibitem{Planck:2018vyg}
N.~Aghanim \textit{et al.} [Planck],
Astron. Astrophys. \textbf{641} (2020), A6
[erratum: Astron. Astrophys. \textbf{652} (2021), C4]
doi:10.1051/0004-6361/201833910
[arXiv:1807.06209 [astro-ph.CO]].

\bibitem{Kajita:2016cak}
T.~Kajita,
Rev. Mod. Phys. \textbf{88} (2016) no.3, 030501
doi:10.1103/RevModPhys.88.030501

\bibitem{McDonald:2016ixn}
A.~B.~McDonald,
Rev. Mod. Phys. \textbf{88} (2016) no.3, 030502
doi:10.1103/RevModPhys.88.030502

\bibitem{Minkowski:1977sc}
P.~Minkowski,
Phys. Lett. B \textbf{67} (1977), 421-428
doi:10.1016/0370-2693(77)90435-X

\bibitem{Yanagida:1979as}
T.~Yanagida,
Conf. Proc. C \textbf{7902131} (1979), 95-99
KEK-79-18-95.

\bibitem{Mohapatra:1979ia}
R.~N.~Mohapatra and G.~Senjanovic,
Phys. Rev. Lett. \textbf{44} (1980), 912
doi:10.1103/PhysRevLett.44.912

\bibitem{Konetschny:1977bn}
W.~Konetschny and W.~Kummer,
Phys. Lett. B \textbf{70} (1977), 433-435
doi:10.1016/0370-2693(77)90407-5

\bibitem{Magg:1980ut}
M.~Magg and C.~Wetterich,
Phys. Lett. B \textbf{94} (1980), 61-64
doi:10.1016/0370-2693(80)90825-4

\bibitem{Lazarides:1980nt}
G.~Lazarides, Q.~Shafi and C.~Wetterich,
Nucl. Phys. B \textbf{181} (1981), 287-300
doi:10.1016/0550-3213(81)90354-0

\bibitem{Schechter:1980gr}
J.~Schechter and J.~W.~F.~Valle,
Phys. Rev. D \textbf{22} (1980), 2227
doi:10.1103/PhysRevD.22.2227

\bibitem{Cheng:1980qt}
T.~P.~Cheng and L.~F.~Li,
Phys. Rev. D \textbf{22} (1980), 2860
doi:10.1103/PhysRevD.22.2860

\bibitem{Bilenky:1980cx}
S.~M.~Bilenky, J.~Hosek and S.~T.~Petcov,
Phys. Lett. B \textbf{94} (1980), 495-498
doi:10.1016/0370-2693(80)90927-2

\bibitem{Casas:2017jjg}
J.~A.~Casas, D.~G.~Cerde\~no, J.~M.~Moreno and J.~Quilis,
JHEP \textbf{05} (2017), 036
doi:10.1007/JHEP05(2017)036
[arXiv:1701.08134 [hep-ph]].

\bibitem{Bhattacharya:2017fid}
S.~Bhattacharya, P.~Ghosh, T.~N.~Maity and T.~S.~Ray,
JHEP \textbf{10} (2017), 088
doi:10.1007/JHEP10(2017)088
[arXiv:1706.04699 [hep-ph]].

\bibitem{Aprile:2018dbl}
E.~Aprile \textit{et al.} [XENON],
Phys. Rev. Lett. \textbf{121} (2018) no.11, 111302
doi:10.1103/PhysRevLett.121.111302
[arXiv:1805.12562 [astro-ph.CO]].

\bibitem{PandaX-4T:2021bab}
Y.~Meng \textit{et al.} [PandaX-4T],
Phys. Rev. Lett. \textbf{127} (2021) no.26, 261802
doi:10.1103/PhysRevLett.127.261802
[arXiv:2107.13438 [hep-ex]].

\bibitem{Fermi-LAT:2017bpc}
S.~Abdollahi \textit{et al.} [Fermi-LAT],
Phys. Rev. D \textbf{95} (2017) no.8, 082007
doi:10.1103/PhysRevD.95.082007
[arXiv:1704.07195 [astro-ph.HE]].

\bibitem{MAGIC:2016xys}
M.~L.~Ahnen \textit{et al.} [MAGIC and Fermi-LAT],
JCAP \textbf{02} (2016), 039
doi:10.1088/1475-7516/2016/02/039
[arXiv:1601.06590 [astro-ph.HE]].

\bibitem{Apreda:2001us}
R.~Apreda, M.~Maggiore, A.~Nicolis and A.~Riotto,
Nucl. Phys. B \textbf{631} (2002), 342-368
doi:10.1016/S0550-3213(02)00264-X
[arXiv:gr-qc/0107033 [gr-qc]].

\bibitem{Grojean:2004xa}
C.~Grojean, G.~Servant and J.~D.~Wells,
Phys. Rev. D \textbf{71} (2005), 036001
doi:10.1103/PhysRevD.71.036001
[arXiv:hep-ph/0407019 [hep-ph]].

\bibitem{Weir:2017wfa}
D.~J.~Weir,
Phil. Trans. Roy. Soc. Lond. A \textbf{376} (2018) no.2114, 20170126
doi:10.1098/rsta.2017.0126
[arXiv:1705.01783 [hep-ph]].

\bibitem{Alves:2018oct}
A.~Alves, T.~Ghosh, H.~K.~Guo and K.~Sinha,
JHEP \textbf{12} (2018), 070
doi:10.1007/JHEP12(2018)070
[arXiv:1808.08974 [hep-ph]].

\bibitem{Alves:2018jsw}
A.~Alves, T.~Ghosh, H.~K.~Guo, K.~Sinha and D.~Vagie,
JHEP \textbf{04} (2019), 052
doi:10.1007/JHEP04(2019)052
[arXiv:1812.09333 [hep-ph]].

\bibitem{Alves:2019igs}
A.~Alves, D.~Gon\c{c}alves, T.~Ghosh, H.~K.~Guo and K.~Sinha,
JHEP \textbf{03} (2020), 053
doi:10.1007/JHEP03(2020)053
[arXiv:1909.05268 [hep-ph]].

\bibitem{Alves:2020bpi}
A.~Alves, D.~Gon\c{c}alves, T.~Ghosh, H.~K.~Guo and K.~Sinha,
Phys. Lett. B \textbf{818} (2021), 136377
doi:10.1016/j.physletb.2021.136377
[arXiv:2007.15654 [hep-ph]].

\bibitem{Chatterjee:2022pxf}
A.~Chatterjee, A.~Datta and S.~Roy,
JHEP \textbf{06} (2022), 108
doi:10.1007/JHEP06(2022)108
[arXiv:2202.12476 [hep-ph]].

\bibitem{Caprini:2019egz}
C.~Caprini, M.~Chala, G.~C.~Dorsch, M.~Hindmarsh, S.~J.~Huber, T.~Konstandin, J.~Kozaczuk, G.~Nardini, J.~M.~No and K.~Rummukainen, \textit{et al.}
JCAP \textbf{03} (2020), 024
doi:10.1088/1475-7516/2020/03/024
[arXiv:1910.13125 [astro-ph.CO]].

\bibitem{Witten:1984rs}
E.~Witten,
Phys. Rev. D \textbf{30} (1984), 272-285
doi:10.1103/PhysRevD.30.272

\bibitem{Hogan:1986qda}
C.~J.~Hogan,
Mon. Not. Roy. Astron. Soc. \textbf{218} (1986), 629-636

\bibitem{Ellis:2018mja}
J.~Ellis, M.~Lewicki and J.~M.~No,
JCAP \textbf{04} (2019), 003
doi:10.1088/1475-7516/2019/04/003
[arXiv:1809.08242 [hep-ph]].

\bibitem{Alanne:2019bsm}
T.~Alanne, T.~Hugle, M.~Platscher and K.~Schmitz,
JHEP \textbf{03} (2020), 004
doi:10.1007/JHEP03(2020)004
[arXiv:1909.11356 [hep-ph]].

\bibitem{LISA:2017pwj}
P.~Amaro-Seoane \textit{et al.} [LISA],
[arXiv:1702.00786 [astro-ph.IM]].

\bibitem{Gong:2014mca}
X.~Gong, Y.~K.~Lau, S.~Xu, P.~Amaro-Seoane, S.~Bai, X.~Bian, Z.~Cao, G.~Chen, X.~Chen and Y.~Ding, \textit{et al.}
J. Phys. Conf. Ser. \textbf{610} (2015) no.1, 012011
doi:10.1088/1742-6596/610/1/012011
[arXiv:1410.7296 [gr-qc]].

\bibitem{Hu:2017mde}
W.~R.~Hu and Y.~L.~Wu,
Natl. Sci. Rev. \textbf{4} (2017) no.5, 685-686
doi:10.1093/nsr/nwx116

\bibitem{TianQin:2015yph}
J.~Luo \textit{et al.} [TianQin],
Class. Quant. Grav. \textbf{33} (2016) no.3, 035010
doi:10.1088/0264-9381/33/3/035010
[arXiv:1512.02076 [astro-ph.IM]].

\bibitem{Harry:2010zz}
G.~M.~Harry [LIGO Scientific],
Class. Quant. Grav. \textbf{27} (2010), 084006
doi:10.1088/0264-9381/27/8/084006

\bibitem{Corbin:2005ny}
V.~Corbin and N.~J.~Cornish,
Class. Quant. Grav. \textbf{23} (2006), 2435-2446
doi:10.1088/0264-9381/23/7/014
[arXiv:gr-qc/0512039 [gr-qc]].

\bibitem{Kudoh:2005as}
H.~Kudoh, A.~Taruya, T.~Hiramatsu and Y.~Himemoto,
Phys. Rev. D \textbf{73} (2006), 064006
doi:10.1103/PhysRevD.73.064006
[arXiv:gr-qc/0511145 [gr-qc]].

\bibitem{LIGOScientific:2016aoc}
B.~P.~Abbott \textit{et al.} [LIGO Scientific and Virgo],
Phys. Rev. Lett. \textbf{116} (2016) no.6, 061102
doi:10.1103/PhysRevLett.116.061102
[arXiv:1602.03837 [gr-qc]].

\bibitem{LIGOScientific:2017vwq}
B.~P.~Abbott \textit{et al.} [LIGO Scientific and Virgo],
Phys. Rev. Lett. \textbf{119} (2017) no.16, 161101
doi:10.1103/PhysRevLett.119.161101
[arXiv:1710.05832 [gr-qc]].

\bibitem{LIGOScientific:2018mvr}
B.~P.~Abbott \textit{et al.} [LIGO Scientific and Virgo],
Phys. Rev. X \textbf{9} (2019) no.3, 031040
doi:10.1103/PhysRevX.9.031040
[arXiv:1811.12907 [astro-ph.HE]].

\bibitem{LIGOScientific:2020ibl}
R.~Abbott \textit{et al.} [LIGO Scientific and Virgo],
Phys. Rev. X \textbf{11} (2021), 021053
doi:10.1103/PhysRevX.11.021053
[arXiv:2010.14527 [gr-qc]].

\bibitem{NANOGrav:2023gor}
G.~Agazie \textit{et al.} [NANOGrav],
Astrophys. J. Lett. \textbf{951} (2023) no.1, L8
doi:10.3847/2041-8213/acdac6
[arXiv:2306.16213 [astro-ph.HE]].

\bibitem{EPTA:2023fyk}
J.~Antoniadis \textit{et al.} [EPTA],
[arXiv:2306.16214 [astro-ph.HE]].

\bibitem{Yang:2021buu}
X.~Qi and H.~Sun,
Phys. Rev. D \textbf{107} (2023) no.9, 095026
doi:10.1103/PhysRevD.107.095026
[arXiv:2104.01045 [hep-ph]].

\bibitem{Arhrib:2011uy}
A.~Arhrib, R.~Benbrik, M.~Chabab, G.~Moultaka, M.~C.~Peyranere, L.~Rahili and J.~Ramadan,
Phys. Rev. D \textbf{84} (2011), 095005
doi:10.1103/PhysRevD.84.095005
[arXiv:1105.1925 [hep-ph]].

\bibitem{FileviezPerez:2008jbu}
P.~Fileviez Perez, T.~Han, G.~y.~Huang, T.~Li and K.~Wang,
Phys. Rev. D \textbf{78} (2008), 015018
doi:10.1103/PhysRevD.78.015018
[arXiv:0805.3536 [hep-ph]].

\bibitem{Kolb:1990vq}
E.~W.~Kolb and M.~S.~Turner,
Front. Phys. \textbf{69} (1990), 1-547
doi:10.1201/9780429492860

\bibitem{Hektor:2019ote}
A.~Hektor, A.~Hryczuk and K.~Kannike,
JHEP \textbf{03} (2019), 204
doi:10.1007/JHEP03(2019)204
[arXiv:1901.08074 [hep-ph]].

\bibitem{Alarcon:2012nr}
J.~M.~Alarcon, L.~S.~Geng, J.~Martin Camalich and J.~A.~Oller,
Phys. Lett. B \textbf{730} (2014), 342-346
doi:10.1016/j.physletb.2014.01.065
[arXiv:1209.2870 [hep-ph]].

\bibitem{Alloul:2013bka}
A.~Alloul, N.~D.~Christensen, C.~Degrande, C.~Duhr and B.~Fuks,
Comput. Phys. Commun. \textbf{185} (2014), 2250-2300
doi:10.1016/j.cpc.2014.04.012
[arXiv:1310.1921 [hep-ph]].

\bibitem{Belanger:2006is}
G.~Belanger, F.~Boudjema, A.~Pukhov and A.~Semenov,
Comput. Phys. Commun. \textbf{176} (2007), 367-382
doi:10.1016/j.cpc.2006.11.008
[arXiv:hep-ph/0607059 [hep-ph]].

\bibitem{Vaskonen:2016yiu}
V.~Vaskonen,
Phys. Rev. D \textbf{95} (2017) no.12, 123515
doi:10.1103/PhysRevD.95.123515
[arXiv:1611.02073 [hep-ph]].

\bibitem{Ellis:2022lft}
J.~Ellis, M.~Lewicki, M.~Merchand, J.~M.~No and M.~Zych,
JHEP \textbf{01} (2023), 093
doi:10.1007/JHEP01(2023)093
[arXiv:2210.16305 [hep-ph]].

\bibitem{Coleman:1973jx}
S.~R.~Coleman and E.~J.~Weinberg,
Phys. Rev. D \textbf{7} (1973), 1888-1910
doi:10.1103/PhysRevD.7.1888

\bibitem{Martin:2014bca}
S.~P.~Martin,
Phys. Rev. D \textbf{90} (2014) no.1, 016013
doi:10.1103/PhysRevD.90.016013
[arXiv:1406.2355 [hep-ph]].

\bibitem{Elias-Miro:2014pca}
J.~Elias-Miro, J.~R.~Espinosa and T.~Konstandin,
JHEP \textbf{08} (2014), 034
doi:10.1007/JHEP08(2014)034
[arXiv:1406.2652 [hep-ph]].

\bibitem{Baum:2020vfl}
S.~Baum, M.~Carena, N.~R.~Shah, C.~E.~M.~Wagner and Y.~Wang,
JHEP \textbf{03} (2021), 055
doi:10.1007/JHEP03(2021)055
[arXiv:2009.10743 [hep-ph]].

\bibitem{Dolan:1973qd}
L.~Dolan and R.~Jackiw,
Phys. Rev. D \textbf{9} (1974), 3320-3341
doi:10.1103/PhysRevD.9.3320

\bibitem{Weinberg:1974hy}
S.~Weinberg,
Phys. Rev. D \textbf{9} (1974), 3357-3378
doi:10.1103/PhysRevD.9.3357

\bibitem{Kirzhnits:1976ts}
D.~A.~Kirzhnits and A.~D.~Linde,
Annals Phys. \textbf{101} (1976), 195-238
doi:10.1016/0003-4916(76)90279-7

\bibitem{Espinosa:1992kf}
J.~R.~Espinosa, M.~Quiros and F.~Zwirner,
Phys. Lett. B \textbf{314} (1993), 206-216
doi:10.1016/0370-2693(93)90450-V
[arXiv:hep-ph/9212248 [hep-ph]].

\bibitem{Parwani:1991gq}
R.~R.~Parwani,
Phys. Rev. D \textbf{45} (1992), 4695
[erratum: Phys. Rev. D \textbf{48} (1993), 5965]
doi:10.1103/PhysRevD.45.4695
[arXiv:hep-ph/9204216 [hep-ph]].

\bibitem{Nielsen:1975fs}
N.~K.~Nielsen,
Nucl. Phys. B \textbf{101} (1975), 173-188
doi:10.1016/0550-3213(75)90301-6

\bibitem{Fukuda:1975di}
R.~Fukuda and T.~Kugo,
Phys. Rev. D \textbf{13} (1976), 3469
doi:10.1103/PhysRevD.13.3469

\bibitem{Laine:1994zq}
M.~Laine,
Phys. Rev. D \textbf{51} (1995), 4525-4532
doi:10.1103/PhysRevD.51.4525
[arXiv:hep-ph/9411252 [hep-ph]].

\bibitem{Baacke:1993aj}
J.~Baacke and S.~Junker,
Phys. Rev. D \textbf{49} (1994), 2055-2073
doi:10.1103/PhysRevD.49.2055
[arXiv:hep-ph/9308310 [hep-ph]].

\bibitem{Baacke:1994ix}
J.~Baacke and S.~Junker,
Phys. Rev. D \textbf{50} (1994), 4227-4228
doi:10.1103/PhysRevD.50.4227
[arXiv:hep-th/9402078 [hep-th]].

\bibitem{Garny:2012cg}
M.~Garny and T.~Konstandin,
JHEP \textbf{07} (2012), 189
doi:10.1007/JHEP07(2012)189
[arXiv:1205.3392 [hep-ph]].

\bibitem{Espinosa:2016nld}
J.~R.~Espinosa, M.~Garny, T.~Konstandin and A.~Riotto,
Phys. Rev. D \textbf{95} (2017) no.5, 056004
doi:10.1103/PhysRevD.95.056004
[arXiv:1608.06765 [hep-ph]].

\bibitem{Patel:2011th}
H.~H.~Patel and M.~J.~Ramsey-Musolf,
JHEP \textbf{07} (2011), 029
doi:10.1007/JHEP07(2011)029
[arXiv:1101.4665 [hep-ph]].

\bibitem{Arunasalam:2021zrs}
S.~Arunasalam and M.~J.~Ramsey-Musolf,
JHEP \textbf{08} (2022), 138
doi:10.1007/JHEP08(2022)138
[arXiv:2105.07588 [hep-ph]].

\bibitem{Lofgren:2021ogg}
J.~L\"ofgren, M.~J.~Ramsey-Musolf, P.~Schicho and T.~V.~I.~Tenkanen,
Phys. Rev. Lett. \textbf{130} (2023) no.25, 251801
doi:10.1103/PhysRevLett.130.251801
[arXiv:2112.05472 [hep-ph]].

\bibitem{Laine:2017hdk}
M.~Laine, M.~Meyer and G.~Nardini,
Nucl. Phys. B \textbf{920} (2017), 565-600
doi:10.1016/j.nuclphysb.2017.04.023
[arXiv:1702.07479 [hep-ph]].

\bibitem{Allen:1997ad}
B.~Allen and J.~D.~Romano,
Phys. Rev. D \textbf{59} (1999), 102001
doi:10.1103/PhysRevD.59.102001
[arXiv:gr-qc/9710117 [gr-qc]].

\bibitem{Caprini:2015zlo}
C.~Caprini, M.~Hindmarsh, S.~Huber, T.~Konstandin, J.~Kozaczuk, G.~Nardini, J.~M.~No, A.~Petiteau, P.~Schwaller and G.~Servant, \textit{et al.}
JCAP \textbf{04} (2016), 001
doi:10.1088/1475-7516/2016/04/001
[arXiv:1512.06239 [astro-ph.CO]].

\bibitem{Cai:2017cbj}
R.~G.~Cai, Z.~Cao, Z.~K.~Guo, S.~J.~Wang and T.~Yang,
Natl. Sci. Rev. \textbf{4} (2017) no.5, 687-706
doi:10.1093/nsr/nwx029
[arXiv:1703.00187 [gr-qc]].

\bibitem{Caprini:2018mtu}
C.~Caprini and D.~G.~Figueroa,
Class. Quant. Grav. \textbf{35} (2018) no.16, 163001
doi:10.1088/1361-6382/aac608
[arXiv:1801.04268 [astro-ph.CO]].

\bibitem{Romano:2016dpx}
J.~D.~Romano and N.~J.~Cornish,
Living Rev. Rel. \textbf{20} (2017) no.1, 2
doi:10.1007/s41114-017-0004-1
[arXiv:1608.06889 [gr-qc]].

\bibitem{Christensen:2018iqi}
N.~Christensen,
Rept. Prog. Phys. \textbf{82} (2019) no.1, 016903
doi:10.1088/1361-6633/aae6b5
[arXiv:1811.08797 [gr-qc]].

\bibitem{Espinosa:2010hh}
J.~R.~Espinosa, T.~Konstandin, J.~M.~No and G.~Servant,
JCAP \textbf{06} (2010), 028
doi:10.1088/1475-7516/2010/06/028
[arXiv:1004.4187 [hep-ph]].

\bibitem{Turner:1992tz}
M.~S.~Turner, E.~J.~Weinberg and L.~M.~Widrow,
Phys. Rev. D \textbf{46} (1992), 2384-2403
doi:10.1103/PhysRevD.46.2384

\bibitem{Wainwright:2011kj}
C.~L.~Wainwright,
Comput. Phys. Commun. \textbf{183} (2012), 2006-2013
doi:10.1016/j.cpc.2012.04.004
[arXiv:1109.4189 [hep-ph]].

\bibitem{Chiang:2019oms}
C.~W.~Chiang and B.~Q.~Lu,
JHEP \textbf{07} (2020), 082
doi:10.1007/JHEP07(2020)082
[arXiv:1912.12634 [hep-ph]].

\bibitem{Hindmarsh:2015qta}
M.~Hindmarsh, S.~J.~Huber, K.~Rummukainen and D.~J.~Weir,
Phys. Rev. D \textbf{92} (2015) no.12, 123009
doi:10.1103/PhysRevD.92.123009
[arXiv:1504.03291 [astro-ph.CO]].

\bibitem{Kosowsky:1992rz}
A.~Kosowsky, M.~S.~Turner and R.~Watkins,
Phys. Rev. Lett. \textbf{69} (1992), 2026-2029
doi:10.1103/PhysRevLett.69.2026

\bibitem{Kosowsky:1991ua}
A.~Kosowsky, M.~S.~Turner and R.~Watkins,
Phys. Rev. D \textbf{45} (1992), 4514-4535
doi:10.1103/PhysRevD.45.4514

\bibitem{Kosowsky:1992vn}
A.~Kosowsky and M.~S.~Turner,
Phys. Rev. D \textbf{47} (1993), 4372-4391
doi:10.1103/PhysRevD.47.4372
[arXiv:astro-ph/9211004 [astro-ph]].

\bibitem{Bodeker:2017cim}
D.~Bodeker and G.~D.~Moore,
JCAP \textbf{05} (2017), 025
doi:10.1088/1475-7516/2017/05/025
[arXiv:1703.08215 [hep-ph]].

\bibitem{Hindmarsh:2013xza}
M.~Hindmarsh, S.~J.~Huber, K.~Rummukainen and D.~J.~Weir,
Phys. Rev. Lett. \textbf{112} (2014), 041301
doi:10.1103/PhysRevLett.112.041301
[arXiv:1304.2433 [hep-ph]].

\bibitem{Giblin:2013kea}
J.~T.~Giblin, Jr. and J.~B.~Mertens,
JHEP \textbf{12} (2013), 042
doi:10.1007/JHEP12(2013)042
[arXiv:1310.2948 [hep-th]].

\bibitem{Giblin:2014qia}
J.~T.~Giblin and J.~B.~Mertens,
Phys. Rev. D \textbf{90} (2014) no.2, 023532
doi:10.1103/PhysRevD.90.023532
[arXiv:1405.4005 [astro-ph.CO]].

\bibitem{Schmitz:2020syl}
K.~Schmitz,
JHEP \textbf{01} (2021), 097
doi:10.1007/JHEP01(2021)097
[arXiv:2002.04615 [hep-ph]].

\bibitem{Caprini:2006jb}
C.~Caprini and R.~Durrer,
Phys. Rev. D \textbf{74} (2006), 063521
doi:10.1103/PhysRevD.74.063521
[arXiv:astro-ph/0603476 [astro-ph]].

\bibitem{Kahniashvili:2008pf}
T.~Kahniashvili, A.~Kosowsky, G.~Gogoberidze and Y.~Maravin,
Phys. Rev. D \textbf{78} (2008), 043003
doi:10.1103/PhysRevD.78.043003
[arXiv:0806.0293 [astro-ph]].

\bibitem{Kahniashvili:2008pe}
T.~Kahniashvili, L.~Campanelli, G.~Gogoberidze, Y.~Maravin and B.~Ratra,
Phys. Rev. D \textbf{78} (2008), 123006
[erratum: Phys. Rev. D \textbf{79} (2009), 109901]
doi:10.1103/PhysRevD.78.123006
[arXiv:0809.1899 [astro-ph]].

\bibitem{Kahniashvili:2009mf}
T.~Kahniashvili, L.~Kisslinger and T.~Stevens,
Phys. Rev. D \textbf{81} (2010), 023004
doi:10.1103/PhysRevD.81.023004
[arXiv:0905.0643 [astro-ph.CO]].

\bibitem{Caprini:2009yp}
C.~Caprini, R.~Durrer and G.~Servant,
JCAP \textbf{12} (2009), 024
doi:10.1088/1475-7516/2009/12/024
[arXiv:0909.0622 [astro-ph.CO]].

\bibitem{Kisslinger:2015hua}
L.~Kisslinger and T.~Kahniashvili,
Phys. Rev. D \textbf{92} (2015) no.4, 043006
doi:10.1103/PhysRevD.92.043006
[arXiv:1505.03680 [astro-ph.CO]].

\bibitem{DES:2017txv}
T.~M.~C.~Abbott \textit{et al.} [DES],
Mon. Not. Roy. Astron. Soc. \textbf{480} (2018) no.3, 3879-3888
doi:10.1093/mnras/sty1939
[arXiv:1711.00403 [astro-ph.CO]].

\bibitem{Hindmarsh:2019phv}
M.~Hindmarsh and M.~Hijazi,
JCAP \textbf{12} (2019), 062
doi:10.1088/1475-7516/2019/12/062
[arXiv:1909.10040 [astro-ph.CO]].

\bibitem{Guo:2020grp}
H.~K.~Guo, K.~Sinha, D.~Vagie and G.~White,
JCAP \textbf{01} (2021), 001
doi:10.1088/1475-7516/2021/01/001
[arXiv:2007.08537 [hep-ph]].

\bibitem{Hindmarsh:2020hop}
M.~B.~Hindmarsh, M.~L\"uben, J.~Lumma and M.~Pauly,
SciPost Phys. Lect. Notes \textbf{24} (2021), 1
doi:10.21468/SciPostPhysLectNotes.24
[arXiv:2008.09136 [astro-ph.CO]].

\bibitem{Pen:2015qta}
U.~L.~Pen and N.~Turok,
Phys. Rev. Lett. \textbf{117} (2016) no.13, 131301
doi:10.1103/PhysRevLett.117.131301
[arXiv:1510.02985 [astro-ph.CO]].

\bibitem{Hindmarsh:2017gnf}
M.~Hindmarsh, S.~J.~Huber, K.~Rummukainen and D.~J.~Weir,
Phys. Rev. D \textbf{96} (2017) no.10, 103520
[erratum: Phys. Rev. D \textbf{101} (2020) no.8, 089902]
doi:10.1103/PhysRevD.96.103520
[arXiv:1704.05871 [astro-ph.CO]].

\bibitem{No:2011fi}
J.~M.~No,
Phys. Rev. D \textbf{84} (2011), 124025
doi:10.1103/PhysRevD.84.124025
[arXiv:1103.2159 [hep-ph]].

\bibitem{Huitu:1996su}
K.~Huitu, J.~Maalampi, A.~Pietila and M.~Raidal,
Nucl. Phys. B \textbf{487} (1997), 27-42
doi:10.1016/S0550-3213(97)87466-4
[arXiv:hep-ph/9606311 [hep-ph]].

\bibitem{Chakrabarti:1998qy}
S.~Chakrabarti, D.~Choudhury, R.~M.~Godbole and B.~Mukhopadhyaya,
Phys. Lett. B \textbf{434} (1998), 347-353
doi:10.1016/S0370-2693(98)00743-6
[arXiv:hep-ph/9804297 [hep-ph]].

\bibitem{Chun:2003ej}
E.~J.~Chun, K.~Y.~Lee and S.~C.~Park,
Phys. Lett. B \textbf{566} (2003), 142-151
doi:10.1016/S0370-2693(03)00770-6
[arXiv:hep-ph/0304069 [hep-ph]].

\bibitem{Akeroyd:2005gt}
A.~G.~Akeroyd and M.~Aoki,
Phys. Rev. D \textbf{72} (2005), 035011
doi:10.1103/PhysRevD.72.035011
[arXiv:hep-ph/0506176 [hep-ph]].

\bibitem{Garayoa:2007fw}
J.~Garayoa and T.~Schwetz,
JHEP \textbf{03} (2008), 009
doi:10.1088/1126-6708/2008/03/009
[arXiv:0712.1453 [hep-ph]].

\bibitem{Kadastik:2007yd}
M.~Kadastik, M.~Raidal and L.~Rebane,
Phys. Rev. D \textbf{77} (2008), 115023
doi:10.1103/PhysRevD.77.115023
[arXiv:0712.3912 [hep-ph]].

\bibitem{Akeroyd:2007zv}
A.~G.~Akeroyd, M.~Aoki and H.~Sugiyama,
Phys. Rev. D \textbf{77} (2008), 075010
doi:10.1103/PhysRevD.77.075010
[arXiv:0712.4019 [hep-ph]].


\bibitem{delAguila:2008cj}
F.~del Aguila and J.~A.~Aguilar-Saavedra,
Nucl. Phys. B \textbf{813} (2009), 22-90
doi:10.1016/j.nuclphysb.2008.12.029
[arXiv:0808.2468 [hep-ph]].

\bibitem{Akeroyd:2009hb}
A.~G.~Akeroyd and C.~W.~Chiang,
Phys. Rev. D \textbf{80} (2009), 113010
doi:10.1103/PhysRevD.80.113010
[arXiv:0909.4419 [hep-ph]].

\bibitem{Melfo:2011nx}
A.~Melfo, M.~Nemevsek, F.~Nesti, G.~Senjanovic and Y.~Zhang,
Phys. Rev. D \textbf{85} (2012), 055018
doi:10.1103/PhysRevD.85.055018
[arXiv:1108.4416 [hep-ph]].

\bibitem{Aoki:2011pz}
M.~Aoki, S.~Kanemura and K.~Yagyu,
Phys. Rev. D \textbf{85} (2012), 055007
doi:10.1103/PhysRevD.85.055007
[arXiv:1110.4625 [hep-ph]].

\bibitem{Akeroyd:2011zza}
A.~G.~Akeroyd and H.~Sugiyama,
Phys. Rev. D \textbf{84} (2011), 035010
doi:10.1103/PhysRevD.84.035010
[arXiv:1105.2209 [hep-ph]].

\bibitem{Chiang:2012dk}
C.~W.~Chiang, T.~Nomura and K.~Tsumura,
Phys. Rev. D \textbf{85} (2012), 095023
doi:10.1103/PhysRevD.85.095023
[arXiv:1202.2014 [hep-ph]].

\bibitem{Chun:2012jw}
E.~J.~Chun, H.~M.~Lee and P.~Sharma,
JHEP \textbf{11} (2012), 106
doi:10.1007/JHEP11(2012)106
[arXiv:1209.1303 [hep-ph]].

\bibitem{Akeroyd:2012nd}
A.~G.~Akeroyd, S.~Moretti and H.~Sugiyama,
Phys. Rev. D \textbf{85} (2012), 055026
doi:10.1103/PhysRevD.85.055026
[arXiv:1201.5047 [hep-ph]].

\bibitem{Chun:2012zu}
E.~J.~Chun and P.~Sharma,
JHEP \textbf{08} (2012), 162
doi:10.1007/JHEP08(2012)162
[arXiv:1206.6278 [hep-ph]].

\bibitem{Dev:2013ff}
P.~S.~Bhupal Dev, D.~K.~Ghosh, N.~Okada and I.~Saha,
JHEP \textbf{03} (2013), 150
[erratum: JHEP \textbf{05} (2013), 049]
doi:10.1007/JHEP03(2013)150
[arXiv:1301.3453 [hep-ph]].

\bibitem{Banerjee:2013hxa}
S.~Banerjee, M.~Frank and S.~K.~Rai,
Phys. Rev. D \textbf{89} (2014) no.7, 075005
doi:10.1103/PhysRevD.89.075005
[arXiv:1312.4249 [hep-ph]].

\bibitem{delAguila:2013mia}
F.~del \'Aguila and M.~Chala,
JHEP \textbf{03} (2014), 027
doi:10.1007/JHEP03(2014)027
[arXiv:1311.1510 [hep-ph]].

\bibitem{Chun:2013vma}
E.~J.~Chun and P.~Sharma,
Phys. Lett. B \textbf{728} (2014), 256-261
doi:10.1016/j.physletb.2013.11.056
[arXiv:1309.6888 [hep-ph]].

\bibitem{Kanemura:2013vxa}
S.~Kanemura, K.~Yagyu and H.~Yokoya,
Phys. Lett. B \textbf{726} (2013), 316-319
doi:10.1016/j.physletb.2013.08.054
[arXiv:1305.2383 [hep-ph]].

\bibitem{Kanemura:2014goa}
S.~Kanemura, M.~Kikuchi, K.~Yagyu and H.~Yokoya,
Phys. Rev. D \textbf{90} (2014) no.11, 115018
doi:10.1103/PhysRevD.90.115018
[arXiv:1407.6547 [hep-ph]].

\bibitem{Kanemura:2014ipa}
S.~Kanemura, M.~Kikuchi, H.~Yokoya and K.~Yagyu,
PTEP \textbf{2015} (2015), 051B02
doi:10.1093/ptep/ptv071
[arXiv:1412.7603 [hep-ph]].

\bibitem{kang:2014jia}
Z.~Kang, J.~Li, T.~Li, Y.~Liu and G.~Z.~Ning,
Eur. Phys. J. C \textbf{75} (2015) no.12, 574
doi:10.1140/epjc/s10052-015-3774-1
[arXiv:1404.5207 [hep-ph]].

\bibitem{Han:2015hba}
Z.~L.~Han, R.~Ding and Y.~Liao,
Phys. Rev. D \textbf{91} (2015), 093006
doi:10.1103/PhysRevD.91.093006
[arXiv:1502.05242 [hep-ph]].

\bibitem{Han:2015sca}
Z.~L.~Han, R.~Ding and Y.~Liao,
Phys. Rev. D \textbf{92} (2015) no.3, 033014
doi:10.1103/PhysRevD.92.033014
[arXiv:1506.08996 [hep-ph]].

\bibitem{Das:2016bir}
D.~Das and A.~Santamaria,
Phys. Rev. D \textbf{94} (2016) no.1, 015015
doi:10.1103/PhysRevD.94.015015
[arXiv:1604.08099 [hep-ph]].

\bibitem{Babu:2016rcr}
K.~S.~Babu and S.~Jana,
Phys. Rev. D \textbf{95} (2017) no.5, 055020
doi:10.1103/PhysRevD.95.055020
[arXiv:1612.09224 [hep-ph]].

\bibitem{Mitra:2016wpr}
M.~Mitra, S.~Niyogi and M.~Spannowsky,
Phys. Rev. D \textbf{95} (2017) no.3, 035042
doi:10.1103/PhysRevD.95.035042
[arXiv:1611.09594 [hep-ph]].

\bibitem{Cai:2017mow}
Y.~Cai, T.~Han, T.~Li and R.~Ruiz,
Front. in Phys. \textbf{6} (2018), 40
doi:10.3389/fphy.2018.00040
[arXiv:1711.02180 [hep-ph]].

\bibitem{Ghosh:2017pxl}
D.~K.~Ghosh, N.~Ghosh, I.~Saha and A.~Shaw,
Phys. Rev. D \textbf{97} (2018) no.11, 115022
doi:10.1103/PhysRevD.97.115022
[arXiv:1711.06062 [hep-ph]].

\bibitem{Crivellin:2018ahj}
A.~Crivellin, M.~Ghezzi, L.~Panizzi, G.~M.~Pruna and A.~Signer,
Phys. Rev. D \textbf{99} (2019) no.3, 035004
doi:10.1103/PhysRevD.99.035004
[arXiv:1807.10224 [hep-ph]].

\bibitem{Du:2018eaw}
Y.~Du, A.~Dunbrack, M.~J.~Ramsey-Musolf and J.~H.~Yu,
JHEP \textbf{01} (2019), 101
doi:10.1007/JHEP01(2019)101
[arXiv:1810.09450 [hep-ph]].

\bibitem{Dev:2018kpa}
P.~S.~Bhupal Dev and Y.~Zhang,
JHEP \textbf{10} (2018), 199
doi:10.1007/JHEP10(2018)199
[arXiv:1808.00943 [hep-ph]].

\bibitem{Antusch:2018svb}
S.~Antusch, O.~Fischer, A.~Hammad and C.~Scherb,
JHEP \textbf{02} (2019), 157
doi:10.1007/JHEP02(2019)157
[arXiv:1811.03476 [hep-ph]].

\bibitem{Aboubrahim:2018tpf}
A.~Aboubrahim and P.~Nath,
Phys. Rev. D \textbf{98} (2018) no.9, 095024
doi:10.1103/PhysRevD.98.095024
[arXiv:1810.12868 [hep-ph]].

\bibitem{deMelo:2019asm}
T.~B.~de Melo, F.~S.~Queiroz and Y.~Villamizar,
Int. J. Mod. Phys. A \textbf{34} (2019) no.27, 1950157
doi:10.1142/S0217751X19501574
[arXiv:1909.07429 [hep-ph]].

\bibitem{Primulando:2019evb}
R.~Primulando, J.~Julio and P.~Uttayarat,
JHEP \textbf{08} (2019), 024
doi:10.1007/JHEP08(2019)024
[arXiv:1903.02493 [hep-ph]].

\bibitem{Padhan:2019jlc}
R.~Padhan, D.~Das, M.~Mitra and A.~Kumar Nayak,
Phys. Rev. D \textbf{101} (2020) no.7, 075050
doi:10.1103/PhysRevD.101.075050
[arXiv:1909.10495 [hep-ph]].

\bibitem{Chun:2019hce}
E.~J.~Chun, S.~Khan, S.~Mandal, M.~Mitra and S.~Shil,
Phys. Rev. D \textbf{101} (2020) no.7, 075008
doi:10.1103/PhysRevD.101.075008
[arXiv:1911.00971 [hep-ph]].

\bibitem{Ashanujjaman:2021txz}
S.~Ashanujjaman and K.~Ghosh,
JHEP \textbf{03} (2022), 195
doi:10.1007/JHEP03(2022)195
[arXiv:2108.10952 [hep-ph]].

\bibitem{Mandal:2022zmy}
S.~Mandal, O.~G.~Miranda, G.~Sanchez Garcia, J.~W.~F.~Valle and X.~J.~Xu,
Phys. Rev. D \textbf{105} (2022) no.9, 095020
doi:10.1103/PhysRevD.105.095020
[arXiv:2203.06362 [hep-ph]].

\bibitem{Dutta:2014dba}
B.~Dutta, R.~Eusebi, Y.~Gao, T.~Ghosh and T.~Kamon,
Phys. Rev. D \textbf{90} (2014), 055015
doi:10.1103/PhysRevD.90.055015
[arXiv:1404.0685 [hep-ph]].

\bibitem{ATLAS:2012hi}
G.~Aad \textit{et al.} [ATLAS],
Eur. Phys. J. C \textbf{72} (2012), 2244
doi:10.1140/epjc/s10052-012-2244-2
[arXiv:1210.5070 [hep-ex]].

\bibitem{Chatrchyan:2012ya}
S.~Chatrchyan \textit{et al.} [CMS],
Eur. Phys. J. C \textbf{72} (2012), 2189
doi:10.1140/epjc/s10052-012-2189-5
[arXiv:1207.2666 [hep-ex]].

\bibitem{ATLAS:2014kca}
G.~Aad \textit{et al.} [ATLAS],
JHEP \textbf{03} (2015), 041
doi:10.1007/JHEP03(2015)041
[arXiv:1412.0237 [hep-ex]].

\bibitem{Khachatryan:2014sta}
V.~Khachatryan \textit{et al.} [CMS],
Phys. Rev. Lett. \textbf{114} (2015) no.5, 051801
doi:10.1103/PhysRevLett.114.051801
[arXiv:1410.6315 [hep-ex]].

\bibitem{CMS:2016cpz}
 [CMS],
CMS-PAS-HIG-14-039.

\bibitem{CMS:2017pet}
 [CMS],
CMS-PAS-HIG-16-036.

\bibitem{Aaboud:2017qph}
M.~Aaboud \textit{et al.} [ATLAS],
Eur. Phys. J. C \textbf{78} (2018) no.3, 199
doi:10.1140/epjc/s10052-018-5661-z
[arXiv:1710.09748 [hep-ex]].

\bibitem{CMS:2017fhs}
A.~M.~Sirunyan \textit{et al.} [CMS],
Phys. Rev. Lett. \textbf{120} (2018) no.8, 081801
doi:10.1103/PhysRevLett.120.081801
[arXiv:1709.05822 [hep-ex]].

\bibitem{Aaboud:2018qcu}
M.~Aaboud \textit{et al.} [ATLAS],
Eur. Phys. J. C \textbf{79} (2019) no.1, 58
doi:10.1140/epjc/s10052-018-6500-y
[arXiv:1808.01899 [hep-ex]].

\bibitem{Aad:2021lzu}
G.~Aad \textit{et al.} [ATLAS],
JHEP \textbf{06} (2021), 146
doi:10.1007/JHEP06(2021)146
[arXiv:2101.11961 [hep-ex]].

\bibitem{Ghosh:2018drw}
T.~Ghosh, S.~Jana and S.~Nandi,
Phys. Rev. D \textbf{97} (2018) no.11, 115037
doi:10.1103/PhysRevD.97.115037
[arXiv:1802.09251 [hep-ph]].

\bibitem{Baer:2014kya}
H.~Baer, A.~Mustafayev and X.~Tata,
Phys. Rev. D \textbf{90} (2014) no.11, 115007
doi:10.1103/PhysRevD.90.115007
[arXiv:1409.7058 [hep-ph]].

\bibitem{Dutta:2012xe}
B.~Dutta, A.~Gurrola, W.~Johns, T.~Kamon, P.~Sheldon and K.~Sinha,
Phys. Rev. D \textbf{87} (2013) no.3, 035029
doi:10.1103/PhysRevD.87.035029
[arXiv:1210.0964 [hep-ph]].

\bibitem{Dutta:2014jda}
B.~Dutta, T.~Ghosh, A.~Gurrola, W.~Johns, T.~Kamon, P.~Sheldon, K.~Sinha, K.~Wang and S.~Wu,
Phys. Rev. D \textbf{91} (2015) no.5, 055025
doi:10.1103/PhysRevD.91.055025
[arXiv:1411.6043 [hep-ph]].

\bibitem{Ajaib:2015yma}
M.~A.~Ajaib, B.~Dutta, T.~Ghosh, I.~Gogoladze and Q.~Shafi,
Phys. Rev. D \textbf{92} (2015) no.7, 075033
doi:10.1103/PhysRevD.92.075033
[arXiv:1505.05896 [hep-ph]].

\bibitem{Dutta:2017nqv}
B.~Dutta, K.~Fantahun, A.~Fernando, T.~Ghosh, J.~Kumar, P.~Sandick, P.~Stengel and J.~W.~Walker,
Phys. Rev. D \textbf{96} (2017) no.7, 075037
doi:10.1103/PhysRevD.96.075037
[arXiv:1706.05339 [hep-ph]].

\bibitem{ATLAS:2016neq}
G.~Aad \textit{et al.} [ATLAS and CMS],
JHEP \textbf{08} (2016), 045
doi:10.1007/JHEP08(2016)045
[arXiv:1606.02266 [hep-ex]].

\bibitem{CMS:2018uag}
A.~M.~Sirunyan \textit{et al.} [CMS],
Eur. Phys. J. C \textbf{79} (2019) no.5, 421
doi:10.1140/epjc/s10052-019-6909-y
[arXiv:1809.10733 [hep-ex]].

\bibitem{CMS:2018yfx}
A.~M.~Sirunyan \textit{et al.} [CMS],
Phys. Lett. B \textbf{793} (2019), 520-551
doi:10.1016/j.physletb.2019.04.025
[arXiv:1809.05937 [hep-ex]].

\bibitem{Arhrib:2011vc}
A.~Arhrib, R.~Benbrik, M.~Chabab, G.~Moultaka and L.~Rahili,
JHEP \textbf{04} (2012), 136
doi:10.1007/JHEP04(2012)136
[arXiv:1112.5453 [hep-ph]].

\bibitem{Gunion:1989we}
J.~F.~Gunion, H.~E.~Haber, G.~L.~Kane and S.~Dawson,
Front. Phys. \textbf{80} (2000), 1-404
SCIPP-89/13.

\bibitem{ATLAS:2022tnm}
G.~Aad \textit{et al.} [ATLAS],
JHEP \textbf{07} (2023), 088
doi:10.1007/JHEP07(2023)088
[arXiv:2207.00348 [hep-ex]].

\bibitem{CMS:2021kom}
A.~M.~Sirunyan \textit{et al.} [CMS],
JHEP \textbf{07} (2021), 027
doi:10.1007/JHEP07(2021)027
[arXiv:2103.06956 [hep-ex]].

\bibitem{Carilli:2004nx}
C.~L.~Carilli and S.~Rawlings,
New Astron. Rev. \textbf{48} (2004), 979
doi:10.1016/j.newar.2004.09.001
[arXiv:astro-ph/0409274 [astro-ph]].

\bibitem{Janssen:2014dka}
G.~Janssen, G.~Hobbs, M.~McLaughlin, C.~Bassa, A.~T.~Deller, M.~Kramer, K.~Lee, C.~Mingarelli, P.~Rosado and S.~Sanidas, \textit{et al.}
PoS \textbf{AASKA14} (2015), 037
doi:10.22323/1.215.0037
[arXiv:1501.00127 [astro-ph.IM]].

\bibitem{Weltman:2018zrl}
A.~Weltman, P.~Bull, S.~Camera, K.~Kelley, H.~Padmanabhan, J.~Pritchard, A.~Raccanelli, S.~Riemer-S\o{}rensen, L.~Shao and S.~Andrianomena, \textit{et al.}
Publ. Astron. Soc. Austral. \textbf{37} (2020), e002
doi:10.1017/pasa.2019.42
[arXiv:1810.02680 [astro-ph.CO]].

\bibitem{McLaughlin:2013ira}
M.~A.~McLaughlin,
Class. Quant. Grav. \textbf{30} (2013), 224008
doi:10.1088/0264-9381/30/22/224008
[arXiv:1310.0758 [astro-ph.IM]].

\bibitem{Ghorbani:2018yfr}
K.~Ghorbani and P.~H.~Ghorbani,
J. Phys. G \textbf{47} (2020) no.1, 015201
doi:10.1088/1361-6471/ab4823
[arXiv:1804.05798 [hep-ph]].

\bibitem{Comelli:1996vm}
D.~Comelli and J.~R.~Espinosa,
Phys. Rev. D \textbf{55} (1997), 6253-6263
doi:10.1103/PhysRevD.55.6253
[arXiv:hep-ph/9606438 [hep-ph]].

\bibitem{Basler:2018cwe}
P.~Basler and M.~M\"uhlleitner,
Comput. Phys. Commun. \textbf{237} (2019), 62-85
doi:10.1016/j.cpc.2018.11.006
[arXiv:1803.02846 [hep-ph]].

\bibitem{Carrington:1991hz}
M.~E.~Carrington,
Phys. Rev. D \textbf{45} (1992), 2933-2944
doi:10.1103/PhysRevD.45.2933

\bibitem{Kannike:2012pe}
K.~Kannike,
Eur. Phys. J. C \textbf{72} (2012), 2093
doi:10.1140/epjc/s10052-012-2093-z
[arXiv:1205.3781 [hep-ph]].

\bibitem{Belanger:2012zr}
G.~Belanger, K.~Kannike, A.~Pukhov and M.~Raidal,
JCAP \textbf{01} (2013), 022
doi:10.1088/1475-7516/2013/01/022
[arXiv:1211.1014 [hep-ph]].

\bibitem{Lerner:2009xg}
R.~N.~Lerner and J.~McDonald,
Phys. Rev. D \textbf{80} (2009), 123507
doi:10.1103/PhysRevD.80.123507
[arXiv:0909.0520 [hep-ph]].

\bibitem{Lee:1977eg}
B.~W.~Lee, C.~Quigg and H.~B.~Thacker,
Phys. Rev. D \textbf{16} (1977), 1519
doi:10.1103/PhysRevD.16.1519

\bibitem{Esteban:2020cvm}
I.~Esteban, M.~C.~Gonzalez-Garcia, M.~Maltoni, T.~Schwetz and A.~Zhou,
JHEP \textbf{09} (2020), 178
doi:10.1007/JHEP09(2020)178
[arXiv:2007.14792 [hep-ph]].

\bibitem{Dev:2021axj}
P.~S.~B.~Dev, B.~Dutta, T.~Ghosh, T.~Han, H.~Qin and Y.~Zhang,
JHEP \textbf{03} (2022), 068
doi:10.1007/JHEP03(2022)068
[arXiv:2109.04490 [hep-ph]].

\bibitem{Kakizaki:2003jk}
M.~Kakizaki, Y.~Ogura and F.~Shima,
Phys. Lett. B \textbf{566} (2003), 210-216
doi:10.1016/S0370-2693(03)00833-5
[arXiv:hep-ph/0304254 [hep-ph]].

\bibitem{Akeroyd:2009nu}
A.~G.~Akeroyd, M.~Aoki and H.~Sugiyama,
Phys. Rev. D \textbf{79} (2009), 113010
doi:10.1103/PhysRevD.79.113010
[arXiv:0904.3640 [hep-ph]].

\bibitem{Dinh:2012bp}
D.~N.~Dinh, A.~Ibarra, E.~Molinaro and S.~T.~Petcov,
JHEP \textbf{08} (2012), 125
[erratum: JHEP \textbf{09} (2013), 023]
doi:10.1007/JHEP08(2012)125
[arXiv:1205.4671 [hep-ph]].

\bibitem{Bellgardt:1987du}
U.~Bellgardt \textit{et al.} [SINDRUM],
Nucl. Phys. B \textbf{299} (1988), 1-6
doi:10.1016/0550-3213(88)90462-2

\bibitem{TheMEG:2016wtm}
A.~M.~Baldini \textit{et al.} [MEG],
Eur. Phys. J. C \textbf{76} (2016) no.8, 434
doi:10.1140/epjc/s10052-016-4271-x
[arXiv:1605.05081 [hep-ex]].

\bibitem{Aoki:2012jj}
M.~Aoki, S.~Kanemura, M.~Kikuchi and K.~Yagyu,
Phys. Rev. D \textbf{87} (2013) no.1, 015012
doi:10.1103/PhysRevD.87.015012
[arXiv:1211.6029 [hep-ph]].

\bibitem{ParticleDataGroup:2020ssz}
P.~A.~Zyla \textit{et al.} [Particle Data Group],
PTEP \textbf{2020} (2020) no.8, 083C01
doi:10.1093/ptep/ptaa104





\end{thebibliography}
\end{document}